\documentclass[12pt,a4paper]{article}
\usepackage{amssymb}
\usepackage{amsmath}
\usepackage{latexsym}
\usepackage{cite}
\usepackage{physics}
\usepackage{float}
\usepackage{graphicx}
\usepackage{url}
\begin{document}

\title{\vspace{-1cm}\bf Remarks on quantum scalar field theory\\ for a uniformly accelerated reference frame\\ and a Schwarzschild black hole}

\author{
Mikhail~N.~Smolyakov
\\
{\small{\em Skobeltsyn Institute of Nuclear Physics, Lomonosov Moscow
State University,
}}\\
{\small{\em Moscow 119991, Russia}}}

\date{}
\maketitle

\begin{abstract}
In the present paper, the cases of a uniformly accelerated reference frame and a Schwarzschild black hole in two and four dimensions are discussed focusing on formal aspects of the corresponding quantum scalar field theories. The main emphasis is made on building consistent quantum field theories and highlighting the problems that are not considered elsewhere. In particular, some pathologies in the case of a uniformly accelerated reference frame are discussed in detail: it is shown that some quantum states that are well defined in Minkowski spacetime possess an infinite energy when considered in Rindler spacetime. In the case of a Schwarzschild black hole in four dimensions, it is shown that the standard theory providing the Unruh vacuum state turns out to be inconsistent. The problem of consistency of quantum scalar field theory in Kruskal-Szekeres spacetime is also discussed.
\end{abstract}

\tableofcontents

\section{Introduction}
The problem of quantization of fields in curved backgrounds is widely discussed in modern scientific literature. After the classical papers \cite{Boulware:1974dm,HH}, the main focus is made on various types of black holes. However, there exist other objects and systems which deserve examination. One of such systems is the uniformly accelerated reference frame in flat background metric leading to the well-known Fulling-Davies-Unruh effect \cite{Fulling:1972md,Davies:1974th,Unruh:1976db}. In many ways the cases of a uniformly accelerated reference frame and a Schwarzschild black hole are very similar, moreover, they represent the simplest systems allowing one to build quantum field theories in a quite standard manner (except for the four-dimensional Schwarzschild black hole in Kruskal-Szekeres spacetime, this problem will be discusses separately). The present paper is devoted exactly to these systems.

The number of papers on this subject is very large, so foundations of the corresponding quantum field theories can be found in many books and reviews; see, for example, \cite{BD,Traschen:1999zr,Jacobson:2003vx,Crispino:2007eb,Buoninfante:2024oxl}. However, some less obvious aspects of these theories are still not discussed, at least in the most known papers and reviews. Several such problems are considered in the present paper. The focus is made on the quantum field theories proper without going to phenomenological consequences, experimental manifestations and problems of interaction of particles with detectors, as well as to adjacent areas like thermodynamics. I do not consider collapsing objects, only simple eternal systems like a uniformly accelerated reference frame and a Schwarzschild black hole in two and four dimensions are discussed. The simplest real massless scalar field is chosen as the object of study. The approach of canonical quantization is used throughout the present paper because it seems to be more transparent and illustrative for the purposes of the analysis than the path integral formalism.

Since the case of a uniformly accelerated reference frame is somewhat simpler than the case of a Schwarzschild black hole (in particular, there exist exact analytical solutions of the corresponding equations of motion in both Minkowski and Rindler spacetimes in two and four dimensions), the analysis begins with a detailed description of the Fulling-Davies-Unruh effect including all necessary calculations. Even though descriptions of this effect can be found elsewhere, it was decided anyway to reproduce these well-known results. The reasons for this are the following. First, it is convenient to have all the necessary basic results at hand and in a form that, at least in my opinion, seems to be more convenient or transparent for the purposes of the present paper. Indeed, the analysis of problems and pathologies existing in the quantum and even classical theories under consideration (in particular, a seeming paradox arising when one uses the standard description of particles by plane waves) is based on the use of some formulas which arise when one describes the Fulling-Davies-Unruh effect, so it is better to have the corresponding formulas with derivation at hand in the notations used in the present paper. In particular, I have failed to find a straightforward calculation of the Bogolyubov coefficients for the four-dimensional Fulling-Davies-Unruh effect, even though the coefficients proper and some indirect methods of calculation (namely, on the horizon \cite{Crispino:2007eb}) can be found in the literature. Second, some of the known results can be obtained in a different way than it is usually done. For example, this applies to the derivation of coordinate-dependent vacuum energy densities using specific representation of delta function, which seems to be quite illustrative. And third, the attention is paid to the points that are important when one considers formal aspects of quantum field theory. In particular, the Hamiltonian of the system is considered from the very beginning instead of the particle number or just the energy-momentum tensor, and the canonical commutation relations are discussed in more detail than it is usually done providing some important observations. The latter turns out to be useful for the analysis of quantum field theory in Kruskal-Szekeres spacetime.

Except the reproduction of the standard results for uniformly accelerated reference frame, a detailed discussion of the Unruh vacuum state in the cases of a two-dimensional uniformly accelerated reference frame and a two-dimensional Schwarzschild black hole is presented. Since the quantum field theories in both cases are almost identical, the calculations are performed for the case of a uniformly accelerated reference frame because this setup turns out to be more mathematically rigorous, even though one cannot consider such a theory as a physically relevant one. An important result is that the canonical commutation relations, whose fulfillment is necessary for a quantum field theory to be correct and consistent, are indeed fulfilled in such a theory, which is not obvious taking into account how the theory is constructed. This result is an important and illustrative example of a well-defined theory providing the Unruh vacuum state, in contrast to the case of a four-dimensional Schwarzschild black hole in which at least the standard theory providing the Unruh vacuum state turns out to be inconsistent. It is interesting to note that, as will be also demonstrated below, it is in principle impossible to construct a theory providing the Unruh vacuum state in the case of a four-dimensional uniformly accelerated reference frame.

Even though the quantum scalar field theories for two-dimensional and four-dimensional uniformly accelerated reference frames are well defined, there arise already mentioned pathologies in Rindler spacetime when one considers somewhat more complicated states than the vacuum one. Namely, it turns out that the standard one-particle states and some normalized quantum states in Minkowski spacetime have infinite energies in Rindler spacetime. These problems are discussed in detail in the corresponding part of the paper.

The part of the paper devoted to the four-dimensional Schwarzschild black hole starts with a description of the quantum scalar field theory in Schwarzschild spacetime. Even though the results presented in this section are well-known, some of them will be used in the subsequent sections, so again it is better to have the corresponding description at hand than just to refer to other papers. Then a more detailed examination of the theory providing the Unruh vacuum state is performed, explicitly demonstrating that the canonical commutation relations are not satisfied in this theory thus leading to a conclusion that such a quantum field theory cannot be considered as a correct one.

Finally, a quantum scalar field theory in Kruskal-Szekeres spacetime is considered. It is well known that there exist problems with building scalar field theories in this spacetime in the standard way, which are consequences of the form of the metric. In particular, the Hamiltonian constructed in the standard way is not conserved over time, whereas the equation of motion for the scalar field does not allow one to perform the usual separation of variables and to get solutions describing stationary states. However, the analysis presented in this part of the paper suggests that there is a good reason to believe that there exists a consistent quantum scalar field theory satisfying the necessary canonical commutation relations even though the form of the modes in such a theory (including that on the horizon) still is not clear.

\section{Fulling-Davies-Unruh effect for a massless scalar field in two-dimensional spacetime}
Let us start with the simplest case of a uniformly accelerated reference frame in two dimensions, leading to the Fulling-Davies-Unruh effect \cite{Fulling:1972md,Davies:1974th,Unruh:1976db}.\footnote{Very often this effect is named simply the ``Unruh effect'', however, here the title ``Fulling-Davies-Unruh effect'' will be used, among other things, in order to avoid possible confusion with the term ``Unruh vacuum''.} In this section, foundations of the Fulling-Davies-Unruh effect, including all necessary calculations, will be presented. A more detailed analysis, including various consequences such as interaction of particles with detectors, can be found elsewhere; see, for example, book \cite{BD} and review \cite{Crispino:2007eb}.

\subsection{Setup}
Let us consider a two-dimensional massless scalar field with the action
\begin{equation}\label{scalaction}
S=\int\mathcal{L}\,dx^{0}dx^{1}=\frac{1}{2}\int\sqrt{-g}\,g^{\mu\nu}\partial_{\mu}\phi\,\partial_{\nu}\phi\,dx^{0}dx^{1}.
\end{equation}
The covariant conservation law
\begin{equation}\label{tcons}
\nabla_{\mu}T^{\mu}_{\nu}=\frac{1}{\sqrt{-g}}\frac{\partial\left(\sqrt{-g}\,T^{\mu}_{\nu}\right)}{\partial x^{\mu}}-\frac{1}{2}\frac{\partial g_{\mu\sigma}}{\partial x^{\nu}}T^{\mu\sigma}=0
\end{equation}
is satisfied for any energy-momentum tensor \cite{LL-FT}. In the cases with static metric, relation \eqref{tcons} for $\nu=0$ leads to
\begin{equation}\label{tconsstatic}
\frac{d}{dx^{0}}\left(\int\sqrt{-g}\,T^{0}_{0}dx^{1}\right)=0.
\end{equation}
Thus we can define the Hamiltonian of the system as
\begin{equation}\label{Hamilt}
H=\int\sqrt{-g}\,g^{00}T_{00}dx^{1}.
\end{equation}
We suppose that the quantized field satisfies the standard canonical commutation relations
\begin{align}\label{CCR}
&[\phi(x^{0},x^{1}),\pi(x^{0},{x^{1}}')]=i\delta(x^{1}-{x^{1}}'),\\
&[\phi(x^{0},x^{1}),\phi(x^{0},{x^{1}}')]=0,\\
&[\pi(x^{0},x^{1}),\pi(x^{0},{x^{1}}')]=0,
\end{align}
where $\pi(x^{0},x^{1})$ is the canonically conjugate momentum. In what follows, for the two-dimen\-sional case the notations $x^{0}=T$, $x^{1}=X$ and $x^{0}=t$, $x^{1}=x$ will be used.

First, let us consider the case of the Minkowski metric
\begin{equation}\label{metric-M}
ds^2=dT^2-dX^2,
\end{equation}
leading to the following equation of motion for the scalar field:
\begin{equation}\label{eqm-M}
\frac{\partial^{2}\phi}{\partial T^{2}}-\frac{\partial^{2}\phi}{\partial X^{2}}=0.
\end{equation}
In Minkowski spacetime, the scalar field will be denoted as $\phi_{M}(T,X)$. We can take the standard expansion of quantized scalar field
\begin{equation}\label{scfieldM}
\phi_{M}(T,X)=\int\limits_{-\infty}^{\infty}\frac{dk}{\sqrt{4\pi|k|}}\left(e^{-i|k|T}e^{ikX}a^{}(k)+
e^{i|k|T}e^{-ikX}a^{\dagger}(k)\right),
\end{equation}
where the creation and annihilation operators satisfy the standard commutation relations
\begin{equation}\label{CCRa}
[a^{}(k),a^{\dagger}(k')]=\delta(k-k'),
\end{equation}
all other commutators being equal to zero. The vacuum of the theory in Minkowski spacetime is defined as
\begin{equation}\label{Mvacuum}
a^{}(k)\ket{0_{M}}=0.
\end{equation}
From definition of Hamiltonian \eqref{Hamilt} one gets
\begin{equation}\label{H-M}
H_{M}=\frac{1}{2}\int\left(\left(\frac{\partial\phi}{\partial T}\right)^{2}+\left(\frac{\partial\phi}{\partial X}\right)^{2}\right)dX.
\end{equation}
Substituting \eqref{scfieldM} into \eqref{H-M}, we arrive at
\begin{equation}\label{HamiltMinkdef}
H_{M}=\int\limits_{-\infty}^{\infty}|k|a^{\dagger}(k)a^{}(k)dk,
\end{equation}
where the infinite $c$-number term is omitted.

By definition, the canonically conjugate momentum is
\begin{equation}\label{ccm1}
\pi(T,X)=\frac{\partial\mathcal{L}}{\partial\left(\frac{\partial\phi(T,X)}{\partial T}\right)}=\frac{\partial\phi(T,X)}{\partial T}.
\end{equation}
It is clear that in such a theory the canonical commutation relations
\begin{align}\label{CCR-M}
&\left[\phi_{M}(T,X),\frac{\partial\phi_{M}(T,X')}{\partial T}\right]=i\delta(X-X'),\\
&\left[\phi_{M}(T,X),\phi_{M}(T,X')\right]=0,\\\label{CCR-M3}
&\left[\frac{\partial\phi_{M}(T,X)}{\partial T},\frac{\partial\phi_{M}(T,X')}{\partial T}\right]=0
\end{align}
are satisfied. Although all these steps of canonical quantization are well known and can be found in any textbook on quantum field theory, the explicit formulas presented above will be used in the subsequent analysis.

\begin{figure}[ht]
\centering
\includegraphics[width=0.5\linewidth]{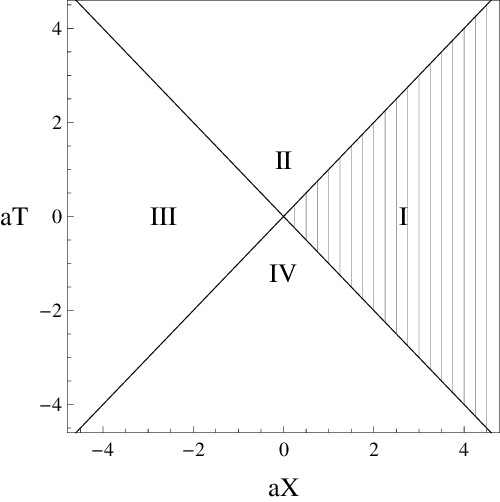}
\caption{Wedge I (hatched area) corresponds to the Rindler coordinates $t$ and $x$ in \eqref{TRind} and \eqref{XRind}.}\label{fig1}
\end{figure}

Now let us turn to Rindler spacetime. The coordinate transformations, leading to the Rindler metric, have the form
\begin{align}\label{TRind}
&T=\frac{1}{a}\,e^{ax}\sinh(at),\\\label{XRind}
&X=\frac{1}{a}\,e^{ax}\cosh(at),
\end{align}
resulting in
\begin{equation}\label{metric-R}
ds^2=e^{2ax}\left(dt^2-dx^2\right).
\end{equation}
The coordinates $t$, $x$ in \eqref{TRind} and \eqref{XRind} are defined in wedge I in Fig.~\Ref{fig1}.
Two-dimensional metric \eqref{metric-R} is conformally flat, so the field equation also has the simple form
\begin{equation}\label{eqm-R}
\frac{\partial^{2}\phi}{\partial t^{2}}-\frac{\partial^{2}\phi}{\partial x^{2}}=0.
\end{equation}
Thus we can also use the standard expansion of quantized scalar field
\begin{equation}\label{scfieldR}
\phi_{R}(t,x)=\int\limits_{-\infty}^{\infty}\frac{dp}{\sqrt{4\pi|p|}}\left(e^{-i|p|t}e^{ipx}b^{}(p)+
e^{i|p|t}e^{-ipx}b^{\dagger}(p)\right),
\end{equation}
where $\phi_{R}(t,x)$ denotes the scalar field in Rindler spacetime and
\begin{equation}
[b^{}(p),b^{\dagger}(p')]=\delta(p-p'),
\end{equation}
all other commutators being equal to zero. The vacuum of the theory in Rindler spacetime is defined as
\begin{equation}\label{Rindvac2D}
b^{}(p)\ket{0_{R}}=0.
\end{equation}
From definition of Hamiltonian \eqref{Hamilt} one gets
\begin{equation}\label{H-R}
H_{R}=\frac{1}{2}\int\left(\left(\frac{\partial\phi}{\partial t}\right)^{2}+\left(\frac{\partial\phi}{\partial x}\right)^{2}\right)dx.
\end{equation}
Substituting \eqref{scfieldR} into \eqref{H-R}, we arrive at
\begin{equation}\label{H-Rb+b}
H_{R}=\frac{1}{2}\int\limits_{-\infty}^{\infty}|p|\left(b^{\dagger}(p)b^{}(p)+b^{}(p)b^{\dagger}(p)\right)dp\to
\int\limits_{-\infty}^{\infty}|p|b^{\dagger}(p)b^{}(p)dp,
\end{equation}
where the infinite $c$-number term
\begin{equation}\label{vacERind2D}
\frac{1}{2}\int\limits_{-\infty}^{\infty}|p|\delta(p-p)\,dp
\end{equation}
is omitted.

By definition, the canonically conjugate momentum is
\begin{equation}\label{ccm2}
\pi(t,x)=\frac{\partial\mathcal{L}}{\partial\left(\frac{\partial\phi(t,x)}{\partial t}\right)}=\sqrt{-g(x)}\,g^{00}(x)\frac{\partial\phi(t,x)}{\partial t}=\frac{\partial\phi(t,x)}{\partial t},
\end{equation}
The standard canonical commutation relations
\begin{align}\label{CCR-R}
&\left[\phi_{R}(t,x),\frac{\partial\phi_{R}(t,x')}{\partial t}\right]=i\delta(x-x'),\\
&\left[\phi_{R}(t,x),\phi_{R}(t,x')\right]=0,\\\label{CCR-R3}
&\left[\frac{\partial\phi_{R}(t,x)}{\partial t},\frac{\partial\phi_{R}(t,x')}{\partial t}\right]=0
\end{align}
are also satisfied in Rindler spacetime.

\subsection{Two representations of a massless scalar field in the Rindler coordinates and Bogolyubov coefficients}
Since the value of the scalar field remains the same in different coordinate systems, the relation
\begin{equation}\label{equivMR}
\phi_{R}(t,x)\equiv\phi_{M}(T(t,x),X(t,x)),
\end{equation}
where $T$ and $X$ are defined by \eqref{TRind} and \eqref{XRind} respectively, holds in wedge I (see Fig.~\Ref{fig1}). Equality \eqref{equivMR} implies that the scalar field in Rindler spacetime can be represented in two different forms. Let us define the light cone coordinates
\begin{equation}\label{light-cone-uv}
u=t-x,\qquad v=t+x.
\end{equation}
With \eqref{light-cone-uv}, representation \eqref{scfieldR} can be rewritten as
\begin{equation}\label{phiRpm}
\phi_{R}(t,x)=\int\limits_{0}^{\infty}\frac{dp}{\sqrt{4\pi p}}\left(e^{-ipu}b_{+}^{}(p)+
e^{ipu}b_{+}^{\dagger}(p)\right)+\int\limits_{-\infty}^{0}\frac{dp}{\sqrt{4\pi|p|}}\left(e^{ipv}b_{-}^{}(p)+
e^{-ipv}b_{-}^{\dagger}(p)\right),
\end{equation}
where
\begin{equation}
b_{+}^{}(p)=b^{}(p)\theta(p),\qquad b_{-}^{}(p)=b^{}(p)\theta(-p).
\end{equation}
On the other hand, with \eqref{TRind} and \eqref{XRind} scalar field \eqref{scfieldM} in wedge I can be represented as
\begin{align}\nonumber
\phi_{M}(T(t,x),X(t,x))&=\int\limits_{0}^{\infty}\frac{dk}{\sqrt{4\pi k}}\left(e^{i\frac{k}{a}e^{-au}}a_{+}^{}(k)+
e^{-i\frac{k}{a}e^{-au}}a_{+}^{\dagger}(k)\right)\\\label{PhiMRind2D}
&+\int\limits_{-\infty}^{0}\frac{dk}{\sqrt{4\pi|k|}}\left(e^{i\frac{k}{a}e^{av}}a_{-}^{}(k)+
e^{-i\frac{k}{a}e^{av}}a_{-}^{\dagger}(k)\right),
\end{align}
where
\begin{equation}
a_{+}^{}(k)=a^{}(k)\theta(k),\qquad a_{-}^{}(k)=a^{}(k)\theta(-k).
\end{equation}

The corresponding Bogolyubov transformations connecting the operators $a^{}(k), a^{\dagger}(k)$ with the operators $b^{}(p), b^{\dagger}(p)$ take the form
\begin{align}\label{ba1final}
&b_{+}^{}(p)=\int\limits_{0}^{\infty}\left(a_{+}^{}(k)A_{+}^{}(k,p)+a_{+}^{\dagger}(k)A_{+}^{*}(k,-p)\right)\sqrt{\frac{p}{k}}\,dk,\quad p>0,\\\label{ba2final}
&b_{-}^{}(p)=\int\limits_{-\infty}^{0}\left(a_{-}^{}(k)A_{-}^{}(k,p)+a_{-}^{\dagger}(k)A_{-}^{*}(k,-p)\right)\sqrt{\frac{|p|}{|k|}}\,dk,\quad p<0
\end{align}
with
\begin{align}\label{A+kp}
&A_{+}^{}(k,p)=\frac{1}{2\pi a}\,e^{-i\frac{p}{a}\ln\left(\frac{a}{k}\right)+\frac{\pi p}{2a}}\,\Gamma\left(-i\frac{p}{a}\right),\\\label{A-kp}
&A_{-}^{}(k,p)=\frac{1}{2\pi a}\,e^{-i\frac{p}{a}\ln\left(\frac{a}{|k|}\right)-\frac{\pi p}{2a}}\,\Gamma\left(-i\frac{p}{a}\right),
\end{align}
see detailed calculations in Appendix~A (of course, these Bogolyubov coefficients were calculated earlier, see, for example, \cite{Lee:1985rp}).

\subsection{The Minkowski spacetime vacuum energy in Rindler spacetime}\label{Sec2-4}
Suppose that the physical vacuum is the Minkowski one defined by \eqref{Mvacuum}. Then the vacuum energy in Rindler spacetime can be calculated as (strictly speaking, it is the mean energy)
\begin{equation}
E_{vac}=\frac{\bra{0_{M}}H_{R}\ket{0_{M}}}{\bra{0_{M}}\ket{0_{M}}}.
\end{equation}
Note that from here and below the Hamiltonian of the system, which is well defined for a consistent quantum field theory, will be considered. The energy defined by the Hamiltonian seems to be a more appropriate value for calculation than the particle number or similar quantities which are used in many papers on this subject. In particular, the operator of particle number cannot be constructed from the field itself and its derivatives, unlike the Hamiltonian defined by \eqref{H-R}. Since the Hamiltonian is conserved over time and accounts for the energy in the whole space at a fixed moment of time, this is its advantage that simplifies interpretation of the obtained results and makes some of them clearer.

Since $\bra{0_{M}}\ket{0_{M}}=1$, one gets
\begin{align}\nonumber
E_{vac}=\int\limits_{-\infty}^{\infty}|p|\bra{0_{M}}b^{\dagger}(p)b(p)\ket{0_{M}}dp\\
=\int\limits_{0}^{\infty}p\bra{0_{M}}b_{+}^{\dagger}(p)b_{+}^{}(p)\ket{0_{M}}dp
-\int\limits_{-\infty}^{0}p\bra{0_{M}}b_{-}^{\dagger}(p)b_{-}^{}(p)\ket{0_{M}}dp.
\end{align}
Using formulas \eqref{ba1final} and \eqref{ba2final}, we get for the vacuum energy
\begin{align}\nonumber
E_{vac}
=&\int\limits_{0}^{\infty}\int\limits_{0}^{\infty}A_{+}^{*}(k,-p)A_{+}^{}(k,-p)\frac{p^{2}}{k}\,dkdp\\\label{vacenergy}
-&\int\limits_{-\infty}^{0}\int\limits_{-\infty}^{0}A_{-}^{*}(k,-p)A_{-}^{}(k,-p)\frac{p^{2}}{k}\,dkdp.
\end{align}
First, let us calculate the term
\begin{equation}\label{vacenergy+}
\bra{0_{M}}H_{R+}\ket{0_{M}}
=\int\limits_{0}^{\infty}\int\limits_{0}^{\infty}A_{+}^{*}(k,-p)A_{+}^{}(k,-p)\frac{p^{2}}{k}\,dkdp.
\end{equation}
At this step, it is convenient to rewrite \eqref{vacenergy+} as
\begin{equation}\label{vacenergy+delta}
\int\limits_{0}^{\infty}\int\limits_{0}^{\infty}A_{+}^{*}(k,-p)A_{+}^{}(k,-p)\frac{p^{2}}{k}\,dkdp
=\int\limits_{0}^{\infty}\int\limits_{0}^{\infty}\int\limits_{0}^{\infty}A_{+}^{*}(k,-p)A_{+}^{}(k,-p')\delta(p-p')\frac{p^{2}}{k}\,dk\,dp\,dp'.
\end{equation}
Substituting \eqref{A+kp} into the rhs of \eqref{vacenergy+delta},
one gets
\begin{equation}\label{vacenergy+delta2}
\bra{0_{M}}H_{R+}\ket{0_{M}}
=\int\limits_{0}^{\infty}\int\limits_{0}^{\infty}\int\limits_{0}^{\infty}
\,e^{i\frac{p'-p}{a}\ln\left(\frac{a}{k}\right)}e^{-\frac{\pi(p+p')}{2a}}\,\Gamma\left(i\frac{p'}{a}\right)\Gamma\left(-i\frac{p}{a}\right)
\delta(p-p')\frac{p^{2}}{4\pi^{2}a^{2}k}\,dk\,dp\,dp'.
\end{equation}
Let us calculate the integral over $k$ in \eqref{vacenergy+delta2}. This can be easily done, resulting in
\begin{equation}\label{vacenergy+intk}
\int\limits_{0}^{\infty}
\frac{e^{i\frac{p'-p}{a}\ln\left(\frac{a}{k}\right)}}{k}\,dk=2\pi a \delta(p-p').
\end{equation}
Using \eqref{vacenergy+intk} and the relation \cite{Korn-Korn}
\begin{equation}
\Gamma\left(\xi\right)\Gamma\left(-\xi\right)=-\frac{\pi}{\xi\sin(\pi\xi)},
\end{equation}
formula \eqref{vacenergy+delta2} can be rewritten as
\begin{align}\nonumber
&\bra{0_{M}}H_{R+}\ket{0_{M}}
=2\pi a\int\limits_{0}^{\infty}e^{-\frac{\pi p}{a}}\Gamma\left(i\frac{p}{a}\right)\Gamma\left(-i\frac{p}{a}\right)\frac{p^{2}\delta(p-p)}{4\pi^{2}a^{2}}\,dp\\\label{vacenergy+result}
&=\int\limits_{0}^{\infty}e^{-\frac{\pi p}{a}}\left(-\frac{\pi}{i\frac{p}{a}\sin\left(i\pi\frac{p}{a}\right)}\right)\frac{p^{2}\delta(p-p)}{2\pi a}\,dp=\int\limits_{0}^{\infty}\frac{p\,\delta(p-p)}{e^{\frac{2\pi p}{a}}-1}\,dp.
\end{align}

A fully analogous procedure can be performed for the second term in \eqref{vacenergy}:
\begin{equation}\label{vacenergy-result}
\bra{0_{M}}H_{R-}\ket{0_{M}}
=-\int\limits_{-\infty}^{0}\int\limits_{-\infty}^{0}A_{-}^{*}(k,-p)A_{-}^{}(k,-p)\frac{p^{2}}{k}\,dkdp
=\int\limits_{-\infty}^{0}\frac{|p|\delta(p-p)}{e^{\frac{2\pi |p|}{a}}-1}\,dp.
\end{equation}
Combining \eqref{vacenergy+result} and \eqref{vacenergy-result}, we arrive at the well-known result
\begin{equation}\label{vacenergyresult0}
E_{vac}
=\int\limits_{-\infty}^{\infty}\frac{|p|\delta(p-p)}{e^{\frac{2\pi |p|}{a}}-1}\,dp.
\end{equation}
We see that the total energy is infinite because of the existence of $\delta(p-p)$ in \eqref{vacenergyresult0}. According to the orthonormality condition for the modes in \eqref{scfieldR}, we have
\begin{equation}\label{delta-l}
\delta(p-p)=\frac{1}{2\pi}\int\limits_{-\infty}^{\infty}e^{i(p-p)x}dx=\frac{1}{2\pi}\int\limits_{-\infty}^{\infty}dx,
\end{equation}
which does not depend on $p$. We see that $2\pi\delta(p-p)$ is just the volume $\int_{-\infty}^{\infty}dx$, see also \cite{Lee:1985rp} (as we will see below, in the four-dimensional case the situation turns out to be more involved and such a delta function should be treated more accurately). Thus \eqref{vacenergyresult0} can be rewritten as
\begin{equation}\label{vacenergyresult}
E_{vac}
=\frac{1}{2\pi}\int\limits_{-\infty}^{\infty}dx\int\limits_{-\infty}^{\infty}\frac{|p|}{e^{\frac{2\pi |p|}{a}}-1}\,dp.
\end{equation}
We can define the Minkowski spacetime vacuum energy density as
\begin{equation}\label{vacdens2D}
\rho_{vac\,\,2D}=\frac{1}{2\pi}\int\limits_{-\infty}^{\infty}\frac{|p|}{e^{\frac{2\pi |p|}{a}}-1}\,dp=\frac{a^{2}}{24\pi},
\end{equation}
which turns out to be finite. One can check that formula \eqref{vacdens2D} represents the one-dimensional Stefan-Boltzmann law \cite{DeVos,LandsbergDeVos}. Note that definition of energy density \eqref{vacdens2D} corresponds to a ``field-theoretical'' picture with effective flat metric in Eq.~\eqref{eqm-R}. Of course, actual metric \eqref{metric-R} is non-flat, so the energy density defined with respect to the metric has a different form.

Note that the structure of \eqref{vacenergyresult0} is the same as the structure of the vacuum energy described by the infinite $c$-number term represented by \eqref{vacERind2D}. The difference is that the integral over $p$ in \eqref{vacenergyresult} is finite, whereas the integral over $p$ in \eqref{vacERind2D} diverges. However, even though usually the focus is made on the thermal spectrum in \eqref{vacenergyresult0}, the corresponding delta function is also very important, the latter will become clearer in the four-dimensional case.

It should be mentioned that instead of calculating the Bogolyubov coefficients explicitly, very often a different approach is used when one considers uniformly accelerated reference frames or black holes. This approach was proposed in \cite{Unruh:1976db}, it is based on the use of solutions provided by two different theories in Rindler (Schwarzschild) spacetimes corresponding to wedges I and III of Fig.~\ref{fig1}. The idea is that certain linear combinations of the solutions corresponding to different theories possess the same analytic properties as the actual solutions in Minkowski (Kruskal-Szekeres) spacetime, thus providing the same vacuum state as the one in Minkowski (Kruskal-Szekeres) spacetime. The essence of the method is nicely explained in review \cite{Jacobson:2003vx}, a detailed discussion of the method in connection with the Fulling-Davies-Unruh effect can be found in \cite{Crispino:2007eb,Unruh:1976db,Unruh:1983ms} (see also \cite{BD} for a more general discussion of the method). Even though these linear combinations do not correspond to actual wave functions of the modes in Minkowski (Kruskal-Szekeres) spacetime, they allow one to properly define the vacuum state of the theory at least in the cases of a uniformly accelerated reference frame or a two-dimensional Schwarzschild black hole (as will be shown below, the case of a four-dimensional Schwarzschild black hole turns out to be more involved) using only the creation and annihilation operators of the theories in Rindler (Schwarzschild) spacetimes corresponding to wedges I and III of Fig.~\Ref{fig1}. In what follows, I will not use this method preferring to use Bogolyubov transformations whenever possible.

\subsection{Canonical commutation relations}\label{CCR2Dsection}
It is easy to check that if commutation relations \eqref{CCRa} for the operators $a^{}(k)$, $a^{\dagger}(k)$ are satisfied, the commutation relations
\begin{align}\label{crb1}
&[b^{}(p),b^{}(p')]=0,\\
&[b^{\dagger}(p),b^{\dagger}(p')]=0,\\\label{crb3}
&[b^{}(p),b^{\dagger}(p')]=\delta(p-p')
\end{align}
are also satisfied for the operators $b^{}(p)$, $b^{\dagger}(p)$ defined by \eqref{ba1final}, \eqref{ba2final} with \eqref{A+kp}, \eqref{A-kp}. This leads to the following observation, which will be useful when we move on to discussion of the four-dimensional Schwarzschild black hole. Indeed, since commutation relations \eqref{CCR-M}--\eqref{CCR-M3} and commutation relations \eqref{CCR-R}--\eqref{CCR-R3} are satisfied, equality \eqref{equivMR} suggests that the commutation relations in Rindler spacetime
\begin{align}\label{CCR-MR}
&\left[\phi_{M}(T(t,x),X(t,x)),\frac{\partial\phi_{M}(T(t,x'),X(t,x'))}{\partial t}\right]=i\delta(x-x'),\\\label{CCR-MR2a}
&\left[\phi_{M}(T(t,x),X(t,x)),\phi_{M}(T(t,x'),X(t,x'))\right]=0,\\\label{CCR-MR3}
&\left[\frac{\partial\phi_{M}(T(t,x),X(t,x))}{\partial t},\frac{\partial\phi_{M}(T(t,x'),X(t,x'))}{\partial t}\right]=0
\end{align}
and the commutation relations in Minkowski spacetime
\begin{align}\label{CCR-MR2}
&\left[\phi_{R}(t(T,X),x(T,X)),\frac{\partial\phi_{R}(t(T,X'),x(T,X'))}{\partial T}\right]=i\delta(X-X'),\\
&\left[\phi_{R}(t(T,X),x(T,X)),\phi_{R}(t(T,X'),x(T,X'))\right]=0,\\\label{CCR-MR2-3}
&\left[\frac{\partial\phi_{R}(t(T,X),x(T,X))}{\partial T},\frac{\partial\phi_{R}(t(T,X'),x(T,X'))}{\partial T}\right]=0
\end{align}
also hold. Although the fulfillment of \eqref{CCR-MR}--\eqref{CCR-MR3} or \eqref{CCR-MR2}--\eqref{CCR-MR2-3} is quite obvious, the explicit check of these commutation relations can be illustrative. One can find verification of \eqref{CCR-MR}--\eqref{CCR-MR3} in Appendix~B. Analogous commutation relations will be useful in examining the quantum scalar field theory in Kruskal-Szekeres spacetime.

\section[Connection with a two-dimensional Schwarzschild black hole]{Connection with a two-dimensional Schwarzschild\\ black hole}
A two-dimensional Schwarzschild black hole is described by the metric
\begin{equation}\label{metric-Schw}
ds^2=\left(1-\frac{r_{s}}{r(r_{*})}\right)\left(dt^2-d{r_{*}}^2\right),
\end{equation}
where
\begin{equation}\label{tortoise}
r_{*}=r+r_{s}\ln\left(\frac{r}{r_{s}}-1\right),\qquad r_{*}\in (-\infty,\infty)
\end{equation}
is the standard tortoise coordinate for $r>r_{s}$ and $r_{s}$ is the Schwarzschild radius. It is obtained from the standard four-dimensional Schwarzschild solution by omitting the angular part of the metric.
\begin{figure}[ht]
\centering
\includegraphics[width=0.5\linewidth]{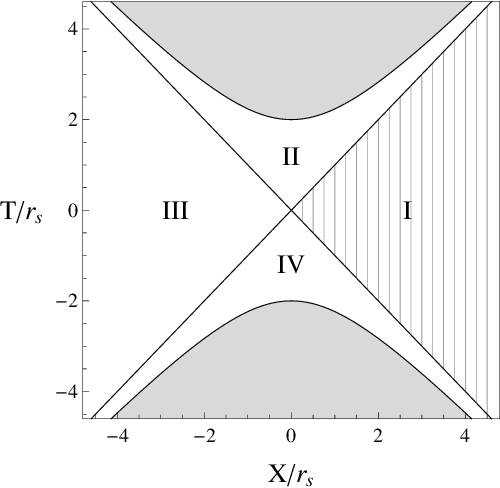}
\caption{The Kruskal-Szekeres plane. Hatched area corresponds to Schwarzschild spacetime for $r>r_{s}$, filled areas correspond to forbidden regions with $T^{2}-X^{2}>4r_{s}^{2}$ (the lines $T^{2}-X^{2}=4r_{s}^{2}$ correspond to $r=0$).}\label{fig4}
\end{figure}
Metric in the Kruskal–Szekeres coordinates \cite{Kruskal,Szekeres} in wedge I (see Fig.~\ref{fig4}) has the form
\begin{equation}\label{KS}
ds^2=\frac{r_{s}e^{-\frac{r(T,X)}{r_{s}}}}{r(T,X)}\left(dT^2-dX^2\right),
\end{equation}
where
\begin{align}\label{Trstart}
&T=2\,r_{s}e^{\frac{r_{*}}{2r_{s}}}\sinh\left(\frac{t}{2\,r_{s}}\right),\\\label{Xrstart}
&X=2\,r_{s}e^{\frac{r_{*}}{2r_{s}}}\cosh\left(\frac{t}{2\,r_{s}}\right),
\end{align}
and
\begin{equation}\label{rstar}
r_{*}=r_{s}\ln\left(\frac{X^{2}-T^{2}}{4r_{s}^{2}}\right).
\end{equation}
Note that, contrary to the case of Rindler spacetime, metric \eqref{KS} explicitly depends on time $T$.

Since the metrics in both the Kruskal-Szekeres plane and Schwarzschild spacetime (wedge I in Fig.~\ref{fig4}) in two dimensions are conformally flat, the equations of motion for the two-dimensional massless scalar field take the form
\begin{equation}
\frac{\partial^{2}\phi}{\partial T^{2}}-\frac{\partial^{2}\phi}{\partial X^{2}}=0
\end{equation}
in the Kruskal-Szekeres plane and
\begin{equation}
\frac{\partial^{2}\phi}{\partial t^{2}}-\frac{\partial^{2}\phi}{\partial {r_{*}}^{2}}=0
\end{equation}
in Schwarzschild spacetime, like Eqs.~\eqref{eqm-M} and \eqref{eqm-R}. Thus naively one can use the same plane wave solutions like those in \eqref{scfieldM} and \eqref{scfieldR}. However, contrary to the case of Minkowski spacetime (see Fig.~\ref{fig1}), in the Kruskal-Szekeres plane there exist forbidden regions $T^{2}-X^{2}>4r_{s}^{2}$ (see Fig.~\ref{fig4}), where the lines $T^{2}-X^{2}=4r_{s}^{2}$ correspond to $r=0$. The latter implies imposition of a boundary condition for the scalar field in the Kruskal-Szekeres plane at $T^{2}-X^{2}=4r_{s}^{2}$. For example, one can consider
\begin{equation}\label{phirzero}
\frac{\partial\phi(t,r)}{\partial r}\biggl|_{r=0}=0.
\end{equation}
Passing to the tortoise coordinate for $r<r_{s}$
\begin{equation}
y=-r-r_{s}\ln\left(1-\frac{r}{r_{s}}\right),\qquad r\in (0,r_{s}),\qquad y\in (0,\infty),
\end{equation}
boundary condition \eqref{phirzero} can be transformed into
\begin{equation}
\frac{\partial\phi(t,r)}{\partial r}=\frac{\partial\phi(t,r(y))}{\partial y}\frac{\partial y}{\partial r}
=\frac{\partial\phi(t,r(y))}{\partial y}\frac{r}{\left(r_{s}-r\right)}.
\end{equation}
If
\begin{equation}
\frac{\partial\phi(t,r(y))}{\partial y}\biggl|_{y=0}
\end{equation}
is finite, then \eqref{phirzero} holds automatically. Let us consider the coordinate transformations
\begin{align}
&T=2r_{s}e^{-\frac{y}{2r_{s}}}\cosh\left(\frac{t}{2r_{s}}\right),\\
&X=-2r_{s}e^{-\frac{y}{2r_{s}}}\sinh\left(\frac{t}{2r_{s}}\right),
\end{align}
which are valid in wedge II. Then
\begin{equation}
\frac{\partial\phi(t,r(y))}{\partial y}\biggl|_{y=0}
=-\frac{1}{2r_{s}}\left(T\,\frac{\partial\phi(T,X)}{\partial T}+X\,\frac{\partial\phi(T,X)}{\partial X}\right)\biggl|_{T^{2}-X^{2}=4r_{s}^{2}}.
\end{equation}
Analogous calculations can be performed for wedge IV. Thus regular solutions of the field $\phi(T,X)$ lead to \eqref{phirzero}. So we can formally continue the equation to the regions $T^{2}-X^{2}>4r_{s}^{2}$ (i.e., we expand the spacetime for the equation of motion for the scalar field to these regions) and take the same solution that was used in the case of two-dimensional Minkowski spacetime as if there were no forbidden regions on the Kruskal-Szekeres plane. Since plane wave solutions are regular at $T^{2}-X^{2}=4r_{s}^{2}$, boundary condition \eqref{phirzero} is fulfilled. Note that this is just a mathematical trick, in this case the formal existence of solutions in the forbidden regions $T^{2}-X^{2}>4r_{s}^{2}$ can be considered just as a mathematical artifact. Then in the physical area, which does not include the forbidden regions,\footnote{Note that technically for our analysis we need explicit solutions only in wedge I (both in Rindler and Kruskal–Szekeres planes, see Figs.~\ref{fig1} and \ref{fig4}), not in wedges II--IV.} the solution for the scalar field is supposed to have the form of plane waves. Even though these reasonings are not rather rigorous from a physical point of view,\footnote{It is quite possible that the actual solution to the problem of forbidden regions is much more complicated, resulting in solutions for the modes that differ significantly from simple plane waves even in wedge I.} technically with additional condition \eqref{phirzero} all the results for the two-dimensional Rindler spacetime obtained above (and those that will be obtained below in the two-dimensional case) can be applied to the case of a two-dimensional Schwarzschild black hole just by changing
\begin{equation}
a\rightarrow\frac{1}{2\,r_{s}}.
\end{equation}

It should be noted that even though the metric \eqref{KS} depends on time $T$ and thus $\frac{\partial g_{\mu\sigma}}{\partial T}\neq 0$, for a massless scalar field in two dimensions the term
\begin{equation}\label{unwanted}
\frac{1}{2}\frac{\partial g_{\mu\sigma}}{\partial T}\,T^{\mu\sigma}
\end{equation}
in \eqref{tcons} turns out to be equal to zero, which allows one to use Hamiltonian \eqref{Hamilt} that is conserved over time.\footnote{It does not work in the four-dimensional case.} Indeed,
\begin{equation}
\frac{1}{2}\frac{\partial g_{\mu\sigma}}{\partial T}\,T^{\mu\sigma}=\frac{1}{2}\left(\frac{\partial g_{00}}{\partial T}\,T^{00}+\frac{\partial g_{11}}{\partial T}\,T^{11}\right).
\end{equation}
Given that $g_{11}=-g_{00}$, we get
\begin{equation}\label{tmunuextraterm}
\frac{1}{2}\frac{\partial g_{00}}{\partial T}\left(T^{00}-T^{11}\right).
\end{equation}
With
\begin{align}\label{T00scalar}
&T_{00}=\frac{1}{2}\left(\frac{\partial\phi}{\partial T}\right)^{2}+\frac{1}{2}\left(\frac{\partial\phi}{\partial X}\right)^{2},
\\\label{T11scalar}
&T_{11}=\frac{1}{2}\left(\frac{\partial\phi}{\partial T}\right)^{2}+\frac{1}{2}\left(\frac{\partial\phi}{\partial X}\right)^{2}=T_{00},
\end{align}
for \eqref{tmunuextraterm} we get
\begin{equation}
\frac{1}{2}\frac{\partial g_{00}}{\partial T}\left(g^{00}g^{00}T_{00}-g^{11}g^{11}T_{11}\right)=\frac{1}{2}\frac{\partial g_{00}}{\partial T}g^{00}g^{00}\left(T_{00}-T_{11}\right)=0.
\end{equation}
The latter is necessary to build a consistent scalar quantum field theory. Meanwhile, due to the cancellation of ``unwanted'' term \eqref{unwanted}, the two-dimensional case turns out to be very specific, whereas a more realistic case of the four-dimensional Schwarzschild black hole appears to be more involved.

Now let us discuss different vacuum states connected with the Schwarzschild black holes. There are three vacuum states that are considered for black holes \cite{BD}: it is the Boulware vacuum state \cite{Boulware:1974dm} which is analogous to the Rindler vacuum $\ket{0_{R}}$ \eqref{Rindvac2D}; the Hartle-Hawking vacuum state \cite{HH} which is analogous to the Minkowski vacuum $\ket{0_{M}}$ \eqref{Mvacuum}; and the Unruh vacuum state \cite{Unruh:1976db}. The latter is used to model Hawking radiation in the case of eternal black holes. For a two-dimensional massless scalar field the theory providing the Unruh vacuum state can be easily constructed: technically, one should take the field in the form (see, for example, \cite{Anderson:2022icz,Balbinot:2023vcm})
\begin{equation}\label{Unruhtheordef2DBH}
\phi_{U}(T,X,t,x)=\int\limits_{0}^{\infty}\frac{dk}{\sqrt{4\pi k}}\left(e^{-ikU}c_{+}^{}(k)+
e^{ikU}c_{+}^{\dagger}(k)\right)+\int\limits_{-\infty}^{0}\frac{dk}{\sqrt{4\pi|k|}}\left(e^{ikv}c_{-}^{}(k)+
e^{-ikv}c_{-}^{\dagger}(k)\right),
\end{equation}
where
\begin{equation}
U=T-X,\qquad v=t+r_{*}.
\end{equation}
We see that the field consists of modes from completely different sets. The Unruh vacuum state is defined as
\begin{equation}
c_{+}^{}(k)\ket{0_{U}}=c_{-}^{}(k)\ket{0_{U}}=0
\end{equation}
for all $k$.

Detailed calculations (for a two-dimensional uniformly accelerated reference frame) will be presented in the next section, so here only the final result is presented. The Unruh vacuum energy has the form
\begin{equation}\label{Uvacfinal2DSchw}
E_{vacU}=\frac{1}{2\pi}\int\limits_{-\infty}^{\infty}dr_{*}\int\limits_{0}^{\infty}\frac{p}{e^{4\pi r_{s}p}-1}\,dp.
\end{equation}
The total energy \eqref{Uvacfinal2DSchw} is infinite and remains the same at any moment of time.

\section{The Unruh vacuum state for a uniformly accelerated reference frame (and a Schwarz\-schild black hole) in two-dimen\-sional spacetime}
As has already been noted, in the case of a Schwarzschild black hole three different vacua can be considered, one of which is the Unruh vacuum. Below a full analogue of the Unruh vacuum in the case of a uniformly accelerated reference frame will be constructed in the two-dimensional case and consistency of the resulting quantum field theory will be checked. Since in the case of a massless scalar field in two dimensions the field theories in Rindler and Schwarzschild spacetimes are almost identical, it is clear that all calculations that will be performed below in Rindler spacetime are applicable to the case of a two-dimensional Schwarzschild black hole too.

An important remark is in order here. Technically, the Unruh vacuum state is defined within the framework of a theory that unifies modes from different spacetimes, see \eqref{Unruhtheordef2DBH}. In connection with the uniformly accelerated reference frame, the latter means that in fact definition of the Unruh vacuum state depends on the acceleration of the reference frame with respect to the initial Minkowski spacetime. Of course, from a physical point of view it is hard to believe that the vacuum state of a quantum theory depends on the acceleration of an observer, so the Unruh vacuum state is not considered here as a physical one (in such a case, the only physically reasonable vacuum state is the one provided by the quantum theory in Minkowski spacetime, it is defined by \eqref{Mvacuum}). However, even though the Unruh vacuum state is inherent to black holes, from a purely mathematical point of view the Unruh vacuum state for a uniformly accelerated reference frame turns out to be more rigorous (recall the problem with forbidden regions in Kruskal-Szekeres spacetime, see Fig.~\ref{fig4}) and illustrative. Indeed, later it will allow us not only to see the differences between the two-dimensional and four-dimensional cases but to apply some observations to the case of a four-dimensional Schwarzschild black hole.

\subsection{Two representations of a two-dimensional massless scalar field in the Rindler coordinates}
Using the light cone coordinates
\begin{equation}
U=T-X,\qquad V=T+X
\end{equation}
the scalar field in Minkowski spacetime can be represented as
\begin{equation}\label{phiMpm}
\phi_{M}(T,X)=\int\limits_{0}^{\infty}\frac{dk}{\sqrt{4\pi k}}\left(e^{-ikU}a_{+}^{}(k)+
e^{ikU}a_{+}^{\dagger}(k)\right)+\int\limits_{-\infty}^{0}\frac{dk}{\sqrt{4\pi|k|}}\left(e^{ikV}a_{-}^{}(k)+
e^{-ikV}a_{-}^{\dagger}(k)\right),
\end{equation}
where
\begin{equation}
a_{+}^{}(k)=a^{}(k)\theta(k),\qquad a_{-}^{}(k)=a^{}(k)\theta(-k).
\end{equation}
The idea of a theory providing the Unruh vacuum state \cite{Unruh:1976db} is that the modes coming from the past horizon are positive frequency modes with respect to $U$, whereas the modes coming from infinity are positive frequency modes with respect to $v$ (see, for example, \cite{Sciama:1981hr}). Technically, the scalar field providing the Unruh vacuum state is constructed by taking solutions from both Rindler \eqref{phiRpm} and Minkowski \eqref{phiMpm} spacetimes \cite{Anderson:2022icz,Balbinot:2023vcm}. In explicit form it can be represented as \eqref{Unruhtheordef2DBH}
\begin{equation}\label{phiUdef}
\phi_{U}(T,X,t,x)=\int\limits_{0}^{\infty}\frac{dk}{\sqrt{4\pi k}}\left(e^{-ikU}c_{+}^{}(k)+
e^{ikU}c_{+}^{\dagger}(k)\right)+\int\limits_{-\infty}^{0}\frac{dk}{\sqrt{4\pi|k|}}\left(e^{ikv}c_{-}^{}(k)+
e^{-ikv}c_{-}^{\dagger}(k)\right),
\end{equation}
where $U=T-X$, $v=t+x$,
\begin{equation}
c_{+}^{}(k)=c^{}(k)\theta(k),\qquad c_{-}^{}(k)=c^{}(k)\theta(-k)
\end{equation}
and
\begin{equation}
[c^{}(k),c^{\dagger}(k')]=\delta(k-k'),
\end{equation}
all other commutators being equal to zero. The Unruh vacuum state is defined in the standard way as
\begin{equation}
c^{}(k)\ket{0_{U}}=0.
\end{equation}

In order to work with field \eqref{phiUdef}, one should pass to a single coordinate system. It is convenient to pass to the Rindler coordinates, which can be done by substituting \eqref{TRind} and \eqref{XRind} into \eqref{phiUdef}, resulting in
\begin{equation}
\phi_{U}(t,x)=\int\limits_{0}^{\infty}\frac{dk}{\sqrt{4\pi k}}\left(e^{i\frac{k}{a}e^{-au}}c_{+}^{}(k)+
e^{-i\frac{k}{a}e^{-au}}c_{+}^{\dagger}(k)\right)+\int\limits_{-\infty}^{0}\frac{dk}{\sqrt{4\pi|k|}}\left(e^{ikv}c_{-}^{}(k)+
e^{-ikv}c_{-}^{\dagger}(k)\right).
\end{equation}
On the other hand, we know that in Rindler spacetime the scalar field can be represented as \eqref{phiRpm}. Using the relation
\begin{equation}
\phi_{R}(t,x)\equiv\phi_{U}(t,x)
\end{equation}
and the results obtained above, we can get the following relation between the creation and annihilation operators in both representations:
\begin{align}
&b_{+}^{}(p)=\int\limits_{0}^{\infty}\left(c_{+}^{}(k)A_{+}^{}(k,p)+c_{+}^{\dagger}(k)A_{+}^{*}(k,-p)\right)\sqrt{\frac{p}{k}}\,dk,\\
&b_{-}^{}(p)=c_{-}^{}(p),
\end{align}
where $A_{+}^{}(k,p)$ is defined by \eqref{A+kp}.

\subsection{The Unruh vacuum energy in Rindler spacetime}
The Unruh vacuum energy in Rindler spacetime can be easily calculated, resulting in
\begin{equation}\label{Uvacenergy}
E_{vacU}=\bra{0_{U}}H_{R}\ket{0_{U}}=
\int\limits_{0}^{\infty}p\bra{0_{U}}b_{+}^{\dagger}(p)b_{+}^{}(p)\ket{0_{U}}dp
-\int\limits_{-\infty}^{0}p\bra{0_{U}}b_{-}^{\dagger}(p)b_{-}^{}(p)\ket{0_{U}}dp.
\end{equation}
Since $b_{-}^{}(p)=c_{-}^{}(p)$, the second integral in \eqref{Uvacenergy} is equal to zero, so we are left with
\begin{equation}
E_{vacU}=
\int\limits_{0}^{\infty}p\bra{0_{U}}b_{+}^{\dagger}(p)b_{+}^{}(p)\ket{0_{U}}dp
=\int\limits_{0}^{\infty}\int\limits_{0}^{\infty}A_{+}^{*}(k,-p)A_{+}^{}(k,-p)\frac{p^{2}}{k}\,dkdp.
\end{equation}
Repeating the calculations presented in Section~\Ref{Sec2-4}, one can easily get
\begin{equation}\label{Uvacfinal}
E_{vacU}=\frac{1}{2\pi}\int\limits_{-\infty}^{\infty}dx\int\limits_{0}^{\infty}\frac{p}{e^{\frac{2\pi p}{a}}-1}\,dp.
\end{equation}
The latter formula describes scalar radiation with thermal spectrum, but, contrary to the case of Minkowski vacuum \eqref{vacenergyresult}, here all particles describing the radiation have positive momenta, thus moving from the horizon towards the future infinity. However, since derivation of \eqref{Uvacfinal} is similar to derivation of \eqref{vacenergyresult}, whereas \eqref{vacenergyresult} describes particles that move symmetrically in opposite directions, it seems to be incorrect to describe the flux moving in one direction in \eqref{Uvacfinal} as originating from the horizon. Instead of this, it looks as if the particles with positive momenta just exist at any point of space, which is the consequence of the fact that the vacuum is ``spread'' over the whole space. Indeed, for particles with negative momenta in \eqref{vacenergyresult} there is no ``source'' at $x\to\infty$. The same is valid for the case of a two-dimensional Schwarzschild black hole too.

\subsection{Canonical commutation relations}\label{CCRUvac}
In order to check whether or not the resulting quantum field theory is consistent, we should check the validity of the canonical commutation relations
\begin{align}\label{CCR-UR}
&\left[\phi_{U}(t,x),\frac{\partial\phi_{U}(t,x')}{\partial t}\right]=i\delta(x-x'),\\
&\left[\phi_{U}(t,x),\phi_{U}(t,x')\right]=0,\\
&\left[\frac{\partial\phi_{U}(t,x)}{\partial t},\frac{\partial\phi_{U}(t,x')}{\partial t}\right]=0
\end{align}
(since we are interested in the theory in Rindler spacetime, it is more convenient to choose coordinates $t$, $x$ instead of $T$, $X$). For the second commutation relation one can get
\begin{equation}\label{CCR2DUa1}
\left[\phi_{U}(t,x),\phi_{U}(t,x')\right]
=\frac{i}{2\pi}\int\limits_{0}^{\infty}\frac{dk}{k}\,\sin\left(\frac{k}{a}e^{-at}\left(e^{ax}-e^{ax'}\right)\right)
+\int\limits_{-\infty}^{0}\frac{dk}{4\pi k}\,e^{-ik(x-x')}-\int\limits_{-\infty}^{0}\frac{dk}{4\pi k}\,e^{ik(x-x')}.
\end{equation}
Here the second and third integral in \eqref{CCR2DUa1} correspond to the contribution of the modes with $k<0$. The latter formula can be rewritten as
\begin{equation}\label{CCR2DUa2}
\left[\phi_{U}(t,x),\phi_{U}(t,x')\right]=\frac{i}{2\pi}\int\limits_{0}^{\infty}\frac{dk}{k}\,\sin\left(\frac{k}{a}e^{-at}\left(e^{ax}-e^{ax'}\right)\right)
-\frac{i}{2\pi}\int\limits_{-\infty}^{0}\frac{dk}{k}\,\sin\left(k\left(x-x'\right)\right).
\end{equation}
By changing $k\to-k$ in the second integral of \eqref{CCR2DUa2}, one gets
\begin{align}\nonumber
\left[\phi_{U}(t,x),\phi_{U}(t,x')\right]&=\frac{i}{2\pi}\int\limits_{0}^{\infty}\frac{dk}{k}\,\sin\left(\frac{k}{a}e^{-at}\left(e^{ax}-e^{ax'}\right)\right)
-\frac{i}{2\pi}\int\limits_{0}^{\infty}\frac{dk}{k}\,\sin\left(k\left(x-x'\right)\right)\\\label{CCR2DUa2b}
&=\frac{i}{4}\,\textrm{sign}\left(\frac{e^{-at}}{a}\left(e^{ax}-e^{ax'}\right)\right)-\frac{i}{4}\,\textrm{sign}(x-x')=0.
\end{align}
We see that this commutation relation is satisfied.

The commutator from the third commutation relation takes the form
\begin{align}\nonumber
&\left[\frac{\partial\phi_{U}(t,x)}{\partial t},\frac{\partial\phi_{U}(t,x')}{\partial t}\right]\\\nonumber
&=\frac{e^{-2at}e^{a(x+x')}}{4\pi}\left(\int\limits_{0}^{\infty}dk\,k\,e^{i\frac{k}{a}e^{-at}\left(e^{ax}-e^{ax'}\right)}
-\int\limits_{0}^{\infty}dk\,k\,e^{-i\frac{k}{a}e^{-at}\left(e^{ax}-e^{ax'}\right)}\right)\\\label{CCR2DUa2b2}
&-\frac{1}{4\pi}\left(\int\limits_{-\infty}^{0}dk\,k\,e^{ik(x-x')}-\int\limits_{-\infty}^{0}dk\,k\,e^{-ik(x-x')}\right),
\end{align}
where the two last integrals stand for the contribution of the modes with $k<0$. It is clear that for $x=x'$
\begin{equation}
\left[\frac{\partial\phi_{U}(t,x)}{\partial t},\frac{\partial\phi_{U}(t,x)}{\partial t}\right]=0.
\end{equation}
For $x\neq x'$, by changing $k\to -k$ in the second and fourth integrals in \eqref{CCR2DUa2b2}, one can get
\begin{equation}\label{CCR2DUb1}
\left[\frac{\partial\phi_{U}(t,x)}{\partial t},\frac{\partial\phi_{U}(t,x')}{\partial t}\right]=\frac{e^{-2at}e^{a(x+x')}}{4\pi}\int\limits_{-\infty}^{\infty}dk\,k\,e^{i\frac{k}{a}e^{-at}\left(e^{ax}-e^{ax'}\right)}
-\frac{1}{4\pi}\int\limits_{-\infty}^{\infty}dk\,k\,e^{ik(x-x')}.
\end{equation}
Using equality \eqref{equalityC}, we get
\begin{equation}
\left[\frac{\partial\phi_{U}(t,x)}{\partial t},\frac{\partial\phi_{U}(t,x')}{\partial t}\right]=0.
\end{equation}
We see that this commutation relation is satisfied too.

The commutator from the remaining commutation relation has the form
\begin{align}\nonumber
&\left[\phi_{U}(t,x),\frac{\partial\phi_{U}(t,x')}{\partial t}\right]\\\label{CCR2DUc1}
&=\frac{ie^{-at}e^{ax'}}{4\pi}\int\limits_{-\infty}^{\infty}dk\,e^{i\frac{k}{a}e^{-at}\left(e^{ax}-e^{ax'}\right)}
+\frac{i}{4\pi}\int\limits_{-\infty}^{0}dk\,e^{ik(x-x')}+\frac{i}{4\pi}\int\limits_{-\infty}^{0}dk\,e^{-ik(x-x')},
\end{align}
where the second and third integrals in \eqref{CCR2DUc1} correspond to the contribution of the modes with $k<0$. By changing $k\to -k$ in the third integral in the latter formula, we get
\begin{equation}\label{CCR2DUc2}
\left[\phi_{U}(t,x),\frac{\partial\phi_{U}(t,x')}{\partial t}\right]=\frac{ie^{-at}e^{ax'}}{4\pi}\int\limits_{-\infty}^{\infty}dk\,e^{i\frac{k}{a}e^{-at}\left(e^{ax}-e^{ax'}\right)}
+\frac{i}{4\pi}\int\limits_{-\infty}^{\infty}dk\,e^{ik(x-x')}.
\end{equation}
Applying the calculations presented in \eqref{CCR2Dc2} to the first integral in \eqref{CCR2DUc2}, we can rewrite \eqref{CCR2DUc2} as
\begin{equation}\label{CCR2DUc3}
\left[\phi_{U}(t,x),\frac{\partial\phi_{U}(t,x')}{\partial t}\right]
=\frac{ie^{-at}e^{ax'}}{2}\,\delta\left(\frac{e^{-at}}{a}\left(e^{ax}-e^{ax'}\right)\right)+\frac{i}{2}\,\delta(x-x')=i\delta(x-x').
\end{equation}

We see that all canonical commutation relations are exactly satisfied, which means that the corresponding quantum field theory providing the Unruh vacuum state is consistent. From a purely mathematical point of view this fact is quite surprising, because the quantized field is composed of two completely different sets of modes, namely, of sets that are inherent to different coordinate systems.

As has already been mentioned, the case of a uniformly accelerated reference frame in two-dimensions is almost identical to the case of a two-dimensional Schwarzschild black hole. However, the situation becomes more complicated when we move on to four dimensions.

\section{Uniformly accelerated reference frame in four-dimen\-sional spacetime}\label{URF4D}
\subsection{Two representations of a massless scalar field in the Rindler coordinates and Bogolyubov coefficients}
\begin{figure}[ht]
\centering
\includegraphics[width=0.7\linewidth]{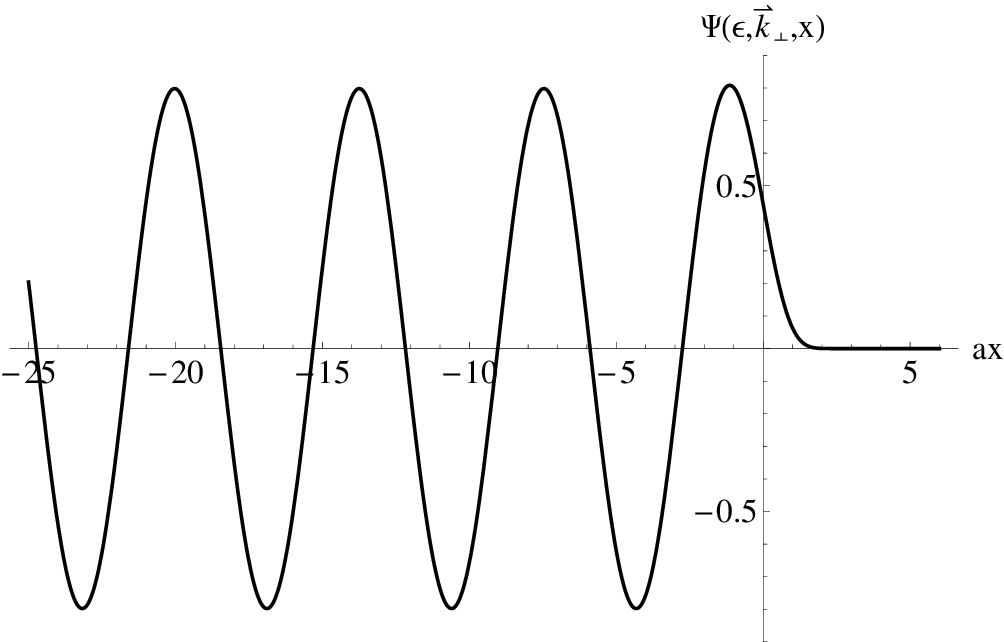}
\caption{$\Psi(\epsilon,\vec k_{\perp},x)$ for $|\vec k_{\perp}|=a$ and $\epsilon=a$.}\label{fig5}
\end{figure}

Now let us consider a more complicated case of a four-dimensional massless scalar field. The corresponding action takes the form
\begin{equation}\label{act4Duarf}
S=\frac{1}{2}\int\sqrt{-g}g^{\mu\nu}\partial_{\mu}\phi\,\partial_{\nu}\phi\,d^{4}x.
\end{equation}
The four-dimensional Rindler metric has the form
\begin{equation}\label{4DRindmetric}
ds^2=e^{2ax}\left(dt^2-(dx^{1})^2\right)-(dx^{2})^{2}-(dx^{3})^{2}.
\end{equation}
The equation of motion following from action \eqref{act4Duarf} for metric \eqref{4DRindmetric} has the form
\begin{equation}\label{eqm-massiveR}
\frac{\partial^{2}\phi}{\partial t^{2}}-\frac{\partial^{2}\phi}{\partial {x^{1}}^{2}}-e^{2ax}\left(\frac{\partial^{2}\phi}{\partial {x^{2}}^{2}}+\frac{\partial^{2}\phi}{\partial {x^{3}}^{2}}\right)=0.
\end{equation}
Using the notations
\begin{equation}\label{xtransdef}
\vec x=(x,\vec x_{\perp}),\qquad \vec x_{\perp}\equiv\{x^{2},x^{3}\},
\end{equation}
the quantized scalar field in Rindler spacetime can be represented as
\begin{equation}\label{scfieldRmassive}
\phi_{R}(t,\vec x)=\iint d^{2}k_{\perp}\int\limits_{0}^{\infty}d\epsilon\frac{1}{2\pi\sqrt{2\epsilon}}\Psi(\epsilon,\vec k_{\perp},x)\left(e^{-i\epsilon t}e^{i\vec k_{\perp}\vec x_{\perp}}b^{}(\epsilon,\vec k_{\perp})+
e^{i\epsilon t}e^{-i\vec k_{\perp}\vec x_{\perp}}b^{\dagger}(\epsilon,\vec k_{\perp})\right),
\end{equation}
where the functions $\Psi(\epsilon,\vec k_{\perp},x)$ satisfy the Schr\"{o}dinger equation
\begin{equation}\label{eqm-massiveRSch}
\epsilon^{2}\Psi(\epsilon,\vec k_{\perp},x)=-\frac{d^{2}\Psi(\epsilon,\vec k_{\perp},x)}{dx^{2}}+{\vec k_{\perp}}^{2}e^{2ax}\Psi(\epsilon,\vec k_{\perp},x)
\end{equation}
and for $\vec k_{\perp}\neq 0$ have the form \cite{Fulling:1972md}
\begin{equation}\label{Psi4DRind}
\Psi(\epsilon,\vec k_{\perp},x)=\frac{1}{\pi}\sqrt{2\,\frac{\epsilon}{a}\sinh\left(\pi\frac{\epsilon}{a}\right)}\,K_{i\frac{\epsilon}{a}}\left(\frac{|\vec k_{\perp}|}{a}\,e^{ax}\right),
\end{equation}
see also Fig.~\ref{fig5}.
These functions obey the orthonormality condition
\begin{equation}\label{Psi4Dorth}
\int\limits_{-\infty}^{\infty}\Psi(\epsilon,\vec k_{\perp},x)\Psi(\epsilon',\vec k_{\perp},x)dx=\delta(\epsilon-\epsilon')
\end{equation}
and the completeness relation
\begin{equation}
\int\limits_{0}^{\infty}\Psi(\epsilon,\vec k_{\perp},x)\Psi(\epsilon,\vec k_{\perp},x')d\epsilon=\delta(x-x').
\end{equation}
The creation and annihilation operators satisfy the standard commutation relations
\begin{equation}
[b^{}(\epsilon,\vec k_{\perp}),b^{\dagger}(\epsilon',\vec k'_{\perp})]=\delta(\epsilon-\epsilon')\delta^{(2)}(\vec k_{\perp}-\vec k'_{\perp}),
\end{equation}
all other commutators being equal to zero.

The Hamiltonian takes the form
\begin{equation}
H_{R}=\frac{1}{2}\int\left[\left(\frac{\partial\phi_{R}}{\partial t}\right)^{2}+\left(\frac{\partial\phi_{R}}{\partial x}\right)^{2}
+e^{2ax}\left(\left(\frac{\partial\phi_{R}}{\partial x^{2}}\right)^{2}+\left(\frac{\partial\phi_{R}}{\partial x^{3}}\right)^{2}\right)\right]d^{3}x.
\end{equation}
Using Eq.~\eqref{eqm-massiveR}, it can be brought to the form
\begin{equation}
H_{R}=\frac{1}{2}\int\left(\left(\frac{\partial\phi_{R}}{\partial t}\right)^{2}-\phi_{R}\frac{\partial^{2}\phi_{R}}{\partial t^{2}}\right)d^{3}x.
\end{equation}
Substituting representation \eqref{scfieldRmassive} into the Hamiltonian, we get
\begin{equation}
H_{R}=\frac{1}{2}\iint d^{2}k_{\perp}\int\limits_{0}^{\infty}d\epsilon\,\epsilon\left(b^{\dagger}(\epsilon,\vec k_{\perp})b^{}(\epsilon,\vec k_{\perp})+b^{}(\epsilon,\vec k_{\perp})b^{\dagger}(\epsilon,\vec k_{\perp})\right),
\end{equation}
which can be brought to the standard form
\begin{equation}
H_{R}=\iint d^{2}k_{\perp}\int\limits_{0}^{\infty}d\epsilon\,\epsilon\,b^{\dagger}(\epsilon,\vec k_{\perp})b^{}(\epsilon,\vec k_{\perp}).
\end{equation}
In the latter formula, the infinite $c$-number term
\begin{equation}\label{cnumber4D}
\frac{1}{2}\iint d^{2}k_{\perp}\int\limits_{0}^{\infty}\delta(\epsilon-\epsilon)\,\delta^{(2)}\left(\vec k_{\perp}-\vec k_{\perp}\right)\epsilon\,d\epsilon
\end{equation}
is omitted.

In Minkowski spacetime, a quantized scalar field has the standard form
\begin{equation}\label{phi4DMink4D}
\phi_{M}(T,\vec X)=\int\frac{d^{3}k}{(2\pi)^{\frac{3}{2}}\sqrt{2|\vec k|}}\left(e^{-i|\vec k|T}e^{i\vec k\vec X}a^{}(\vec k)+
e^{i|\vec k|T}e^{-i\vec k\vec X}a^{\dagger}(\vec k)\right).
\end{equation}
Let us introduce the notations
\begin{equation}
\vec X=(X,\vec X_{\perp}),\qquad \vec X_{\perp}\equiv\vec x_{\perp}.
\end{equation}
Then, using coordinate transformations \eqref{TRind} and \eqref{XRind}, scalar field \eqref{phi4DMink4D} in the Rindler coordinates takes the form
\begin{align}\nonumber
\phi_{M}(t,\vec x)&=\int\limits_{-\infty}^{\infty}dk\iint d^{2}k_{\perp}\frac{1}{(2\pi)^{\frac{3}{2}}\sqrt{2\sqrt{k^{2}+{\vec k_{\perp}}^{2}}}}\\\nonumber &\times\Biggl(e^{-i\frac{\sqrt{k^{2}+{\vec k_{\perp}}^{2}}}{a}\,e^{ax}\sinh(at)}
e^{i\frac{k}{a}\,e^{ax}\cosh(at)}e^{i\vec k_{\perp}\vec x_{\perp}}a^{}(k,\vec k_{\perp})\\\label{phiM4Dtx}&+
e^{i\frac{\sqrt{k^{2}+{\vec k_{\perp}}^{2}}}{a}\,e^{ax}\sinh(at)}
e^{-i\frac{k}{a}\,e^{ax}\cosh(at)}e^{-i\vec k_{\perp}\vec x_{\perp}}a^{\dagger}(k,\vec k_{\perp})\Biggr),
\end{align}
where $a^{}(\vec k)=a^{}(k,\vec k_{\perp})$ with
\begin{equation}
[a^{}(k,\vec k_{\perp}),a^{\dagger}(k',\vec k'_{\perp})]=\delta(k-k')\delta^{(2)}(\vec k_{\perp}-\vec k'_{\perp}).
\end{equation}
Contrary to the two-dimensional case, in the four-dimensional case the light cone coordinates $u$ and $v$ \eqref{light-cone-uv} are not so useful as in the two-dimensional case, so these coordinates will not be used in the subsequent calculations.

By using the relation $\phi_{M}(t,\vec x)\equiv\phi_{R}(t,\vec x)$, one can obtain the corresponding Bogolyubov transformations. They have the form \cite{Fulling:1972md}
\begin{equation}\label{BC4Dfinal}
b^{}(\epsilon,\vec k_{\perp})=
\int\limits_{-\infty}^{\infty}\frac{dk\,e^{i\frac{\epsilon}{a}\,q}}{\sqrt{4\pi\sqrt{k^{2}+{\vec k_{\perp}}^{2}}}\sqrt{a\sinh\left(\pi\frac{\epsilon}{a}\right)}}
\Biggl(e^{\frac{\pi\epsilon}{2a}}a^{}(k,\vec k_{\perp})
+e^{-\frac{\pi\epsilon}{2a}}a^{\dagger}(k,-\vec k_{\perp})\Biggr),
\end{equation}
where $q$ is defined by
\begin{equation}\label{sinhkkperp}
\sinh(q)=\frac{k}{|\vec k_{\perp}|}.
\end{equation}
In \cite{Fulling:1972md}, the Bogolyubov coefficients are presented without derivation, so their detailed derivation can be found in Appendix~C.

One can check that the commutation relations (they are fully analogous to commutation relations \eqref{CCR-MR}--\eqref{CCR-MR3} in the two-dimensional case)
\begin{align}\label{CCRExtra4D1}
&\left[\phi_{M}(t,\vec x),\phi_{M}(t,\vec x')\right]=0,\\
&\left[\frac{\partial\phi_{M}(t,\vec x)}{\partial t},\frac{\partial\phi_{M}(t,\vec x')}{\partial t}\right]=0,\\\label{CCRExtra4D3}
&\left[\phi_{M}(t,\vec x),\frac{\partial\phi_{M}(t,\vec x')}{\partial t}\right]=\delta^{(3)}(\vec x-\vec x')
\end{align}
are also satisfied, see detailed calculations in Appendix~D.

\subsection{The Minkowski spacetime vacuum energy in Rindler spacetime}\label{MvacRind4D}
The Minkowski spacetime vacuum energy in Rindler spacetime is defined as
\begin{align}\nonumber
\bra{0_{M}}H_{R}\ket{0_{M}}
&=\iint d^{2}k_{\perp}\int\limits_{0}^{\infty}d\epsilon\,\epsilon\bra{0_{M}}b^{\dagger}(\epsilon,\vec k_{\perp})b^{}(\epsilon,\vec k_{\perp})\ket{0_{M}}\\\label{vacen4D1}&
=\iint d^{2}k_{\perp}\int\limits_{0}^{\infty}d\epsilon'\int\limits_{0}^{\infty}d\epsilon\,\epsilon\,\delta(\epsilon'-\epsilon)\bra{0_{M}}b^{\dagger}(\epsilon',\vec k_{\perp})b^{}(\epsilon,\vec k_{\perp})\ket{0_{M}},
\end{align}
where the extra integration over $\epsilon'$ is introduced to make the subsequent calculations connected with a divergent integral more transparent, like it was done in \eqref{vacenergy+delta}. Substituting \eqref{BC4Dfinal} into \eqref{vacen4D1}, after some straightforward but trivial calculations one gets
\begin{align}\nonumber
&\bra{0_{M}}H_{R}\ket{0_{M}}\\\label{vacen4D2}
&=\iint d^{2}k_{\perp}\delta^{(2)}(\vec k_{\perp}-\vec k_{\perp})\int\limits_{0}^{\infty}d\epsilon'\int\limits_{0}^{\infty}d\epsilon\,\epsilon\,\delta(\epsilon'-\epsilon)
\int\limits_{-\infty}^{\infty}
\frac{dk\,e^{i\frac{\epsilon-\epsilon'}{a}\,q}e^{-\frac{\pi(\epsilon+\epsilon')}{2a}}}{4\pi a\sqrt{k^{2}+{\vec k_{\perp}}^{2}}\sqrt{\sinh\left(\pi\frac{\epsilon'}{a}\right)}\sqrt{\sinh\left(\pi\frac{\epsilon}{a}\right)}}.
\end{align}
With \eqref{sinhkkperp}, the integral over $k$ in \eqref{vacen4D2} turns out to be
\begin{equation}
\int\limits_{-\infty}^{\infty}
\frac{dk}{\sqrt{k^{2}+{\vec k_{\perp}}^{2}}}\,e^{i\frac{\epsilon-\epsilon'}{a}\,q}=\int\limits_{-\infty}^{\infty}
e^{i\frac{\epsilon-\epsilon'}{a}\,q}\,dq=2\pi a\delta(\epsilon-\epsilon'),
\end{equation}
leading to
\begin{equation}\label{vacen4D3}
E_{vac\,\,4D}=\bra{0_{M}}H_{R}\ket{0_{M}}
=\iint d^{2}k_{\perp}\int\limits_{0}^{\infty}\frac{\epsilon\,\delta(\epsilon-\epsilon)\,\delta^{(2)}(\vec k_{\perp}-\vec k_{\perp})}{e^{\frac{2\pi\epsilon}{a}}-1}\,d\epsilon
\end{equation}
(compare its structure with the one of \eqref{cnumber4D}).

\begin{figure}[ht]
\centering
\begin{minipage}{.49\textwidth}
\centering
\includegraphics[width=0.95\linewidth]{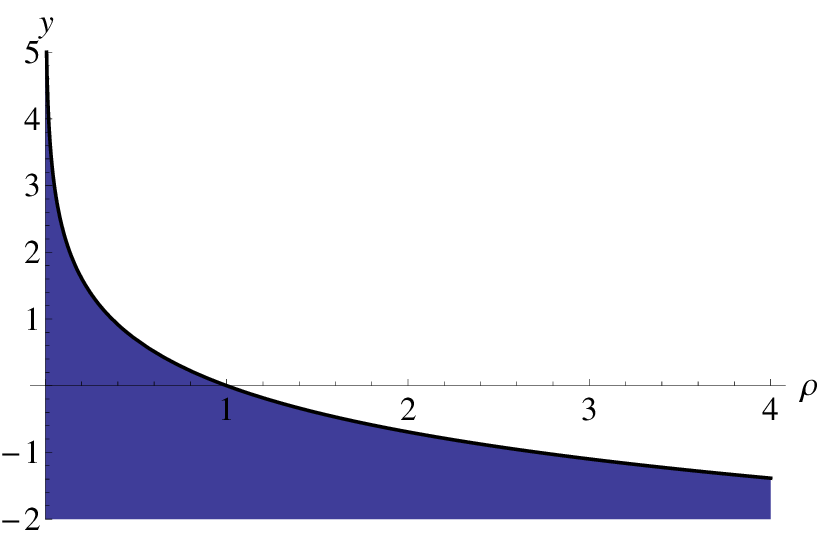}
\end{minipage}
\begin{minipage}{.49\textwidth}
\centering
\includegraphics[width=0.95\linewidth]{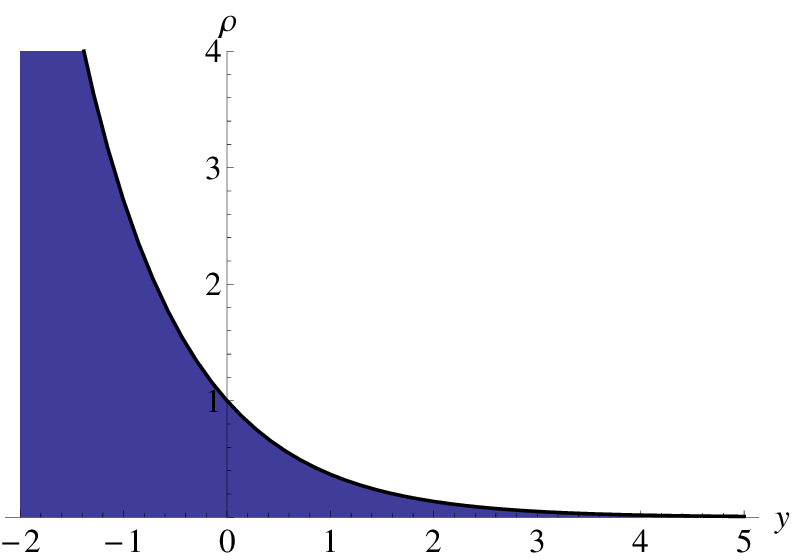}
\end{minipage}
\caption{Filled regions represent one and the same area of integration in \eqref{intarea}: left plot with the curve $-\ln\rho$ corresponds to the lhs of \eqref{intarea}, right plot with the curve $e^{-y}$ corresponds to the rhs of \eqref{intarea}.}\label{fig-IA}
\end{figure}

It is clear that $\delta(\epsilon-\epsilon)\,\delta^{(2)}(\vec k_{\perp}-\vec k_{\perp})$ is somehow connected with the spatial volume. In the simple cases like Minkowski or two-dimensional Rindler spacetimes it is equal (up to a constant) to the integral over the whole space (like in \eqref{delta-l}). As we will see below, in more complicated cases delta functions may provide different factors for different regions of space. For example, contrary to the case of \eqref{delta-l}, the delta function $\delta(\epsilon-\epsilon)$ in \eqref{vacen4D3} does not correspond to the whole space, thus providing a more complicated form of the energy density. To make sure of this, let us start with finding the asymptotic behavior of $\Psi(\epsilon,\vec k_{\perp},x)$. It is not difficult to show that for large negative $x$ solutions \eqref{Psi4DRind} have the form
\begin{equation}\label{Psiasymp}
\Psi(\epsilon,\vec k_{\perp},x)\approx\sqrt{\frac{2}{\pi}}\cos\left(\epsilon x+\sigma(\epsilon,\vec k_{\perp})\right),
\end{equation}
where $\sigma(\epsilon,\vec k_{\perp})$ is some phase. For large positive $x$ the solution totally vanishes, see Fig.~\Ref{fig5}. Practically, the solution changes its behavior from \eqref{Psiasymp} to almost zero on a short segment (again see Fig.~\Ref{fig5}), so the point $x_{0}$ where the wave function begins to rapidly decrease can be approximately defined as a solution of the equation $\epsilon^{2}=\vec k_{\perp}^{2}e^{2ax_{0}}$. Thus in explicit form \eqref{Psi4Dorth} for $\epsilon=\epsilon'$ can be rewritten as
\begin{equation}\label{deltaek}
\delta(\epsilon-\epsilon)\approx\frac{2}{\pi}\int\limits_{-\infty}^{\frac{1}{2a}\ln\left(\frac{\epsilon^{2}}{\vec k_{\perp}^{2}}\right)}\cos^{2}\left(\epsilon x+\sigma(\epsilon)\right)dx\approx\frac{2}{\pi}\int\limits_{-\infty}^{\frac{1}{2a}\ln\left(\frac{\epsilon^{2}}{\vec k_{\perp}^{2}}\right)}\frac{1}{2}\,dx
=\frac{1}{\pi}\int\limits_{-\infty}^{\frac{1}{2a}\ln\left(\frac{\epsilon^{2}}{\vec k_{\perp}^{2}}\right)}dx,
\end{equation}
where the oscillating term in the integrand was omitted. It looks as if \eqref{deltaek} depends on $\epsilon$ and $|\vec k_{\perp}|$ in such a representation. Since the dependence on $\vec x_{\perp}$ is described by just the standard plane waves, the delta functions $\delta^{(2)}(\vec k_{\perp}-\vec k_{\perp})$ can be represented in the standard form as
\begin{equation}
\delta^{(2)}(\vec k_{\perp}-\vec k_{\perp})=\frac{1}{4\pi^{2}}\iint d^{2}x_{\perp}.
\end{equation}
Thus formula \eqref{vacen4D3} can be rewritten as
\begin{equation}\label{vacen4D3a}
E_{vac\,\,4D}
=\frac{1}{4\pi^{3}}\iint d^{2}x_{\perp}\int\limits_{0}^{\infty}\frac{\epsilon\,}{e^{\frac{2\pi\epsilon}{a}}-1}\iint d^{2}k_{\perp}\left(\int\limits_{-\infty}^{\frac{1}{2a}\ln\left(\frac{\epsilon^{2}}{\vec k_{\perp}^{2}}\right)}dx\right)d\epsilon.
\end{equation}
Using the new variables $\rho=\frac{|\vec k_{\perp}|}{\epsilon}$ and $y=ax$, it is convenient to rewrite the integrals in \eqref{vacen4D3a} as
\begin{equation}
\iint d^{2}k_{\perp}\int\limits_{-\infty}^{\frac{1}{2a}\ln\left(\frac{\epsilon^{2}}{\vec k_{\perp}^{2}}\right)}dx
=\frac{2\pi\epsilon^{2}}{a}\int\limits_{0}^{\infty}\rho\,d\rho\int\limits_{-\infty}^{-\ln\rho}dy.
\end{equation}
Then, the latter integrals can be rewritten as (see Fig.~\ref{fig-IA} for additional clarification)
\begin{equation}\label{intarea}
\int\limits_{0}^{\infty}\rho\,d\rho\int\limits_{-\infty}^{-\ln\rho}dy
=\int\limits_{-\infty}^{\infty}dy\int\limits_{0}^{e^{-y}}\rho\,d\rho.
\end{equation}
Restoring the original variables, we get
\begin{equation}
\frac{2\pi\epsilon^{2}}{a}\int\limits_{-\infty}^{\infty}dy\int\limits_{0}^{e^{-y}}\rho\,d\rho
=\frac{2\pi\epsilon^{2}}{a}\int\limits_{-\infty}^{\infty}dy\frac{e^{-2y}}{2}=\pi\epsilon^{2}\int\limits_{-\infty}^{\infty}dx\,e^{-2ax}
\end{equation}
and, finally,
\begin{equation}\label{vacen4D3b}
E_{vac\,\,4D}=\frac{1}{4\pi^{2}}\iint d^{2}x_{\perp}\int\limits_{-\infty}^{\infty}dx\,
e^{-2ax}\int\limits_{0}^{\infty}\frac{\epsilon^{3}}{e^{\frac{2\pi\epsilon}{a}}-1}\,d\epsilon.
\end{equation}
We see that energy \eqref{vacen4D3b} is infinite. The Minkowski spacetime vacuum energy density in Rindler spacetime can be defined as (as in the two-dimensional case, in this definition non-flatness of the metric is not taken into account)
\begin{equation}\label{vacen4D4}
\rho_{vac\,\,4D}(x)=\frac{e^{-2ax}}{4\pi^{2}}\int\limits_{0}^{\infty}\frac{\epsilon^{3}}{e^{\frac{2\pi\epsilon}{a}}-1}\,d\epsilon.
\end{equation}
The integrand in the latter formula corresponds to the standard three-dimensional Planck law. Note that if we just naively put $\delta(\epsilon-\epsilon)\,\delta^{(2)}(\vec k_{\perp}-\vec k_{\perp})\sim\textrm{volume}$, we would get an (incorrect) infinite energy density due to $\iint d^{2}k_{\perp}$ in \eqref{vacen4D3}.

For comparison, for vacuum energy \eqref{cnumber4D} in a fully analogous way one can get
\begin{equation}
\frac{1}{8\pi^{2}}\iint d^{2}x_{\perp}\int\limits_{-\infty}^{\infty}dx\,
e^{-2ax}\int\limits_{0}^{\infty}\epsilon^{3}\,d\epsilon,
\end{equation}
also leading to the coordinate-dependent energy density
\begin{equation}\label{vacen4D40}
\rho_{vac\,\,4D\,0}(x)=\frac{e^{-2ax}}{8\pi^{2}}\int\limits_{0}^{\infty}\epsilon^{3}\,d\epsilon.
\end{equation}
We see that behaviors of $\rho_{vac\,\,4D}(x)$ and $\rho_{vac\,\,4D\,0}(x)$ are very similar. However, usually the infinite vacuum energy corresponding to $\rho_{vac\,\,4D\,0}(x)$ is just omitted, whereas the infinite vacuum energy corresponding $\rho_{vac\,\,4D}(x)$ is supposed to contain real particles. In this connection, it is necessary to recall that there exist some arguments in favour of fundamental unmeasurability of the Fulling-Davies-Unruh effect \cite{Pena:2013zfd}. Of course, the same arguments can be applied to the case of black holes too.

Formula \eqref{vacen4D4} can be obtained in a simpler way. Indeed, according to \eqref{deltaek} at the point $x=x_{0}$ only the modes with $\frac{1}{2a}\ln\left(\frac{\epsilon^{2}}{\vec k_{\perp}^{2}}\right)\gtrsim x_{0}$ can contribute to the vacuum energy because wave functions of the modes with $\frac{1}{2a}\ln\left(\frac{\epsilon^{2}}{\vec k_{\perp}^{2}}\right)\lesssim x_{0}$ are almost zero at $x=x_{0}$. The latter relations lead to $|\vec k_{\perp}|_{max}\approx \epsilon e^{-ax_{0}}$. Thus using \eqref{vacen4D3a} for the energy density one can write
\begin{equation}
\rho_{vac\,\,4D}(x_{0})\approx\frac{1}{4\pi^{3}}\int\limits_{0}^{\infty}\frac{\epsilon}{e^{\frac{2\pi\epsilon}{a}}-1}\left(2\pi\int\limits_{0}^{\epsilon e^{-ax_{0}}}\rho\,d\rho\right)d\epsilon=\frac{e^{-2ax_{0}}}{4\pi^{2}}\int\limits_{0}^{\infty}\frac{\epsilon^{3}}{e^{\frac{2\pi\epsilon}{a}}-1}\,d\epsilon,
\end{equation}
which coincides with \eqref{vacen4D4}. According to \eqref{Hamilt}
\begin{equation}
\bra{0_{M}}T_{0}^{0}(x)\ket{0_{M}}=\frac{\rho_{vac\,\,4D}(x)}{\sqrt{-g}}=\frac{\rho_{vac\,\,4D}(x)}{e^{2ax}}\approx
\frac{e^{-4ax}}{4\pi^{2}}\int\limits_{0}^{\infty}\frac{\epsilon^{3}}{e^{\frac{2\pi\epsilon}{a}}-1}\,d\epsilon.
\end{equation}
This result coincides with the one obtained in \cite{Candelas:1977zza} for $x\to\infty$ in a different way, namely, by considering $\bra{0_{M}}T_{\mu\nu}(x)\ket{0_{M}}$ from the very beginning.

Of course, calculations based on the use of \eqref{deltaek} can not be considered as mathematically rigorous in the full sense, especially taking into account that we deal with infinities, which should be treated accurately. In this connection, calculations based on the use of $\bra{0_{M}}T_{\mu\nu}(x)\ket{0_{M}}$ from the very beginning seem to be more accurate. Even though at first glance the dependence of infinite value in \eqref{deltaek} on some parameters may look rather ridiculous, such non-rigorous calculations allow one to take into account the ``number'' of contributing modes from point to point and to get the corresponding vacuum energy densities in a quite simple way. As an additional crosscheck of the method, let us consider a massive scalar field in two-dimensional Rindler spacetime. In such a case, the potential in \eqref{eqm-massiveRSch} takes the form $m^{2}e^{2ax}$, leading to
\begin{equation}
E_{vac\,\,2D\,m}=\bra{0_{M}}H_{R}\ket{0_{M}}
=\frac{1}{\pi}\int\limits_{0}^{\infty}\frac{\epsilon\,}{e^{\frac{2\pi\epsilon}{a}}-1}
\left(\int\limits_{-\infty}^{\frac{1}{2a}\ln\left(\frac{\epsilon^{2}}{m^{2}}\right)}dx\right)d\epsilon.
\end{equation}
Using the trick described above, the latter formula can be rewritten as
\begin{equation}\label{Evac2Dmassive}
E_{vac\,\,2D\,m}=\frac{1}{\pi}\int\limits_{-\infty}^{\infty}dx\int\limits_{me^{ax}}^{\infty}\frac{\epsilon\,}{e^{\frac{2\pi\epsilon}{a}}-1}\,d\epsilon.
\end{equation}
This result is reasonable. Indeed, at the point $x_{0}$ where the potential is equal to $m^{2}e^{2ax_{0}}$ only the particles with $\epsilon\gtrsim me^{ax_{0}}$ survive, because wave functions of the particles with $\epsilon<me^{ax_{0}}$ are almost zero at $x=x_{0}$. In other words, the farther from the horizon, the more energetic particles may still be in that area. This is exactly what is reflected in formula \eqref{Evac2Dmassive}. As for the four-dimensional case, since the potential in \eqref{eqm-massiveRSch} is proportional to $\vec k_{\perp}^{2}$, for any small $\epsilon$ there exist such small values of $\vec k_{\perp}^{2}$ that at the point $x_{0}$ the corresponding potential $\vec k_{\perp}^{2}e^{2ax_{0}}<\epsilon^{2}$. Thus at any point $x_{0}$ and for any small energy $\epsilon$ there still exist particles of this energy that contribute to $\rho_{vac\,\,4D}(x)$.

\subsection{Absence of the Unruh vacuum state}
Though the structure of the horizons and the structure of equations of motion in the vicinity of the horizons are almost the same in two-dimensional and four-dimensional cases, the structure of solutions in Rindler spacetime is different in two-dimensional and four-dimensional cases. This difference manifests itself when one tries to construct a theory providing the Unruh vacuum state. Indeed, in the two-dimensional case one set of modes is taken from those of Minkowski spacetime, whereas another set of modes is taken from those of Rindler spacetime. In the four-dimensional case, even though there is a doubling of modes for a fixed energy $\sqrt{k^{2}+\vec k_{\perp}^{2}}$ and momentum $\vec k_{\perp}$ in Minkowski spacetime (the doubling is connected with the sign of $k$), in Rindler spacetime we have only one mode for a given energy $\epsilon$ and momentum $\vec k_{\perp}\neq 0$, so it is impossible to mix the modes from different sets: we cannot split a single mode in Rindler spacetime, representing a standing wave, into two running waves like in the two-dimensional space. Thus it is technically impossible to construct a theory providing the Unruh vacuum state in the case of a four-dimensional uniformly accelerated reference frame. This observation will be useful when we move on to the four-dimensional Schwarzschild black hole.

\section{Pathologies in the quantum field theories for uniform\-ly accelerated reference frames}\label{Pathologies}
The quantum scalar field theories in Minkowski and Rindler spacetimes are well defined: the canonical commutation relations are satisfied and there exist conserved Hamiltonians in both spacetimes. Moreover, even a consistent theory providing the Unruh vacuum state in the two-dimensional case can be easily constructed. Usually, when the Fulling-Davies-Unruh effect is considered, the focus is made on the manifestation of the Minkowski vacuum state in Rindler spacetime. However, if we go a little bit further, i.e., if we consider other quantum states like a single particle, even in a free theory there arise problems which deserve additional investigation. They will be discussed below in the two-dimensional and four-dimensional cases.

\subsection{Two-dimensional case}
\subsubsection[Energy and momentum of the Minkowski spacetime one-particle state in Rindler spacetime]{Energy and momentum of the Minkowski spacetime one-particle state\\ in Rindler spacetime}
Let us consider a one-particle state in Minkowski spacetime $\ket{k_{a}}=a^{\dagger}(k_{a})\ket{0_{M}}$ such that
\begin{equation}
H_{M}\ket{k_{a}}=|k_{a}|\ket{k_{a}},
\end{equation}
where $\bra{k_{b}}\ket{k_{a}}=\delta(k_{a}-k_{b})$. The energy of this state can be calculated as
\begin{equation}
\frac{\bra{k_{a}}H_{M}\ket{k_{a}}}{\bra{k_{a}}\ket{k_{a}}}=|k_{a}|.
\end{equation}
It is clear that $\ket{k_{a}}$ is not an eigenfunction of the operator $H_{R}$. In order to estimate the (mean) energy of this particle in Rindler spacetime, it is necessary to calculate the value of
\begin{equation}
\frac{\bra{k_{a}}H_{R}\ket{k_{a}}}{\bra{k_{a}}\ket{k_{a}}}.
\end{equation}
In order to do it, let us calculate the matrix element $\bra{k_{b}}H_{R}\ket{k_{a}}$. Without loss of generality, one can take $k_{a}>0$, $k_{b}>0$. First, it is convenient to consider the matrix element $\bra{k_{a}}b^{\dagger}(p)b^{}(p)\ket{k_{b}}$. In explicit form it looks like
\begin{align}\nonumber
&\bra{k_{a}}b^{\dagger}(p)b^{}(p)\ket{k_{b}}=\bra{k_{a}}
\int\limits_{0}^{\infty}dk_{1}\int\limits_{0}^{\infty}dk_{2}\frac{p}{\sqrt{k_{1}k_{2}}}\\\nonumber&\times
\left(a_{+}^{\dagger}(k_{1})a_{+}^{}(k_{2})A_{+}^{*}(k_{1},p)A_{+}^{}(k_{2},p)+
a_{+}^{}(k_{1})a_{+}^{\dagger}(k_{2})A_{+}^{}(k_{1},-p)A_{+}^{*}(k_{2},-p)\right)\ket{k_{b}}\\\nonumber&
+\bra{k_{a}}
\int\limits_{-\infty}^{0}dk_{1}\int\limits_{-\infty}^{0}dk_{2}\frac{|p|}{\sqrt{|k_{1}||k_{2}|}}\\&\times
\left(a_{-}^{\dagger}(k_{1})a_{-}^{}(k_{2})A_{-}^{*}(k_{1},p)A_{-}^{}(k_{2},p)+
a_{-}^{}(k_{1})a_{-}^{\dagger}(k_{2})A_{-}^{}(k_{1},-p)A_{-}^{*}(k_{2},-p)\right)\ket{k_{b}},
\end{align}
where we have used the fact that $\bra{k_{a}}a_{\pm}^{}(k_{1})a_{\pm}^{}(k_{2})\ket{k_{b}}=0$. Since some of the terms with $A_{-}(k,\pm p)$ and $A_{-}^{*}(k,\pm p)$ do not contribute because $k_{a}>0$, $k_{b}>0$\ (see \eqref{ba1final} and \eqref{ba2final}), after some straightforward algebra we get
\begin{align}\nonumber
&\bra{k_{a}}b^{\dagger}(p)b^{}(p)\ket{k_{b}}=
\int\limits_{0}^{\infty}dk_{1}\int\limits_{0}^{\infty}dk_{2}\frac{p}{\sqrt{k_{1}k_{2}}}\bra{k_{a}}a_{+}^{\dagger}(k_{1})a_{+}^{}(k_{2})\ket{k_{b}}
\left(A_{+}^{*}(k_{1},p)A_{+}^{}(k_{2},p)\right.\\\nonumber&\left.+A_{+}^{*}(k_{1},-p)A_{+}^{}(k_{2},-p)\right)\\&
+\int\limits_{0}^{\infty}A_{+}^{*}(k,-p)A_{+}^{}(k,-p)\frac{p}{k}\,dk\bra{k_{a}}\ket{k_{b}}
+\int\limits_{-\infty}^{0}A_{-}^{*}(k,-p)A_{-}^{}(k,-p)\frac{|p|}{|k|}\,dk\bra{k_{a}}\ket{k_{b}}.
\end{align}
Since $\bra{k_{a}}a_{+}^{\dagger}(k_{1})a_{+}^{}(k_{2})\ket{k_{b}}=\delta(k_{1}-k_{a})\delta(k_{2}-k_{b})$, we get
\begin{align}\nonumber
&\bra{k_{a}}b^{\dagger}(p)b^{}(p)\ket{k_{b}}=
\frac{p}{\sqrt{k_{a}k_{b}}}
\left(A_{+}^{*}(k_{a},p)A_{+}^{}(k_{b},p)+A_{+}^{*}(k_{a},-p)A_{+}^{}(k_{b},-p)\right)\\&
+\int\limits_{0}^{\infty}A_{+}^{*}(k,-p)A_{+}^{}(k,-p)\frac{p}{k}\,dk\bra{k_{a}}\ket{k_{b}}
+\int\limits_{-\infty}^{0}A_{-}^{*}(k,-p)A_{-}^{}(k,-p)\frac{|p|}{|k|}\,dk\bra{k_{a}}\ket{k_{b}}.
\end{align}
With the help of \eqref{A+kp}, the first term in the rhs of the latter formula can be rewritten as
\begin{align}\nonumber
&A_{+}^{*}(k_{a},p)A_{+}^{}(k_{b},p)+A_{+}^{*}(k_{a},-p)A_{+}^{}(k_{b},-p)\\&
=\frac{1}{2\pi ap\left(e^{\frac{\pi p}{a}}-e^{-\frac{\pi p}{a}}\right)}\left(e^{\frac{\pi p}{a}}e^{i\frac{p}{a}\left(\ln\left(\frac{a}{k_{a}}\right)-\ln\left(\frac{a}{k_{b}}\right)\right)}+
e^{-\frac{\pi p}{a}}e^{-i\frac{p}{a}\left(\ln\left(\frac{a}{k_{a}}\right)-\ln\left(\frac{a}{k_{b}}\right)\right)}\right).
\end{align}
Using this relation and taking into account \eqref{vacenergy}, for the energy of our particle in Rindler spacetime one gets
\begin{align}\nonumber
&\frac{\bra{k_{a}}H_{R}\ket{k_{a}}}{\bra{k_{a}}\ket{k_{a}}}=
\frac{\bra{k_{a}}\int\limits_{-\infty}^{\infty}|p|b^{\dagger}(p)b^{}(p)dp\ket{k_{b}}\Big|_{\ket{k_{b}}=\ket{k_{a}}}}{\bra{k_{a}}\ket{k_{a}}}
\\\nonumber&=\frac{1}{2\pi a\bra{k_{a}}\ket{k_{a}}}\int\limits_{0}^{\infty}\frac{p\, dp}{\sqrt{k_{a}k_{b}}}\,e^{i\frac{p}{a}\left(\ln\left(\frac{a}{k_{a}}\right)-\ln\left(\frac{a}{k_{b}}\right)\right)}\Bigg|_{\ket{k_{b}}=\ket{k_{a}}}\\\label{partenergy1}&
+\frac{1}{\pi a\bra{k_{a}}\ket{k_{a}}}\int\limits_{0}^{\infty}\frac{p\,e^{-\frac{\pi p}{a}} dp}{\sqrt{k_{a}k_{b}}\left(e^{\frac{\pi p}{a}}-e^{-\frac{\pi p}{a}}\right)}
\,\cos\left(\frac{p}{a}\left(\ln\left(\frac{a}{k_{a}}\right)-\ln\left(\frac{a}{k_{b}}\right)\right)\right)\Bigg|_{\ket{k_{b}}=\ket{k_{a}}}+E_{vac},
\end{align}
where $E_{vac}$ is the infinite vacuum energy defined by \eqref{vacenergyresult}, which will be omitted in the subsequent calculations.\footnote{Appearance of the term $E_{vac}$ in the energy of a single particle or a wave packet (will be considered in the next section) is an additional argument in favor of similarity between $E_{vac}$ and the standard vacuum energy \eqref{vacERind2D}, the latter is usually omitted.} The last but one term in the latter formula takes the form
\begin{equation}\label{lastbutoneterm}
\frac{1}{\pi ak_{a}\delta(k_{a}-k_{a})}\int\limits_{0}^{\infty}\frac{p\,e^{-\frac{\pi p}{a}} dp}{e^{\frac{\pi p}{a}}-e^{-\frac{\pi p}{a}}}=0.
\end{equation}
It is equal to zero because the integral in \eqref{lastbutoneterm} converges, whereas the factor $\delta(k_{a}-k_{a})$ in the denominator of \eqref{lastbutoneterm} is infinite. Thus the only relevant contribution in \eqref{partenergy1} is
\begin{equation}\label{partenergy2}
\frac{1}{2\pi a\bra{k_{a}}\ket{k_{a}}}\int\limits_{0}^{\infty}\frac{p\, dp}{\sqrt{k_{a}k_{b}}}\,e^{i\frac{p}{a}\left(\ln\left(\frac{a}{k_{a}}\right)-\ln\left(\frac{a}{k_{b}}\right)\right)}\Bigg|_{\ket{k_{b}}=\ket{k_{a}}}
=\frac{1}{2\pi ak_{a}\delta(k_{a}-k_{a})}\int\limits_{0}^{\infty}p\,dp.
\end{equation}
The integral in the rhs of \eqref{partenergy2} clearly diverges and at first glance it is not clear how to isolate $\delta(k_{a}-k_{a})$ from this integral. In order to do it, one should pay attention to the exponent in the lhs of \eqref{partenergy2}. Indeed, let us consider the auxiliary integral
\begin{equation}
\int\limits_{-\infty}^{\infty}dp\,e^{i\frac{p}{a}\left(\ln\left(\frac{a}{k_{a}}\right)-\ln\left(\frac{a}{k_{b}}\right)\right)}
=2\pi\delta\left(\frac{1}{a}\left(\ln\left(\frac{a}{k_{a}}\right)-\ln\left(\frac{a}{k_{b}}\right)\right)\right)
=2\pi ak_{a}\delta(k_{a}-k_{b}),
\end{equation}
which involves the same exponent as in \eqref{partenergy2} and connects variables $k$ and $p$. Using this formula, we can write
\begin{equation}
\delta(k_{a}-k_{a})=\frac{1}{2\pi ak_{a}}\int\limits_{-\infty}^{\infty}dp=\frac{1}{\pi ak_{a}}\int\limits_{0}^{\infty}dp,
\end{equation}
which allows one to rewrite the rhs of \eqref{partenergy2} as
\begin{equation}\label{partenergy3}
\frac{\bra{k_{a}}H_{R}\ket{k_{a}}}{\bra{k_{a}}\ket{k_{a}}}-E_{vac}=\frac{1}{2\pi ak_{a}\delta(k_{a}-k_{a})}\int\limits_{0}^{\infty}p\,dp=\frac{1}{2}\frac{\int\limits_{0}^{\infty}p\,dp}{\int\limits_{0}^{\infty}dp}\to\infty.
\end{equation}
If we introduce the cutoff $p_{max}=M_{Pl}$, for the particle energy we get
\begin{equation}\label{partenergy3cutoff}
\frac{\bra{k_{a}}H_{R}\ket{k_{a}}}{\bra{k_{a}}\ket{k_{a}}}-E_{vac}=\frac{1}{2}\frac{\int\limits_{0}^{M_{Pl}}p\,dp}{\int\limits_{0}^{M_{Pl}}dp}
=\frac{M_{Pl}}{4},
\end{equation}
which depends only on the cutoff scale and does not depend on the acceleration $a$. So, even for an extremely small acceleration one gets an extremely large mass of the particle, which looks absurd. The origin of this apparent paradox will become clear a little bit later.

Even though metric \eqref{metric-R} explicitly depends on $x$, in the two-dimensional case the momentum is conserved. The latter is the consequence of the fact that the metric is conformally flat. Indeed, for $\nu=1$ relation \eqref{tcons} has the form
\begin{equation}\label{tconsmomentum}
\frac{\partial\left(\sqrt{-g}\,T^{\mu}_{1}\right)}{\partial x^{\mu}}=\frac{1}{2}\sqrt{-g}\,\frac{\partial g_{\mu\sigma}}{\partial x}T^{\mu\sigma},
\end{equation}
so the momentum can be defined as
\begin{equation}\label{momentum}
P=-\int\sqrt{-g}\,g^{00}T_{01}dx.
\end{equation}
Since $g_{11}=-g_{00}$, we get for the rhs of \eqref{tconsmomentum}
\begin{equation}\label{tconsmomentumrhs}
\frac{1}{2}\sqrt{-g}\,\frac{\partial g_{00}}{\partial x}\left(T^{00}-T^{11}\right).
\end{equation}
With
\begin{equation}
T_{00}=T_{11}=\frac{1}{2}\left(\frac{\partial\phi}{\partial t}\right)^{2}+\frac{1}{2}\left(\frac{\partial\phi}{\partial x}\right)^{2}
\end{equation}
formula \eqref{tconsmomentumrhs} takes the form
\begin{equation}
\frac{1}{2}\sqrt{-g}\,\frac{\partial g_{00}}{\partial T}\left(g^{00}g^{00}T_{00}-g^{11}g^{11}T_{11}\right)=\frac{1}{2}\sqrt{-g}\,\frac{\partial g_{00}}{\partial T}g^{00}g^{00}\left(T_{00}-T_{11}\right)=0.
\end{equation}
Thus the momentum defined by \eqref{momentum} is conserved over time. In explicit form momentum \eqref{momentum} in Rindler spacetime looks like
\begin{equation}\label{momentumexpl}
P_{R}=-\int\frac{\partial\phi}{\partial t}\frac{\partial\phi}{\partial x}dx.
\end{equation}
For the quantized field \eqref{scfieldR} the corresponding operator has the standard form
\begin{equation}\label{momentumQ}
P_{R}=\int\limits_{-\infty}^{\infty}p\,b^{\dagger}(p)b^{}(p)dp.
\end{equation}
Since the momentum of our particle is positive (i.e., $k_{a}>0$), according to \eqref{ba1final} only the Bogolyubov coefficients $A_{+}^{}(k,p)$ defined by \eqref{A+kp} contribute to the relevant part of \eqref{momentumQ}, so we can use the results obtained above for $\frac{\bra{k_{a}}H_{R}\ket{k_{a}}}{\bra{k_{a}}\ket{k_{a}}}$. Namely,
\begin{equation}
\frac{\bra{k_{a}}P_{R}\ket{k_{a}}}{\bra{k_{a}}\ket{k_{a}}}=\frac{\bra{k_{a}}H_{R}\ket{k_{a}}}{\bra{k_{a}}\ket{k_{a}}}-E_{vac}
=\frac{1}{2}\frac{\int\limits_{0}^{\infty}p\,dp}{\int\limits_{0}^{\infty}dp}\to\infty.
\end{equation}

In order to understand better the essence of this effect, in the next section a simple normalized quantum state and its classical analogue (i.e., wave packet with finite energy) will be considered explicitly demonstrating that such a nonstandard behaviour is inherent to some other objects too.

\subsubsection[Energy and momentum of the Minkowski spacetime normalized state in\\ Rindler spacetime]{Energy and momentum of the Minkowski spacetime normalized state\\ in Rindler spacetime}
In order to better understand the result of the previous section, it is convenient to consider normalized states that correspond to wave packets of finite energy in the classical field theory. Suppose we have a wave packet in Minkowski spacetime such that it is localized only in wedge I of Fig.~\ref{fig1}, whereas in other wedges the scalar field is identically zero (for example, it is localized strictly in the area $0<X_{1}<X<X_{2}$ for $T=0$). Since the coordinate transformations inside wedge I are non-singular, one expects that the energy of the wave packet in Rindler spacetime is also finite. However, as will be shown below there exist wave packets of finite energy in Minkowski spacetime that have an infinite energy in Rindler spacetime.

A one-particle state of finite norm is defined as
\begin{equation}
\ket{\psi}=\int\limits_{-\infty}^{\infty}f(k)\ket{k}dk,
\end{equation}
where
\begin{equation}
\int\limits_{-\infty}^{\infty}f^{*}(k)f(k)dk=1,
\end{equation}
that is
\begin{equation}
\bra{\psi}\ket{\psi}=1.
\end{equation}
In order to perform all calculations analytically, it is convenient to take a special form of $f(k)$. For example, one can choose
\begin{equation}\label{fk}
f(k)=\frac{\theta(k-k_{m})\,\theta(k_{M}-k)}{\sqrt{k_{M}-k_{m}}},
\end{equation}
where $\theta(x)$ is the Heaviside theta function. In order to use the results of the previous section, let us also take $k_{M}>k_{m}>0$. The energy of the state $\ket{\psi}$ with $f(k)$ defined by \eqref{fk} in Minkowski spacetime is
\begin{equation}\label{EnergypsiM}
\frac{\bra{\psi}H_{M}\ket{\psi}}{\bra{\psi}\ket{\psi}}=\bra{\psi}H_{M}\ket{\psi}=\int\limits_{-\infty}^{\infty}|k|f^{*}(k)f(k)\,dk
=\frac{1}{k_{M}-k_{m}}\int\limits_{k_{m}}^{k_{M}}k\,dk=\frac{k_{m}+k_{M}}{2},
\end{equation}
where $H_{M}$ is defined by \eqref{HamiltMinkdef}. The momentum of this wave packet is $\bra{\psi}P_{M}\ket{\psi}=\bra{\psi}H_{M}\ket{\psi}$.

Using the results of the previous section (namely, formula \eqref{partenergy1}), we get
\begin{align}\nonumber
&\bra{\psi}H_{R}\ket{\psi}=
\int\limits_{-\infty}^{\infty}f^{*}(k_{a})\,dk_{a}\int\limits_{-\infty}^{\infty}f(k_{b})\,dk_{b}\bra{k_{a}}H_{R}\ket{k_{b}}
\\\nonumber&=\frac{1}{2\pi a}\int\limits_{0}^{\infty}p\,dp\int\limits_{-\infty}^{\infty}dk_{a}\int\limits_{-\infty}^{\infty}dk_{b}
\frac{f^{*}(k_{a})f(k_{b})\,e^{i\frac{p}{a}\left(\ln\left(\frac{a}{k_{a}}\right)-\ln\left(\frac{a}{k_{b}}\right)\right)}}{\sqrt{k_{a}k_{b}}}\,\\\label{wpenergy1}&
+
\frac{1}{\pi a}\int\limits_{0}^{\infty}\frac{p\,e^{-\frac{\pi p}{a}}dp}{e^{\frac{\pi p}{a}}-e^{-\frac{\pi p}{a}}}\int\limits_{-\infty}^{\infty}dk_{a}\int\limits_{-\infty}^{\infty}dk_{b}
\,\frac{f^{*}(k_{a})f(k_{b})\cos\left(\frac{p}{a}\left(\ln\left(\frac{a}{k_{a}}\right)-\ln\left(\frac{a}{k_{b}}\right)\right)\right)}{\sqrt{k_{a}k_{b}}}+E_{vac}.
\end{align}
The second triple integral in the rhs of \eqref{wpenergy1} is clearly convergent for $f(k)$ defined by \eqref{fk}, so let us focus on the first triple integral in the rhs of \eqref{wpenergy1}, which can be rewritten as
\begin{equation}\label{wpenergy2}
\frac{1}{2\pi a}\int\limits_{0}^{\infty}p\,dp\int\limits_{-\infty}^{\infty}
\frac{f^{*}(k_{a})\,e^{i\frac{p}{a}\ln\left(\frac{a}{k_{a}}\right)}}{\sqrt{k_{a}}}\,dk_{a}
\int\limits_{-\infty}^{\infty}
\frac{f(k_{b})\,e^{-i\frac{p}{a}\ln\left(\frac{a}{k_{b}}\right)}}{\sqrt{k_{b}}}\,dk_{b}.
\end{equation}
The integrals over $k_{a}$ and $k_{b}$ can be easily calculated, resulting in
\begin{equation}\label{intfk1}
\int\limits_{-\infty}^{\infty}
\frac{f^{*}(k_{a})\,e^{i\frac{p}{a}\ln\left(\frac{a}{k_{a}}\right)}}{\sqrt{k_{a}}}\,dk_{a}
=\frac{a^{i\frac{p}{a}}}{\sqrt{k_{M}-k_{m}}}\int\limits_{k_{m}}^{k_{M}}k_{a}^{-i\frac{p}{a}-\frac{1}{2}}dk_{a}=
\frac{\sqrt{a}\left(e^{\left(\frac{1}{2}-i\frac{p}{a}\right)\ln\left(\frac{k_{M}}{a}\right)}-
e^{\left(\frac{1}{2}-i\frac{p}{a}\right)\ln\left(\frac{k_{m}}{a}\right)}\right)}{\sqrt{k_{M}-k_{m}}\left(\frac{1}{2}-i\frac{p}{a}\right)}
\end{equation}
and
\begin{equation}\label{intfk2}
\int\limits_{-\infty}^{\infty}
\frac{f(k_{b})\,e^{-i\frac{p}{a}\ln\left(\frac{a}{k_{b}}\right)}}{\sqrt{k_{b}}}\,dk_{b}
=\frac{\sqrt{a}\left(e^{\left(\frac{1}{2}+i\frac{p}{a}\right)\ln\left(\frac{k_{M}}{a}\right)}-
e^{\left(\frac{1}{2}+i\frac{p}{a}\right)\ln\left(\frac{k_{m}}{a}\right)}\right)}{\sqrt{k_{M}-k_{m}}\left(\frac{1}{2}+i\frac{p}{a}\right)}.
\end{equation}
With \eqref{intfk1} and \eqref{intfk2}, integral \eqref{wpenergy2} can be brought to the form
\begin{equation}\label{wpenergy3}
\frac{a}{2\pi(k_{M}-k_{m})}\left((k_{M}+k_{m})\int\limits_{0}^{\infty}\frac{p\,dp}{p^{2}+\frac{a^{2}}{4}}
-2\sqrt{k_{m}k_{M}}\int\limits_{0}^{\infty}\frac{p\,\cos\left(\ln\left(\frac{k_{M}}{k_{m}}\right)\frac{p}{a}\right)dp}{p^{2}+\frac{a^{2}}{4}}\right).
\end{equation}
The second integral in \eqref{wpenergy3} can be evaluated analytically \cite{PBM}:
\begin{equation}
\int\limits_{0}^{\infty}\frac{p\,\cos\left(\ln\left(\frac{k_{M}}{k_{m}}\right)\frac{p}{a}\right)dp}{p^{2}+\frac{a^{2}}{4}}=
-\frac{1}{2}\left(\sqrt{\frac{k_{m}}{k_{M}}}\,\textrm{Ei}\left(\frac{1}{2}\ln\left(\frac{k_{M}}{k_{m}}\right)\right)
+\sqrt{\frac{k_{M}}{k_{m}}}\,\textrm{Ei}\left(-\frac{1}{2}\ln\left(\frac{k_{M}}{k_{m}}\right)\right)\right).
\end{equation}
Thus \eqref{wpenergy3} takes the form
\begin{equation}\label{wpenergy4}
\frac{a(k_{M}+k_{m})}{4\pi(k_{M}-k_{m})}\ln\left(\frac{p^{2}}{a^{2}}+\frac{1}{4}\right)\bigg|_{p=0}^{\infty}
+\frac{a\left(k_{m}\textrm{Ei}\left(\frac{1}{2}\ln\left(\frac{k_{M}}{k_{m}}\right)\right)
+k_{M}\textrm{Ei}\left(-\frac{1}{2}\ln\left(\frac{k_{M}}{k_{m}}\right)\right)\right)}{2\pi(k_{M}-k_{m})}\to\infty.
\end{equation}
We see that due to the first integral in \eqref{wpenergy4}, the energy $\bra{\psi}H_{R}\ket{\psi}$ of the state $\ket{\psi}$ in Rindler spacetime is infinite.

The divergence in \eqref{wpenergy4} is logarithmic, so it is interesting to see what happens if we introduce a cutoff. Let $k_{M}\gg k_{m}$ (say, $k_{M}=100\,k_{m}$) and $a=g$, where $g\approx 9.8\,m/s^{2}\approx 2\cdot 10^{-32}\,GeV c/\hbar$ is Earth's standard gravitational acceleration, and the cutoff scale is $M_{Pl}$. One can check that the main contribution in \eqref{wpenergy4} is provided by the first term, resulting in
\begin{equation}\label{wpenergy4cutoff}
\bra{\psi}H_{R}\ket{\psi}\approx
\frac{g}{2\pi}\ln\left(\frac{M_{Pl}}{g}\right)\approx 18.6\,g\approx 3.7\cdot 10^{-22}\,eV.
\end{equation}
We see that with the cutoff applied, the energy of the normalized state in Rindler spacetime can be much smaller that the (initially finite) energy of this state in Minkowski spacetime, compare with \eqref{EnergypsiM}. This situation differs considerably from the case of a single particle. It is clear that for the momentum we also obtain
\begin{equation}
\bra{\psi}P_{R}\ket{\psi}=\bra{\psi}H_{R}\ket{\psi}-E_{vac}.
\end{equation}

This effect can be observed in the classical field theory too (discussion of some manifestations of the Fulling-Davies-Unruh effect in classical field theory can be found in \cite{Higuchi:1993fn}). Let us consider the function defined by \eqref{fk}:
\begin{equation}\label{wavepacket1}
\psi(X)=\frac{1}{\sqrt{2\pi}}\int\limits_{-\infty}^{\infty}f(k)\,e^{ikX}dk=-\frac{i}{\sqrt{2\pi(k_{M}-k_{m})}\,X}\left(e^{ik_{M}X}-e^{ik_{m}X}\right).
\end{equation}
Since we are dealing with the real scalar field, the corresponding solution of the equation of motion can be chosen to be
\begin{equation}
\phi_{cl}(T,X)=\textrm{Re}\left(\psi(X-T)\right)=\frac{\sin\left(k_{M}(X-T)\right)-\sin\left(k_{m}(X-T)\right)}{\sqrt{2\pi(k_{M}-k_{m})}\,(X-T)}.
\end{equation}
It is not difficult to show that the energy of this wave packet in Minkowski spacetime is finite. Indeed, according to \eqref{H-M}, at large $|X|$ the leading contribution to the energy is proportional to the integral
\begin{equation}\label{energylargeMX}
\int\frac{1}{(X-T)^{2}}\,dX,
\end{equation}
which is convergent at $X\to\pm\infty$ (to obtain the latter integral, the relation $\cos^{2}(x)=\frac{1}{2}+\frac{\cos(2x)}{2}$ was used and the oscillating terms, as well as the terms $\sim\frac{1}{(X-T)^{3}}$ and $\sim\frac{1}{(X-T)^{4}}$, were omitted).

In the Rindler coordinates, the part of solution \eqref{wavepacket1} that resides in wedge I has the form
\begin{equation}\label{wavepacket2}
\phi_{cl}(t,x)=\frac{\sin\left(\frac{k_{M}}{a}\,e^{a(x-t)}\right)-\sin\left(\frac{k_{m}}{a}\,e^{a(x-t)}\right)}
{\sqrt{2\pi\frac{k_{M}-k_{m}}{a^{2}}}\,e^{a(x-t)}}.
\end{equation}
According to \eqref{H-R}, at large $x$ the leading contribution of this part of the wave packet to the energy is proportional to the integral
\begin{equation}\label{energylargeRx}
\int\frac{1}{\left(e^{a(x-t)}\right)^{2}}\,\left(e^{a(x-t)}\right)^{2}dx=\int dx,
\end{equation}
which clearly diverges at $x\to\infty$.

Using \eqref{momentumexpl}, it is easy to show (exactly in the same way as it was made for the energy above) that at large $X$ and $x$ the leading contributions to the momenta in Minkowski and Rindler spacetimes also have the forms \eqref{energylargeMX} and \eqref{energylargeRx} respectively.

We see that there exists a wave packet with finite energy in Minkowski spacetime that has an infinite energy in Rindler spacetime. Since wave function of a single particle in Minkowski spacetime is just a plane wave, which is ``wider'' than wave packet \eqref{wavepacket1}, for this state we also get an infinite energy in Rindler spacetime, even for an extremely small acceleration $a$. A small discussion of this effect and its origin will be presented after examination of the four-dimensional case.

\subsection{Four-dimensional case}
Analogous examples can be provided in the four-dimensional case too. Since momentum is not conserved in the four-dimensional Rindler spacetime, only the energy will be considered.

\subsubsection{Energy of the Minkowski spacetime one-particle state in Rindler spacetime}
Let us again start with the standard one-particle state in Minkowski spacetime
\begin{equation}
H_{M}\ket{k_{a},\vec k_{a\perp}}=\sqrt{k_{a}^{2}+{\vec k_{a\perp}}^{2}}\,\ket{k_{a},\vec k_{a\perp}},
\end{equation}
where $\bra{k_{a},\vec k_{a\perp}}\ket{k_{b},\vec k_{b\perp}}=\delta(k_{a}-k_{b})\delta^{(2)}\left(\vec k_{a\perp}-\vec k_{b\perp}\right)$. As usual
\begin{equation}
\frac{\bra{k_{a},\vec k_{a\perp}}H_{M}\ket{k_{a},\vec k_{a\perp}}}{\bra{k_{a},\vec k_{a\perp}}\ket{k_{a},\vec k_{a\perp}}}=\sqrt{k_{a}^{2}+{\vec k_{a\perp}}^{2}}.
\end{equation}
Our aim is to calculate
\begin{equation}
\frac{\bra{k_{a},\vec k_{a\perp}}H_{R}\ket{k_{a},\vec k_{a\perp}}}{\bra{k_{a},\vec k_{a\perp}}\ket{k_{a},\vec k_{a\perp}}}
\end{equation}
taking into account that $\ket{k_{a}}$ is not an eigenfunction of the operator $H_{R}$.

As in the two-dimensional case, first it is convenient to consider the matrix element
\begin{align}\nonumber
&\bra{k_{a},\vec k_{a\perp}}b^{\dagger}(\epsilon,\vec k_{\perp})b^{}(\epsilon,\vec k_{\perp})\ket{k_{b},\vec k_{b\perp}}
=\iint\frac{dk_{1}dk_{2}\,e^{i\epsilon\frac{q_{2}-q_{1}}{a}}}{4\pi a\left(k_{1}^{2}+{\vec k_{\perp}}^{2}\right)^{\frac{1}{4}}\left(k_{2}^{2}+{\vec k_{\perp}}^{2}\right)^{\frac{1}{4}}\sinh\left(\frac{\pi\epsilon}{a}\right)}\\\nonumber
&\times\left(e^{\frac{\pi\epsilon}{a}}\bra{k_{a},\vec k_{a\perp}}a^{\dagger}(k_{1},\vec k_{\perp})a^{}(k_{2},\vec k_{\perp})\ket{k_{b},\vec k_{b\perp}}
+e^{-\frac{\pi\epsilon}{a}}\bra{k_{a},\vec k_{a\perp}}a^{}(k_{1},-\vec k_{\perp})a^{\dagger}(k_{2},-\vec k_{\perp})\ket{k_{b},\vec k_{b\perp}}\right)\\\nonumber
&=\frac{e^{i\epsilon\frac{q_{b}-q_{a}}{a}}e^{\frac{\pi\epsilon}{a}}\delta^{(2)}(\vec k_{\perp}-\vec k_{a\perp})\,\delta^{(2)}(\vec k_{\perp}-\vec k_{b\perp})
+e^{-i\epsilon\frac{q_{b}-q_{a}}{a}}e^{-\frac{\pi\epsilon}{a}}\delta^{(2)}(\vec k_{\perp}+\vec k_{a\perp})\,\delta^{(2)}(\vec k_{\perp}+\vec k_{b\perp})}{4\pi a\left(k_{a}^{2}+{\vec k_{\perp}}^{2}\right)^{\frac{1}{4}}\left(k_{b}^{2}+{\vec k_{\perp}}^{2}\right)^{\frac{1}{4}}\sinh\left(\frac{\pi\epsilon}{a}\right)}
\\\label{4Dbbkakb}&+E_{vac\,\,4D}\bra{k_{a},\vec k_{a\perp}}\ket{k_{b},\vec k_{b\perp}},
\end{align}
where $q_{1,2}$ are defined as $\sinh(q_{1,2})=\frac{k_{1,2}}{|\vec k_{\perp}|}$, the term $E_{vac\,\,4D}$ is defined by \eqref{vacen4D3}, and Bogolyubov transformations \eqref{BC4Dfinal} are used. Using \eqref{4Dbbkakb}, we can obtain
\begin{align}\nonumber
&\bra{k_{a},\vec k_{a\perp}}H_{R}\ket{k_{b},\vec k_{b\perp}}-E_{vac\,\,4D}\bra{k_{a},\vec k_{a\perp}}\ket{k_{b},\vec k_{b\perp}}\\\nonumber
&=\iint d^{2}k_{\perp}\int\limits_{0}^{\infty}d\epsilon\,\epsilon\bra{k_{a},\vec k_{a\perp}}b^{\dagger}(\epsilon,\vec k_{\perp})b^{}(\epsilon,\vec k_{\perp})\ket{k_{b},\vec k_{b\perp}}-E_{vac\,\,4D}\bra{k_{a},\vec k_{a\perp}}\ket{k_{b},\vec k_{b\perp}}\\\nonumber
&=\frac{\delta^{(2)}(\vec k_{a\perp}-\vec k_{b\perp})}{4\pi a\sqrt{k_{a}^{2}+{\vec k_{a\perp}}^{2}}}\int\limits_{0}^{\infty}d\epsilon\,\epsilon\,
\frac{e^{i\epsilon\frac{q_{b}-q_{a}}{a}}e^{\frac{\pi\epsilon}{a}}+e^{-i\epsilon\frac{q_{b}-q_{a}}{a}}e^{-\frac{\pi\epsilon}{a}}}
{\sinh\left(\frac{\pi\epsilon}{a}\right)}\\
&=\frac{\delta^{(2)}(\vec k_{a\perp}-\vec k_{b\perp})}{4\pi a\sqrt{k_{a}^{2}+{\vec k_{a\perp}}^{2}}}
\left(2\int\limits_{0}^{\infty}d\epsilon\,\epsilon\,
e^{i\epsilon\frac{q_{b}-q_{a}}{a}}+4\int\limits_{0}^{\infty}d\epsilon\,\epsilon\,
\frac{\cos\left(\epsilon\frac{q_{b}-q_{a}}{a}\right)}
{e^{\frac{2\pi\epsilon}{a}}-1}\right),
\end{align}
where $q_{a,b}$ are defined as $\sinh(q_{a,b})=\frac{k_{a,b}}{|\vec k_{a\perp}|}$ (note that denominator in the latter formula is $|\vec k_{a\perp}|$ for both subscripts $a$ and $b$). The relevant part of the energy is
\begin{equation}
\frac{\bra{k_{a},\vec k_{a\perp}}H_{R}\ket{k_{a},\vec k_{a\perp}}}{\bra{k_{a},\vec k_{a\perp}}\ket{k_{a},\vec k_{a\perp}}}-E_{vac\,\,4D}
=\frac{1}{2\pi a\sqrt{k_{a}^{2}+{\vec k_{a\perp}}^{2}}\,\delta(k_{a}-k_{a})}\int\limits_{0}^{\infty}\epsilon\,d\epsilon.
\end{equation}
Again, as in the two-dimensional case, we can take the auxiliary integral
\begin{equation}
\int\limits_{-\infty}^{\infty}e^{i\epsilon\frac{q_{b}-q_{a}}{a}}d\epsilon=2\pi\delta\left(\frac{q_{b}-q_{a}}{a}\right)
=2\pi a\sqrt{k_{a}^{2}+{\vec k_{a\perp}}^{2}}\,\delta(k_{a}-k_{b}),
\end{equation}
which gives
\begin{equation}
\delta(k_{a}-k_{a})=\frac{1}{\pi a\sqrt{k_{a}^{2}+{\vec k_{a\perp}}^{2}}}\int\limits_{0}^{\infty}d\epsilon.
\end{equation}
Finally, for the energy of a one-particle state one gets
\begin{equation}
\frac{\bra{k_{a},\vec k_{a\perp}}H_{R}\ket{k_{a},\vec k_{a\perp}}}{\bra{k_{a},\vec k_{a\perp}}\ket{k_{a},\vec k_{a\perp}}}-E_{vac\,\,4D}
=\frac{1}{2}\frac{\int\limits_{0}^{\infty}\epsilon\,d\epsilon}{\int\limits_{0}^{\infty}d\epsilon}\to\infty,
\end{equation}
which looks the same as in the two-dimensional case. Introducing the cutoff $\epsilon_{max}=M_{Pl}$, we get
\begin{equation}
\frac{\bra{k_{a},\vec k_{a\perp}}H_{R}\ket{k_{a},\vec k_{a\perp}}}{\bra{k_{a},\vec k_{a\perp}}\ket{k_{a},\vec k_{a\perp}}}-E_{vac\,\,4D}
=\frac{M_{Pl}}{4},
\end{equation}
which, exactly as in the two-dimensional case \eqref{partenergy3cutoff}, again depends only on the cutoff scale and does not depend on the acceleration $a$.

\subsubsection{Energy of the Minkowski spacetime normalized state in Rindler spacetime}
Let us take a normalized state
\begin{equation}
\ket{\psi}=\int\limits_{-\infty}^{\infty}dk\iint d^{2}k_{\perp}f\left(k,\vec k_{\perp}\right)\ket{k,\vec k_{\perp}}
\end{equation}
with
\begin{equation}\label{fk4D}
f\left(k,\vec k_{\perp}\right)=\frac{\theta(|\vec k_{\perp}|-\tilde k)\,\theta(k-k_{m})\,\theta(k_{M}-k)}{\sqrt{\pi}\,{\tilde k}\,\sqrt{k_{M}-k_{m}}},
\end{equation}
where $k_{M}>k_{m}>0$. One can show that the energy of the state is
\begin{equation}
\bra{\psi}H_{M}\ket{\psi}=\frac{2}{3{\tilde k}^{2}(k_{M}-k_{m})}\left(C(k_{M})-C(k_{m})-\frac{k_{M}^{4}-k_{m}^{4}}{4}\right),
\end{equation}
where \cite{Dwight}
\begin{equation}
C(k)=\int\left(k^{2}+{\tilde k}^{2}\right)^{\frac{3}{2}}dk=\frac{1}{4}k\left(k^{2}+{\tilde k}^{2}\right)^{\frac{3}{2}}+\frac{3}{8}{\tilde k}^{2}k\sqrt{k^{2}+{\tilde k}^{2}}+\frac{3}{8}{\tilde k}^{4}\ln\left(\frac{k}{\tilde k}+\sqrt{\frac{k^{2}}{{\tilde k}^{2}}+1}\right).
\end{equation}
Let $k_{m}\gg\tilde k$ (this condition will be also used in the subsequent calculations). Then up to and including the terms $\sim{\tilde k}^{2}$
\begin{equation}
C(k)\approx\frac{k^{4}}{4}+\frac{3}{4}{\tilde k}^{2}k^{2},
\end{equation}
leading to
\begin{equation}
\bra{\psi}H_{M}\ket{\psi}\approx\frac{k_{M}+k_{m}}{2},
\end{equation}
which looks like \eqref{EnergypsiM}.

The energy of the state $\ket{\psi}$ in Rindler spacetime has the form
\begin{align}\nonumber
\bra{\psi}H_{R}\ket{\psi}&=
\int\limits_{-\infty}^{\infty}dk_{a}\int\limits_{-\infty}^{\infty}dk_{b}\iint d^{2}k_{a\perp}\iint d^{2}k_{b\perp}
f^{*}\left(k_{a},\vec k_{a\perp}\right)f\left(k_{b},\vec k_{b\perp}\right)\bra{k_{a},\vec k_{a\perp}}H_{R}\ket{k_{b},\vec k_{b\perp}}\\\nonumber
&=\int\limits_{0}^{\infty}d\epsilon\,\epsilon\iint d^{2}k_{\perp}
\int\limits_{-\infty}^{\infty}dk_{a}\int\limits_{-\infty}^{\infty}dk_{b}\iint d^{2}k_{a\perp}\iint d^{2}k_{b\perp}
f^{*}\left(k_{a},\vec k_{a\perp}\right)f\left(k_{b},\vec k_{b\perp}\right)\\\label{energypsistate}
&\times\bra{k_{a},\vec k_{a\perp}}b^{\dagger}(\epsilon,\vec k_{\perp})b^{}(\epsilon,\vec k_{\perp})\ket{k_{b},\vec k_{b\perp}},
\end{align}
where $\bra{k_{a},\vec k_{a\perp}}b^{\dagger}(\epsilon,\vec k_{\perp})b^{}(\epsilon,\vec k_{\perp})\ket{k_{b},\vec k_{b\perp}}$ is defined by formula \eqref{4Dbbkakb}. Without vacuum energy $E_{vac\,\,4D}$ formula \eqref{energypsistate} can be brought to the form
\begin{align}\nonumber
\bra{\psi}H_{R}\ket{\psi}-E_{vac\,\,4D}&\approx
\int\limits_{0}^{\infty}d\epsilon\,\epsilon\iint d^{2}k_{\perp}
\int\limits_{-\infty}^{\infty}dk_{a}\int\limits_{-\infty}^{\infty}dk_{b}\iint d^{2}k_{a\perp}\iint d^{2}k_{b\perp}\\\nonumber
&\times f^{*}\left(k_{a},\vec k_{a\perp}\right)f\left(k_{b},\vec k_{b\perp}\right)
\frac{e^{i\epsilon\frac{q_{b}-q_{a}}{a}}\delta^{(2)}(\vec k_{\perp}-\vec k_{a\perp})\,\delta^{(2)}(\vec k_{\perp}-\vec k_{b\perp})}
{2\pi a\left(k_{a}^{2}+{\vec k_{\perp}}^{2}\right)^{\frac{1}{4}}\left(k_{b}^{2}+{\vec k_{\perp}}^{2}\right)^{\frac{1}{4}}}\\\label{energypsistate2}
&=\frac{1}{2\pi a}\int\limits_{0}^{\infty}d\epsilon\,\epsilon\iint d^{2}k_{\perp}F^{*}(\epsilon,\vec k_{\perp})F(\epsilon,\vec k_{\perp})
\end{align}
with
\begin{equation}\label{F4D}
F(\epsilon,\vec k_{\perp})=\int\limits_{-\infty}^{\infty}dk_{b}\iint d^{2}k_{b\perp}
f\left(k_{b},\vec k_{b\perp}\right)\frac{e^{i\epsilon\frac{q_{b}}{a}}\delta^{(2)}(\vec k_{\perp}-\vec k_{b\perp})}
{\left(k_{b}^{2}+{\vec k_{\perp}}^{2}\right)^{\frac{1}{4}}}=
\int\limits_{-\infty}^{\infty}dk_{b}
\frac{e^{i\epsilon\frac{q_{b}}{a}}f\left(k_{b},\vec k_{\perp}\right)}{\left(k_{b}^{2}+{\vec k_{\perp}}^{2}\right)^{\frac{1}{4}}},
\end{equation}
where $\sinh(q_{b})=\frac{k_{b}}{|\vec k_{\perp}|}$. With \eqref{fk4D}, \eqref{F4D} takes the form
\begin{equation}\label{F4D2}
F(\epsilon,\vec k_{\perp})=\frac{\theta(|\vec k_{\perp}|-\tilde k)}{\sqrt{\pi}\,{\tilde k}\,\sqrt{k_{M}-k_{m}}}
\int\limits_{k_{m}}^{k_{M}}dk_{b}\frac{e^{i\epsilon\frac{q_{b}}{a}}}{\left(k_{b}^{2}+{\vec k_{\perp}}^{2}\right)^{\frac{1}{4}}}.
\end{equation}
Passing from $k_{b}$ to $q_{b}$ in the integral in the latter formula, $F(\epsilon,\vec k_{\perp})$ can be rewritten as
\begin{equation}\label{F4D3}
F(\epsilon,\vec k_{\perp})=\frac{\theta(|\vec k_{\perp}|-\tilde k)\sqrt{|\vec k_{\perp}|}}{\sqrt{\pi}\,{\tilde k}\,\sqrt{k_{M}-k_{m}}}
\int\limits_{q_{m}}^{q_{M}}dq_{b}\,e^{i\epsilon\frac{q_{b}}{a}}\sqrt{\cosh(q_{b})}.
\end{equation}

Since we suppose that $k_{m}\gg\tilde k$, then $\cosh(q_{b})\gg 1$, which means that $\cosh(q_{b})\approx\frac{e^{q_{b}}}{2}$ with a good accuracy. In this approximation
\begin{equation}\label{F4D4}
F(\epsilon,\vec k_{\perp})\approx\frac{\theta(|\vec k_{\perp}|-\tilde k)\sqrt{|\vec k_{\perp}|}}
{\sqrt{2\pi}\,{\tilde k}\,\sqrt{k_{M}-k_{m}}\left(\frac{1}{2}+i\frac{\epsilon}{a}\right)}
\left(e^{\left(i\frac{\epsilon}{a}+\frac{1}{2}\right)q_{M}}-e^{\left(i\frac{\epsilon}{a}+\frac{1}{2}\right)q_{m}}\right).
\end{equation}
Then formula \eqref{energypsistate2} takes the form
\begin{align}\nonumber
&\bra{\psi}H_{R}\ket{\psi}-E_{vac\,\,4D}\approx
\frac{1}{4\pi^{2}a{\tilde k}^{2}(k_{M}-k_{m})}\int\limits_{0}^{\infty}\frac{\epsilon\,d\epsilon}{\frac{1}{4}+\frac{\epsilon^{2}}{a^{2}}}\\\label{EWP4D}
&\times\iint d^{2}k_{\perp}\theta(|\vec k_{\perp}|-\tilde k)\,|\vec k_{\perp}|
\left(e^{q_{M}}+e^{q_{m}}-2\,e^{\frac{q_{M}+q_{m}}{2}}\cos\left(\frac{\epsilon}{a}(q_{M}-q_{m})\right)\right).
\end{align}
For $k_{m}\gg\tilde k$ we have $q_{m,M}\approx\ln\left(2\,\frac{k_{m,M}}{|\vec k_{\perp}|}\right)$ with a good accuracy. In this approximation the leading contribution in \eqref{EWP4D} has the form
\begin{align}\nonumber
&\bra{\psi}H_{R}\ket{\psi}-E_{vac\,\,4D}\approx
\frac{1}{4\pi^{2}a{\tilde k}^{2}(k_{M}-k_{m})}\int\limits_{0}^{\infty}\frac{\epsilon\,d\epsilon}{\frac{1}{4}+\frac{\epsilon^{2}}{a^{2}}}\iint d^{2}k_{\perp}\theta(|\vec k_{\perp}|-\tilde k)\left(2k_{M}+2k_{m}\right)\\\label{EWP4D2}
&=\frac{k_{M}+k_{m}}{2\pi a(k_{M}-k_{m})}\int\limits_{0}^{\infty}\frac{\epsilon\,d\epsilon}{\frac{1}{4}+\frac{\epsilon^{2}}{a^{2}}}
=\frac{a(k_{M}+k_{m})}{4\pi(k_{M}-k_{m})}\ln\left(\frac{\epsilon^{2}}{a^{2}}+\frac{1}{4}\right)\bigg|_{\epsilon=0}^{\infty}\to\infty,
\end{align}
which coincides with the first term in \eqref{wpenergy4}. Since the form of the leading term in \eqref{EWP4D2} is the same as in the two-dimensional case, we expect the same behavior of \eqref{EWP4D2} with a cutoff applied as in the two-dimensional case, see \eqref{wpenergy4cutoff}; and in the case of a classical wave packet corresponding to \eqref{fk4D}, see \eqref{energylargeRx}.

\subsection{Small discussion}
We see that in many ways the two-dimensional and four-dimensional cases are similar, including the pathological behavior. The latter in the examples presented above is due to the form of the corresponding solutions. To understand it in more detail, let us consider coordinate transformations \eqref{TRind} and \eqref{XRind} for very small $a$ and $t$. We get
\begin{align}\label{TRinddis}
&T\approx t\,e^{ax},\\\label{XRinddis}
&X\approx\frac{1}{a}\,e^{ax}.
\end{align}
For small $x$ in the leading order in $t$ and $x$ we can write $T\approx t$ and $X\approx\frac{1}{a}+x$. However, wave function of a scalar particle in Minkowski spacetime is just a plane wave distributed over the whole space, so we cannot restrict ourselves only to small values of $x$ --- we should consider large values of $x$ too. But for large values of $x$ the exponents in \eqref{TRinddis} and \eqref{XRinddis} become relevant, even for extremely small accelerations, which changes the picture drastically. This can be explicitly seen in the case of classical wave packet \eqref{wavepacket2}, in which the energy becomes infinite due to contribution of the exponents at large $x$ \eqref{energylargeRx}: even though the wave packet tail tends to zero as $x\to\infty$, it does not tend to zero rapidly enough to compensate the contribution caused by the increasing frequency of oscillations as $x\to\infty$. The same is valid for the normalized states discussed above too. Of course, for highly localized wave packets (like Gaussian wave packets) these pathologies can be avoided. To make sure of this one can consider very small $a$, narrow Gaussian wave packet (i.e., with the width much smaller than $\frac{1}{a}$) located in such a way that at $T=0$ its center is at $X=\frac{1}{a}$; and then calculate the energy in Rindler spacetime using the coordinate transformations \eqref{TRinddis} and \eqref{XRinddis}.

Such a pathological behavior of the integrals of motion obviously poses a problem of interpretation and usage of the corresponding quantum field theory. In the standard quantum field theory calculations are very often based on the use of plane waves. Indeed, when we consider different inertial reference frames, plane waves from one reference frame remain plane waves in another reference frame, which does not cause any problems. However, this is not so in the case of a uniformly accelerated reference frame, in which we can no longer use the familiar formalism based on the use of plane waves --- one should consider highly localized wave packets instead. On the other hand, except the particles with the thermal spectrum provided by the Minkowski vacuum $\ket{0_{M}}$, there may exist other artifacts with infinite energy (but without thermal spectrum) in Rindler spacetime provided by various ``regular'' states in Minkowski spacetime. It is expected that the same thing may happen in the case of black holes too.

In this connection, it is worth reading a still actual discussion of possible problems which may arise when one tries to construct quantum field theories in non-flat static background metrics presented in Section~III of \cite{Fulling:1972md}. The pathologies discussed above seem to be of the same sort as the problems mentioned in \cite{Fulling:1972md}.

\section{Four-dimensional Schwarzschild black hole}
In the two-dimensional case the quantum theories for a uniformly accelerated reference frame and a Schwarzschild black hole are almost identical. Indeed, in both theories one can construct consistent quantum field theories providing Boulware--like, Hartle-Hawking--like and Unruh--like vacuum states. However, the case of a four-dimensional Schwarzschild black hole is much more different from the case of a two-dimensional Schwarzschild black hole than the case of a uniformly accelerated reference frame in four dimensions from the one in two dimensions. Since the similarity of quantum field theories describing a uniformly accelerated reference frame and (at least) a Schwarzschild black hole is widely used in the literature, it is important to point to existing and possible problems in the case of black holes, as well as to take a more precise look at the differences between the cases of a uniformly accelerated reference frame and a four-dimensional Schwarzschild black hole.

\subsection{Quantum scalar field theory in four-dimensional Schwarzschild spacetime}\label{QFTSchw}
\begin{figure}[ht]
\centering
\includegraphics[width=0.9\linewidth]{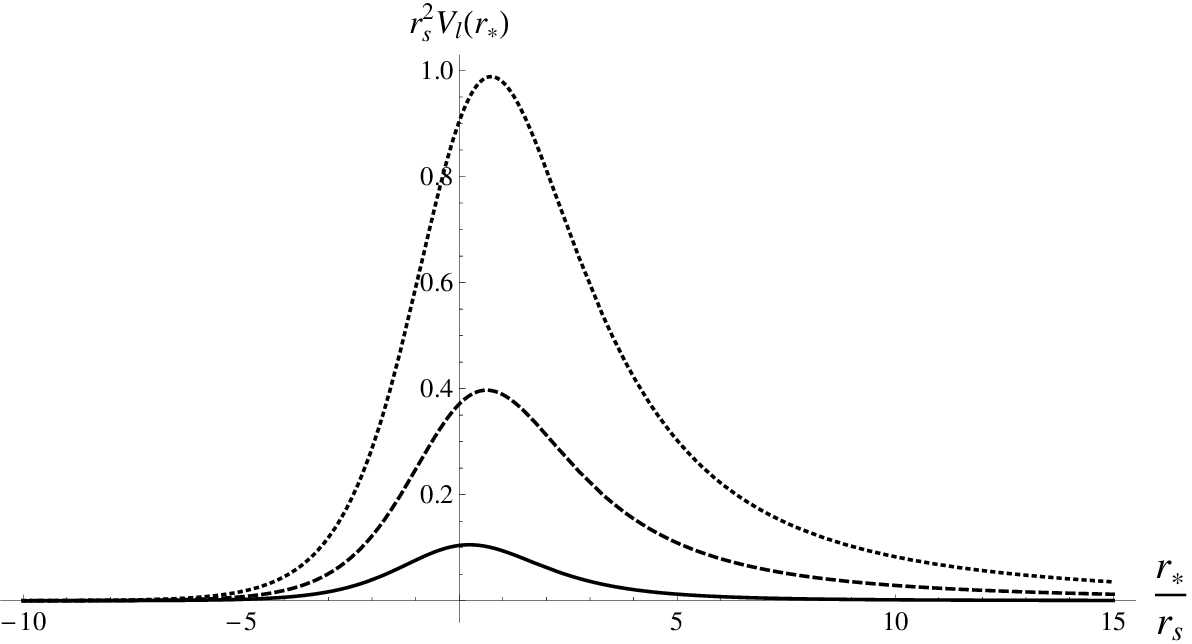}
\caption{$V_{l}(r_{*})$ for $l=0$ (solid line), $l=1$ (dashed line), and $l=2$ (dotted line).}\label{Vl}
\end{figure}

In this section, let us reproduce the well-known results concerning quantum scalar field theory in four-dimensional Schwarzschild spacetime which will be necessary for the subsequent analysis. For $r>r_{s}$ the quantized massless scalar field can be represented as
\begin{equation}\label{operatordecspher}
\phi(t,r,\theta,\varphi)
=\sum\limits_{p=1}^{2}\sum\limits_{l=0}^{\infty}\sum\limits_{m=-l}^{l}\int\limits_{0}^{\infty}\frac{dE}{\sqrt{2E}}\left(e^{-iEt}
\phi_{lmp}^{}(E,r,\theta,\varphi)a_{lmp}^{}(E)+\textrm{h.c.}\right),
\end{equation}
where
\begin{equation}\label{philm}
\phi_{lmp}^{}(E,r,\theta,\varphi)=Y_{lm}(\theta,\varphi)\frac{\psi_{lp}(E,r)}{r}.
\end{equation}
The function $\psi_{lp}(E,r(r_{*}))$, where $r_{*}$ is the tortoise coordinate defined by \eqref{tortoise}, satisfies the Schr\"{o}\-dinger equation
\begin{equation}\label{eqSchr}
-\frac{d^{2}\psi_{lp}(E,r_{*})}{dr_{*}^{2}}+V_{l}(r_{*})\psi_{lp}(E,r_{*})=E^{2}\psi_{lp}(E,r_{*}),
\end{equation}
where the potential has the form \cite{Christensen:1977jc,Barranco:2011eyw}
\begin{equation}\label{VSchr1}
V_{l}(r_{*})=\frac{r(r_{*})-r_{s}}{r(r_{*})}\left(\frac{l(l+1)}{r^{2}(r_{*})}+\frac{r_{s}}{r^{3}(r_{*})}\right),
\end{equation}
see Fig.~\ref{Vl}.
It is known that the functions $\psi_{lp}(E,r_{*})$ are expressed through the Heun functions \cite{Barranco:2011eyw,Zecca3,KRSV}, so in fact we have exact analytical solutions for the wave functions of the modes.

Usually, solutions $\psi_{lp}(E,r_{*})$ are considered in such a form that for $|r_{*}|\to\infty$ they look like \cite{Christensen:1977jc}
\begin{align}\label{psiTR1}
&\psi_{l1}(E,r_{*})=C_{1}(E,l)\left(e^{iEr_{*}}+R_{l1}(E)\,e^{-iEr_{*}}\right),\qquad r_{*}\to -\infty,\\\label{psiTR2}
&\psi_{l1}(E,r_{*})=C_{1}(E,l)\,T_{l1}(E)\,e^{iEr_{*}},\qquad\qquad\qquad\quad\, r_{*}\to\infty
\end{align}
and
\begin{align}\label{psiTR3}
&\psi_{l2}(E,r_{*})=C_{2}(E,l)\left(e^{-iEr_{*}}+R_{l2}(E)\,e^{iEr_{*}}\right),\qquad r_{*}\to\infty,\\\label{psiTR4}
&\psi_{l2}(E,r_{*})=C_{2}(E,l)\,T_{l2}(E)\,e^{-iEr_{*}},\qquad\qquad\qquad\,\,\, r_{*}\to-\infty.
\end{align}
Here $C_{p}(E,l)$ are the normalization constant, $R_{p}(E,l)$ and $T_{p}(E,l)$ are the coefficients responsible for the reflected and transmitted waves respectively. Without loss of generality, $C_{p}(E,l)$ can be chosen to be real. Since the normalization integrals
\begin{equation}\label{orthpsiBH}
\int\limits_{-\infty}^{\infty}\psi_{lp}^{*}(E,r_{*})\psi_{lp}(E',r_{*})dr_{*}=\delta(E-E')
\end{equation}
for solutions \eqref{psiTR1} and \eqref{psiTR2} diverge for $E=E'$, normalization constants for such type of solutions can be determined by the behavior of eigenfunctions in the asymptotic regions; see \cite{LL-QM}. Rewriting \eqref{orthpsiBH} as
\begin{equation}
\int\limits_{-\infty}^{\infty}\psi_{l1}^{*}(E,r_{*})\psi_{l1}(E',r_{*})dr_{*}+\textrm{c.c.}=2\delta(E-E'),
\end{equation}
the leading terms in the limit $E'\to E$ are
\begin{align}\nonumber
C_{1}^{2}(E,l)&\left(\int\limits_{-\infty}^{0}\left(e^{i(E-E')r_{*}}+e^{-i(E-E')r_{*}}
+|R_{l1}(E)|^{2}\left(e^{-i(E-E')r_{*}}+e^{i(E-E')r_{*}}\right)\right)dr_{*}
\right.\\\label{deltaEE}&\left.+
\int\limits_{0}^{\infty}|T_{l1}(E)|^{2}\left(e^{-i(E-E')r_{*}}+e^{i(E-E')r_{*}}\right)dr_{*}\right)=2\delta(E-E'),
\end{align}
resulting in
\begin{equation}
C_{1}^{2}(E,l)=\frac{1}{\pi\left(1+|R_{l1}(E)|^{2}+|T_{l1}(E)|^{2}\right)}=\frac{1}{2\pi},
\end{equation}
where we have used the fact that $|R_{l1}(E)|^{2}+|T_{l1}(E)|^{2}=1$ \cite{Messiah}. Analogously,
\begin{equation}
C_{2}^{2}(E,l)=\frac{1}{2\pi}.
\end{equation}
There exist other relations between the coefficients in solutions \eqref{psiTR1}--\eqref{psiTR4} that will be used in the subsequent calculations \cite{Messiah}:
\begin{align}\label{RT1}
&T_{l1}(E)=T_{l2}(E),\\\label{RT2}
&R_{l2}(E)T_{l1}^{*}(E)=-R_{l1}^{*}(E)T_{l2}(E).
\end{align}
These relations can be obtained by calculating various Wronskians \cite{Messiah} for solutions \eqref{psiTR1}--\eqref{psiTR4}.

The canonically conjugate momentum in this theory is
\begin{equation}\label{ccm}
\pi(t,r_{*},\theta,\varphi)=\frac{\partial\mathcal{L}}{\partial\dot\phi(t,r_{*},\theta,\varphi)}
=\sqrt{-g}\,g^{00}\dot\phi(t,r_{*},\theta,\varphi)=r^{2}(r_{*})\sin\theta\dot\phi(t,r_{*},\theta,\varphi),
\end{equation}
where $\dot{}=\frac{\partial}{\partial t}$. With \eqref{ccm}, it is expected that the canonical commutation relations
\begin{align}\label{CCRSchw1}
&[\phi(t,r_{*},\theta,\varphi),\dot\phi(t,r_{*}',\theta',\varphi')]
=i\frac{\delta(r_{*}-r_{*}')\delta(\theta-\theta')\delta(\varphi-\varphi')}{r^{2}(r_{*})\sin\theta},\\
&[\phi(t,r_{*},\theta,\varphi),\phi(t,r_{*}',\theta',\varphi')]=0,\\\label{CCRSchw3}
&[\dot\phi(t,r_{*},\theta,\varphi),\dot\phi(t,r_{*}',\theta',\varphi')]=0
\end{align}
hold. Let us check that it is indeed so. The first commutator can be brought to the form
\begin{align}\nonumber
&[\phi(t,r_{*},\theta,\varphi),\dot\phi(t,r_{*}',\theta',\varphi')]\\
&=\frac{i}{8\pi rr'}\sum\limits_{l=0}^{\infty}(2l+1)P_{l}(\cos\alpha)
\sum\limits_{p=1}^{2}\int\limits_{0}^{\infty}dE\left(\psi_{lp}(E,r_{*})\psi_{lp}^{*}(E,r_{*}')+\psi_{lp}^{*}(E,r_{*})\psi_{lp}(E,r_{*}')\right),
\end{align}
where we have used the addition theorem for spherical harmonics \cite{Korn-Korn}
\begin{equation}
\sum\limits_{m=-l}^{l}Y_{lm}\left(\theta,\varphi\right)Y_{lm}^{*}\left(\theta',\varphi'\right)
=\frac{2l+1}{4\pi}P_{l}\left(\cos\alpha\right)
\end{equation}
with $\cos\alpha=\cos\theta\cos\theta'+\sin\theta\sin\theta'\cos\left(\varphi-\varphi'\right)$. Using the completeness relation
\begin{equation}
\sum\limits_{p=1}^{2}\int\limits_{0}^{\infty}\psi_{lp}(E,r_{*})\psi_{lp}^{*}(E,r_{*}')dE=\delta(r_{*}-r_{*}')
\end{equation}
and the relation \cite{LL-QM}
\begin{equation}\label{LL-QM-relation}
\frac{1}{4}\sum\limits_{l=0}^{\infty}(2l+1)P_{l}(\cos\alpha)=\delta(1-\cos\alpha),
\end{equation}
one gets
\begin{equation}
[\phi(t,r_{*},\theta,\varphi),\dot\phi(t,r_{*}',\theta',\varphi')]=i\frac{\delta(r_{*}-r_{*}')\delta(1-\cos\alpha)}{\pi r^{2}}.
\end{equation}
The factor $\frac{1}{\pi}\delta(1-\cos\alpha)$ corresponds exactly to $\delta(\theta-\theta')\delta(\varphi-\varphi')/\sin\theta$ (see, for example, \cite{LL-QM} or calculations in the Appendices of \cite{Egorov:2022hgg,Smolyakov:2023pml}). Thus the first commutation relation is exactly satisfied.

Checking the validity of two other commutation relations turns out to be more complicated. In explicit form the corresponding commutators look like
\begin{align}\nonumber
&[\phi(t,r_{*},\theta,\varphi),\phi(t,r_{*}',\theta',\varphi')]\\
&=\frac{i}{8\pi rr'}\sum\limits_{l=0}^{\infty}(2l+1)P_{l}(\cos\alpha)
\sum\limits_{p=1}^{2}\int\limits_{0}^{\infty}\frac{dE}{E}\left(\psi_{lp}(E,r_{*})\psi_{lp}^{*}(E,r_{*}')-\psi_{lp}^{*}(E,r_{*})\psi_{lp}(E,r_{*}')\right),\\\nonumber
&[\dot\phi(t,r_{*},\theta,\varphi),\dot\phi(t,r_{*}',\theta',\varphi')]\\
&=\frac{i}{8\pi rr'}\sum\limits_{l=0}^{\infty}(2l+1)P_{l}(\cos\alpha)
\sum\limits_{p=1}^{2}\int\limits_{0}^{\infty}dE\,E\left(\psi_{lp}(E,r_{*})\psi_{lp}^{*}(E,r_{*}')-\psi_{lp}^{*}(E,r_{*})\psi_{lp}(E,r_{*}')\right).
\end{align}
Let us represent the coefficients in solutions \eqref{psiTR1}--\eqref{psiTR4} as
\begin{align}
&T_{l1}(E)=T_{l2}(E)=|T_{l1}(E)|e^{i\delta_{lT}(E)},\\
&R_{l1}(E)=|R_{l1}(E)|e^{i\delta_{lR}(E)}
\end{align}
and, as follows from \eqref{RT2},
\begin{equation}
R_{l2}(E)=-|R_{l1}(E)|e^{i(2\delta_{lT}(E)-\delta_{lR}(E))}.
\end{equation}
Now let us build the linear combinations
\begin{align}\label{tildepsi1}
&\tilde\psi_{l1}(E,r_{*})=e^{-i\frac{\delta_{lR}(E)}{2}}\cos\left(\gamma_{l}(E)\right)\psi_{l1}(E,r_{*})
+e^{i\left(\frac{\delta_{lR}(E)}{2}-\delta_{lT}(E)\right)}\sin\left(\gamma_{l}(E)\right)\psi_{l2}(E,r_{*}),\\\label{tildepsi2}
&\tilde\psi_{l2}(E,r_{*})=i\left(-e^{-i\frac{\delta_{lR}(E)}{2}}\sin\left(\gamma_{l}(E)\right)\psi_{l1}(E,r_{*})
+e^{i\left(\frac{\delta_{lR}(E)}{2}-\delta_{lT}(E)\right)}\cos\left(\gamma_{l}(E)\right)\psi_{l2}(E,r_{*})\right)
\end{align}
with $\tan\left(\gamma_{l}(E)\right)=\frac{1-|R_{l1}(E)|}{|T_{l1}(E)|}$. One can show that
\begin{align}
&\tilde\psi_{l1}(E,r_{*})=\sqrt{\frac{2}{\pi}}\cos\left(\gamma_{l}(E)\right)\cos\left(Er_{*}-\frac{\delta_{lR}(E)}{2}\right)
\qquad\qquad\qquad\textrm{for}\qquad r_{*}\to-\infty,\\
&\tilde\psi_{l1}(E,r_{*})=\sqrt{\frac{2}{\pi}}\sin\left(\gamma_{l}(E)\right)\cos\left(Er_{*}+\delta_{lT}(E)-\frac{\delta_{lR}(E)}{2}\right)
\qquad\textrm{for}\qquad r_{*}\to\infty,\\
&\tilde\psi_{l2}(E,r_{*})=\sqrt{\frac{2}{\pi}}\sin\left(\gamma_{l}(E)\right)\sin\left(Er_{*}-\frac{\delta_{lR}(E)}{2}\right)
\qquad\qquad\qquad\textrm{for}\qquad r_{*}\to-\infty,\\
&\tilde\psi_{l2}(E,r_{*})=\sqrt{\frac{2}{\pi}}\cos\left(\gamma_{l}(E)\right)\sin\left(Er_{*}+\delta_{lT}(E)-\frac{\delta_{lR}(E)}{2}\right)
\qquad\textrm{for}\qquad r_{*}\to\infty.
\end{align}
It is clear that since $\tilde\psi_{lp}(E,r_{*})$ are real for $r_{*}\to\pm\infty$, they are real for all $r_{*}$. Using \eqref{tildepsi1} and \eqref{tildepsi2}, one can also show that
\begin{equation}
\sum\limits_{p=1}^{2}\tilde\psi_{lp}(E,r_{*})\tilde\psi_{lp}(E,r_{*}')=\sum\limits_{p=1}^{2}\tilde\psi_{lp}(E,r_{*})\tilde\psi_{lp}^{*}(E,r_{*}')
=\sum\limits_{p=1}^{2}\psi_{lp}(E,r_{*})\psi_{lp}^{*}(E,r_{*}').
\end{equation}
Thus
\begin{align}\nonumber
&\sum\limits_{p=1}^{2}\left(\psi_{lp}(E,r_{*})\psi_{lp}^{*}(E,r_{*}')-\psi_{lp}^{*}(E,r_{*})\psi_{lp}(E,r_{*}')\right)\\
=&\sum\limits_{p=1}^{2}\left(\tilde\psi_{lp}(E,r_{*})\tilde\psi_{lp}(E,r_{*}')-\tilde\psi_{lp}(E,r_{*})\tilde\psi_{lp}(E,r_{*}')\right)=0,
\end{align}
which leads to
\begin{align}
&[\phi(t,r_{*},\theta,\varphi),\phi(t,r_{*}',\theta',\varphi')]=0,\\
&[\dot\phi(t,r_{*},\theta,\varphi),\dot\phi(t,r_{*}',\theta',\varphi')]=0.
\end{align}
We see that commutation relations \eqref{CCRSchw1}--\eqref{CCRSchw3} are exactly satisfied.

The Hamiltonian of the theory can be easily obtained and has the form (skipping the irrelevant $c$-number term)
\begin{equation}
H_{Schw}=\sum\limits_{l=0}^{\infty}\sum\limits_{m=-l}^{l}\sum\limits_{p=1}^{2}\int\limits_{0}^{\infty}Ea_{lmp}^{\dagger}(E)a_{lmp}^{}(E)dE.
\end{equation}
The vacuum is defined as
\begin{equation}
a_{lmp}^{}(E)\ket{0_{B}}=0,
\end{equation}
this vacuum state is called the Boulware vacuum.

A more detailed discussion of quantum scalar field theory in Schwarzschild spacetime, including the real solutions $\tilde\psi_{lp}(E,r_{*})$ and transition to the scatteringlike states, some of which resemble properly normalized plane waves far away from the black hole, can be found in \cite{Egorov:2022hgg,Smolyakov:2023pml}.

If there exists a consistent quantum scalar field theory in Kruskal-Szekeres spacetime and the modes in this theory have a certain form on the horizon (namely, $\sim e^{-i\omega(T-X)}$ and $\sim e^{-i\omega(T+X)}$; the coordinates $T$ and $X$ are defined by \eqref{Trstart} and \eqref{Xrstart} respectively), it is possible to describe manifestations of the Hartle-Hawking and Unruh vacuum states in terms of the modes inherent to the theory in Schwarzschild spacetime. These results can be found elsewhere, for example, a detailed analysis was presented in \cite{Candelas:1980zt,Christensen:1977jc}. Below we will focus on formal aspects of the corresponding quantum field theories. As will be shown in the next sections, assumption about the form of the modes on the horizon in Kruskal-Szekeres spacetime is not justified well enough. In particular, it leads to some explicit problems with the Unruh vacuum state, which will be discussed in the next section.

\subsection{Inconsistency of the Unruh vacuum state}
Let us take a closer look at the theory providing the Unruh vacuum state. In paper \cite{Unruh:1976db} it was proposed to use the following solution for the mode on the horizon, which in our notations looks like
\begin{equation}\label{Umode}
\tilde\phi_{l}(T,X,\omega)|_{T+X=0}=\frac{1}{r_{s}\sqrt{4\pi\omega}}\,e^{-i\omega(T-X)}.
\end{equation}
In fact, up to a constant it has the same form as in the two-dimensional case. The first problem with this mode is that if $\tilde\phi_{l}(T,X,\omega)$ has the form \eqref{Umode} not only for $T+X=0$, but also in the vicinity of the horizon, actually it is not a solution to the corresponding equation of motion. To demonstrate it explicitly, let us derive the equation of motion for the scalar field in the vicinity of the horizon. Using explicit form of the metric
\begin{equation}\label{KS4D}
ds^2=\frac{r_{s}e^{-\frac{r(T,X)}{r_{s}}}}{r(T,X)}\left(dT^2-dX^2\right)-r^{2}(T,X)\left(d\theta^{2}+\sin^{2}\theta d\varphi^{2}\right)
\end{equation}
and expanding the scalar field in spherical harmonics, for the ``radial'' solutions in the Kruskal-Szekeres coordinates in wedge I one gets the following equation of motion:
\begin{equation}\label{KSqescf}
\frac{\partial^{2}\phi_{l}}{\partial T^{2}}-\frac{\partial^{2}\phi_{l}}{\partial X^{2}}
+\frac{2}{r}\left(\frac{\partial r}{\partial T}\frac{\partial\phi_{l}}{\partial T}-
\frac{\partial r}{\partial X}\frac{\partial\phi_{l}}{\partial X}\right)+\frac{l(l+1)r_{s}e^{-\frac{r}{r_{s}}}}{r^{3}}\phi_{l}=0,
\end{equation}
where $\phi_{l}=\phi_{l}(T,X)$. Using the fact that
\begin{align}
&\frac{\partial r}{\partial T}=\frac{\partial r}{\partial r_{*}}\frac{\partial r_{*}}{\partial T}=-\frac{2r_{s}(r-r_{s})T}{r(X^{2}-T^{2})},\\
&\frac{\partial r}{\partial X}=\frac{\partial r}{\partial r_{*}}\frac{\partial r_{*}}{\partial X}=\frac{2r_{s}(r-r_{s})X}{r(X^{2}-T^{2})},
\end{align}
where we have used \eqref{tortoise} and \eqref{rstar}, Eq.~\eqref{KSqescf} can be rewritten as
\begin{equation}\label{KSqescf2}
\frac{\partial^{2}\phi_{l}}{\partial T^{2}}-\frac{\partial^{2}\phi_{l}}{\partial X^{2}}
-\frac{4r_{s}(r-r_{s})T}{r^{2}(X^{2}-T^{2})}\frac{\partial\phi_{l}}{\partial T}
-\frac{4r_{s}(r-r_{s})X}{r^{2}(X^{2}-T^{2})}\frac{\partial\phi_{l}}{\partial X}
+\frac{l(l+1)r_{s}e^{-\frac{r}{r_{s}}}}{r^{3}}\phi_{l}=0.
\end{equation}
For $r\to r_{s}$ (i.e., for $r_{*}\to-\infty$) from relation \eqref{tortoise} one can get
\begin{equation}
r-r_{s}\approx e^{-1}r_{s}e^{\frac{r_{*}}{r_{s}}},
\end{equation}
resulting in
\begin{equation}
\frac{4r_{s}(r-r_{s})}{r^{2}(X^{2}-T^{2})}\approx\frac{4r_{s}^{2}e^{-1}e^{\frac{r_{*}}{r_{s}}}}{r_{s}^{2}(X^{2}-T^{2})}=\frac{1}{er_{s}^{2}},
\end{equation}
where we have again used relation \eqref{rstar}. Thus for $r\to r_{s}$ Eq.~\eqref{KSqescf2} takes the form
\begin{equation}\label{KSqescf3}
\frac{\partial^{2}\phi_{l}}{\partial T^{2}}-\frac{\partial^{2}\phi_{l}}{\partial X^{2}}
-\frac{T}{er_{s}^{2}}\frac{\partial\phi_{l}}{\partial T}
-\frac{X}{er_{s}^{2}}\frac{\partial\phi_{l}}{\partial X}
+\frac{l(l+1)}{er_{s}^{2}}\phi_{l}=0.
\end{equation}
Although formally we used coordinate transformations only for wedge I, the metric is smooth on the horizon, so the continuity arguments suggest that Eq.~\eqref{KSqescf3} can be used on both sides of the horizon.

One can check that \eqref{Umode} does not satisfy Eq.~\eqref{KSqescf3}. Indeed, substituting \eqref{Umode} into the lhs of Eq.~\eqref{KSqescf3}, one gets
\begin{equation}
\frac{1}{r_{s}\sqrt{4\pi\omega}}\left(\frac{i\omega}{er_{s}^{2}}(T-X)+\frac{l(l+1)}{er_{s}^{2}}\right)e^{-i\omega(T-X)}\neq 0
\end{equation}
for $T+X=0$. However, one can still get a solution that satisfies Eq.~\eqref{KSqescf3} and is equal to \eqref{Umode} for $T+X=0$. Indeed, the solution
\begin{equation}\label{Umode2}
\hat\phi_{l}(T,X,\omega)|_{T+X\to 0}=\frac{1}{r_{s}\sqrt{4\pi\omega}}\,e^{-i\omega(T-X)}\left(1+\frac{T^{2}-X^{2}}{4er_{s}^{2}}-i\frac{l(l+1)+1}{4\omega er_{s}^{2}}(T+X)\right)
\end{equation}
is a solution to Eq.~\eqref{KSqescf3} at $T+X=0$  (i.e., this solution is valid if we keep only the terms of the zero order in $T+X$ after substituting \eqref{Umode2} into Eq.~\eqref{KSqescf3}). We see that $\hat\phi_{l}(T,X)|_{X+T=0}\equiv\tilde\phi_{l}(T,X)|_{X+T=0}$. In the Schwarzschild coordinates for $r_{*}\to-\infty$ and $t\to-\infty$
\begin{equation}\label{Umode3}
\hat\phi_{l}(t,r_{*},\omega)=\frac{e^{i2r_{s}\omega e^{\frac{r_{*}-t}{2r_{s}}}}}{r_{s}\sqrt{4\pi\omega}}
\left(1-e^{\frac{r_{*}}{r_{s}}-1}-i\frac{l(l+1)+1}{2\omega er_{s}}\,e^{\frac{r_{*}+t}{2r_{s}}}\right).
\end{equation}
However, this solution still has a problem: there is the singular coefficient $\sim\frac{1}{\omega}$ for $\omega\to 0$ in the third term in the brackets of \eqref{Umode2} (the singularity caused by the overall coefficient $\frac{1}{\sqrt{\omega}}$ has a different origin and obviously is not dangerous). It vanishes exactly on the horizon, but it is nonzero for any small value of $T+X$. Formally, for $\omega\to 0$ the corresponding term diverges and ceases to be a small correction, so it is possible that the actual regular solution (if exists) or some different solution (such a solution will be discussed in the next section) does not have such a singularity.

The second problem with the mode \eqref{Umode} is that its use does not lead to the correct canonical commutation relations for the scalar field. According to the structure of the theory providing the Unruh vacuum state \cite{Unruh:1976db} (see, for example, \eqref{Unruhtheordef2DBH}), let us represent the quantized scalar field as
\begin{align}\nonumber
&\phi(t,r,T,X,\theta,\varphi)\\\label{scfUnruhV}
&=\sum\limits_{l=0}^{\infty}\sum\limits_{m=-l}^{l}Y_{lm}(\theta,\varphi)\int\limits_{0}^{\infty}d\omega\left(
\tilde\phi_{l}^{}(T,X,\omega)\,a_{lm1}^{}(\omega)+e^{-i\omega t}\frac{\psi_{l2}(\omega,r)}{\sqrt{2\omega}\,r}\,a_{lm2}^{}(\omega)\right)+\textrm{h.c.},
\end{align}
where $\psi_{l2}(\omega,r)$ is defined by \eqref{psiTR3} and \eqref{psiTR4} and $\tilde\phi_{l}^{}(T,X,\omega)$ is such that condition \eqref{Umode} holds. Let us consider this scalar field in the vicinity of the horizon $T+X=0$ for $T-X\sim r_{s}$. Since
\begin{align}
&T-X=-2r_{s}e^{\frac{r_{*}-t}{2r_{s}}},\\
&T+X=2r_{s}e^{\frac{r_{*}+t}{2r_{s}}},
\end{align}
see \eqref{Trstart} and \eqref{Xrstart}, it corresponds to $r_{*}\to-\infty$, $t\to-\infty$, and $r_{*}-t\sim r_{s}$. With
\begin{equation}\label{Umode4}
\tilde\phi_{l}(t,r_{*},\omega)=\frac{e^{i2r_{s}\omega e^{\frac{r_{*}-t}{2r_{s}}}}}{r_{s}\sqrt{4\pi\omega}}
\end{equation}
in this area (since for a fixed $\omega$ we have $\hat\phi_{l}(t,r_{*},\omega)\to\tilde\phi_{l}(t,r_{*},\omega)$ for $r_{*}\to-\infty$, $t\to-\infty$, and $r_{*}-t\sim r_{s}$, the same is valid for \eqref{Umode3} at least for $\omega\not\to 0$), for $r_{*}\to-\infty$, $t\to-\infty$, and $r_{*}-t\sim r_{s}$ the scalar field takes the form
\begin{align}\nonumber
&\phi(t,r_{*},\theta,\varphi)\\\label{scfUnruhV2}
&=\sum\limits_{l=0}^{\infty}\sum\limits_{m=-l}^{l}Y_{lm}(\theta,\varphi)\int\limits_{0}^{\infty}\frac{d\omega}{r_{s}\sqrt{4\pi\omega}}\left(
e^{i2r_{s}\omega e^{\frac{r_{*}-t}{2r_{s}}}}a_{lm1}^{}(\omega)+e^{-i\omega(t+r_{*})}T_{2}(\omega,l)a_{lm2}^{}(\omega)\right)+\textrm{h.c.}
\end{align}

Let us consider the commutator $[\phi(t,r_{*},\theta,\varphi),\phi(t,r_{*}',\theta',\varphi')]$, where $r_{*}\to-\infty$, $r_{*}'\to-\infty$ and $t\to-\infty$. In explicit form it can be represented as
\begin{align}\nonumber
&[\phi(t,r_{*},\theta,\varphi),\phi(t,r_{*}',\theta',\varphi')]\\\label{CCRUnruh1}
&=\frac{1}{16\pi^{2}r_{s}^{2}}\sum\limits_{l=0}^{\infty}(2l+1)P_{l}(\cos\alpha)
\int\limits_{0}^{\infty}\frac{d\omega}{\omega}\left(
e^{i2r_{s}\omega e^{-\frac{t}{2r_{s}}}\left(e^{\frac{r_{*}}{2r_{s}}}-e^{\frac{r_{*}'}{2r_{s}}}\right)}+|T_{2}(\omega,l)|^{2}
e^{-i\omega(r_{*}-r_{*}')}-\textrm{c.c.}\right).
\end{align}
The integral in \eqref{CCRUnruh1} can be rewritten as
\begin{align}\nonumber
&2i\int\limits_{0}^{\infty}\frac{d\omega}{\omega}\left(
\sin\left(\omega 2r_{s}e^{-\frac{t}{2r_{s}}}\left(e^{\frac{r_{*}}{2r_{s}}}-e^{\frac{r_{*}'}{2r_{s}}}\right)\right)-|T_{2}(\omega,l)|^{2}
\sin(\omega(r_{*}-r_{*}'))\right)\\\nonumber
=&2i\int\limits_{0}^{\infty}\frac{d\omega}{\omega}\left(
\sin\left(\omega 2r_{s}e^{-\frac{t}{2r_{s}}}\left(e^{\frac{r_{*}}{2r_{s}}}-e^{\frac{r_{*}'}{2r_{s}}}\right)\right)-
\sin(\omega(r_{*}-r_{*}'))\right)\\\nonumber
&+2i\int\limits_{0}^{\infty}\frac{d\omega}{\omega}(1-|T_{2}(\omega,l)|^{2})\sin(\omega(r_{*}-r_{*}'))\\\nonumber&=
i\pi\textrm{sign}(r_{*}-r_{*}')-i\pi\textrm{sign}(r_{*}-r_{*}')
+2i\int\limits_{0}^{\infty}\frac{d\omega}{\omega}|R_{2}(\omega,l)|^{2}\sin(\omega(r_{*}-r_{*}'))\\\label{IntCCRUnruh1}
&=2i\int\limits_{0}^{\infty}\frac{d\omega}{\omega}|R_{2}(\omega,l)|^{2}\sin(\omega(r_{*}-r_{*}')).
\end{align}
The resulting integral in \eqref{IntCCRUnruh1} is not equal to zero for all $r_{*}$ and $r_{*}'$. Indeed, $|R_{2}(\omega,l)|^{2}\to 0$ for $\omega\to\infty$, so we can choose $\omega_{max}$ such that
\begin{equation}
\int\limits_{0}^{\infty}\frac{d\omega}{\omega}|R_{2}(\omega,l)|^{2}\sin(\omega(r_{*}-r_{*}'))\approx \int\limits_{0}^{\omega_{max}}\frac{d\omega}{\omega}|R_{2}(\omega,l)|^{2}\sin(\omega(r_{*}-r_{*}'))
\end{equation}
with a good accuracy. Let us take $r_{*}$ and $r_{*}'$ such that $r_{*}-r_{*}'\ll\frac{1}{\omega_{max}}$. Then
\begin{equation}
\int\limits_{0}^{\omega_{max}}\frac{d\omega}{\omega}|R_{2}(\omega,l)|^{2}\sin(\omega(r_{*}-r_{*}'))\approx
(r_{*}-r_{*}')\int\limits_{0}^{\omega_{max}}|R_{2}(\omega,l)|^{2}d\omega\neq 0.
\end{equation}
Thus for \eqref{CCRUnruh1} we get
\begin{equation}
[\phi(t,r_{*},\theta,\varphi),\phi(t,r_{*}',\theta',\varphi')]=\frac{i}{8\pi^{2}r_{s}^{2}}\sum\limits_{l=0}^{\infty}(2l+1)P_{l}(\cos\alpha)
\int\limits_{0}^{\infty}\frac{d\omega}{\omega}|R_{2}(\omega,l)|^{2}\sin(\omega(r_{*}-r_{*}'))\not\equiv 0.
\end{equation}
If $R_{2}(\omega,l)\equiv 0$ for all $\omega$ and $l$ (like in the case of Unruh vacuum state for a uniformly accelerated reference frame, see Section~\ref{CCRUvac}), we would get a correct canonical commutation relation. However, in the case of a four-dimensional Schwarzschild black hole $R_{2}(\omega,l)\not\equiv 0$, so we cannot get a correct canonical commutation relation for all $r_{*}$ and $r_{*}'$ as it should be in a consistent quantum field theory. The use of \eqref{Umode3} instead of \eqref{Umode4} does not solve this problem, because, as has already been noted, $\hat\phi_{l}(t,r_{*},\omega)\to\tilde\phi_{l}(t,r_{*},\omega)$ for $r_{*}\to-\infty$, $t\to-\infty$, and $r_{*}-t\sim r_{s}$.

In principle, non-fulfillment of one canonical commutation relation is sufficient to conclude that the corresponding quantum field theory is inconsistent. However, it is possible to check that the commutation relation with $[\phi(t,r_{*},\theta,\varphi),\dot\phi(t,r_{*}',\theta',\varphi')]$ is not satisfied too. In explicit form, this commutator can be represented as
\begin{align}\nonumber
&[\phi(t,r_{*},\theta,\varphi),\dot\phi(t,r_{*}',\theta',\varphi')]=\frac{i}{16\pi^{2}r_{s}^{2}}\sum\limits_{l=0}^{\infty}(2l+1)P_{l}(\cos\alpha)
\\\label{CCRUnruh2}
&\times\int\limits_{0}^{\infty}d\omega\left(e^{\frac{r_{*}-t}{2r_{s}}}
e^{i2r_{s}\omega e^{-\frac{t}{2r_{s}}}\left(e^{\frac{r_{*}}{2r_{s}}}-e^{\frac{r_{*}'}{2r_{s}}}\right)}+|T_{2}(\omega,l)|^{2}
e^{-i\omega(r_{*}-r_{*}')}+\textrm{c.c.}\right).
\end{align}
The integral in \eqref{CCRUnruh2} can be rewritten as
\begin{align}\nonumber
&\int\limits_{0}^{\infty}d\omega\left(e^{\frac{r_{*}-t}{2r_{s}}}
e^{i2r_{s}\omega e^{-\frac{t}{2r_{s}}}\left(e^{\frac{r_{*}}{2r_{s}}}-e^{\frac{r_{*}'}{2r_{s}}}\right)}+
e^{-i\omega(r_{*}-r_{*}')}-|R_{2}(\omega,l)|^{2}e^{-i\omega(r_{*}-r_{*}')}+\textrm{c.c.}\right)\\\nonumber
&=\int\limits_{-\infty}^{\infty}d\omega\left(e^{\frac{r_{*}-t}{2r_{s}}}
e^{i2r_{s}\omega e^{-\frac{t}{2r_{s}}}\left(e^{\frac{r_{*}}{2r_{s}}}-e^{\frac{r_{*}'}{2r_{s}}}\right)}+
e^{-i\omega(r_{*}-r_{*}')}\right)-2\int\limits_{0}^{\infty}d\omega|R_{2}(\omega,l)|^{2}\cos\left(\omega(r_{*}-r_{*}')\right)
\\\nonumber
&=2\pi e^{\frac{r_{*}-t}{2r_{s}}}\delta\left(2r_{s}e^{-\frac{t}{2r_{s}}}\left(e^{\frac{r_{*}}{2r_{s}}}-e^{\frac{r_{*}'}{2r_{s}}}\right)\right)
+2\pi\delta(r_{*}-r_{*}')-2\int\limits_{0}^{\infty}d\omega|R_{2}(\omega,l)|^{2}\cos\left(\omega(r_{*}-r_{*}')\right)\\\label{IntCCRUnruh2}
&=4\pi\delta(r_{*}-r_{*}')-2\int\limits_{0}^{\infty}d\omega|R_{2}(\omega,l)|^{2}\cos\left(\omega(r_{*}-r_{*}')\right).
\end{align}
Apart from the necessary delta function $\delta(r_{*}-r_{*}')$, we get the extra integral term in \eqref{IntCCRUnruh2}. Again, this integral can be represented as
\begin{equation}
\int\limits_{0}^{\infty}d\omega|R_{2}(\omega,l)|^{2}\cos\left(\omega(r_{*}-r_{*}')\right)\approx
\int\limits_{0}^{\omega_{max}}d\omega|R_{2}(\omega,l)|^{2}\cos\left(\omega(r_{*}-r_{*}')\right).
\end{equation}
For $r_{*}$ and $r_{*}'$ such that $r_{*}-r_{*}'\ll\frac{1}{\omega_{max}}$
\begin{equation}
\int\limits_{0}^{\omega_{max}}d\omega|R_{2}(\omega,l)|^{2}\cos\left(\omega(r_{*}-r_{*}')\right)\approx
\int\limits_{0}^{\omega_{max}}d\omega|R_{2}(\omega,l)|^{2}\neq 0.
\end{equation}
Due to the existence of this nonzero term, the corresponding commutation relation also is not satisfied. Indeed, substituting \eqref{IntCCRUnruh2} into \eqref{CCRUnruh2} and using relation \eqref{LL-QM-relation}, one can get
\begin{align}\nonumber
&[\phi(t,r_{*},\theta,\varphi),\dot\phi(t,r_{*}',\theta',\varphi')]\\\nonumber
&=\frac{i}{\pi r_{s}^{2}}\,\delta(r_{*}-r_{*}')\delta(1-\cos\alpha)-\frac{i}{8\pi^{2}r_{s}^{2}}\sum\limits_{l=0}^{\infty}(2l+1)P_{l}(\cos\alpha)
\int\limits_{0}^{\infty}d\omega|R_{2}(\omega,l)|^{2}\cos\left(\omega(r_{*}-r_{*}')\right)\\
&\neq\frac{i}{\pi r_{s}^{2}}\,\delta(r_{*}-r_{*}')\delta(1-\cos\alpha).
\end{align}
Thus if $R_{2}(\omega,l)\equiv 0$ for all $\omega$ and $l$, we would get the correct canonical commutation relation (of course, for $r_{*}\to-\infty$ and $r_{*}'\to-\infty$, which leads to the factor $\frac{1}{r_{s}^{2}}$ instead of $\frac{1}{r^{2}}$ in front of the delta functions).

Contrary to the cases of a uniformly accelerated reference frame in two dimensions or a two-dimensional Schwarzschild black hole, in which the theories providing the Unruh vacuum state are well defined and do not contain obvious pathologies (see Section~\ref{CCRUvac}), the quantum theory providing the Unruh vacuum state for the four-dimensional Schwarzschild black hole, at least if constructed in the standard manner, is inconsistent. Possible origins of this problem are the following. First, it is possible that correct solutions on the horizon in the Kruskal-Szekeres coordinates have a form different from those of \eqref{Umode} or \eqref{Umode2}. Second, it is also possible that the structure of the modes in Kruskal-Szekeres spacetime is such that it is impossible to construct a theory providing the Unruh vacuum state at all, exactly as in the case of a four-dimensional uniformly accelerated reference frame. In any case, a possible answer could be found if we knew better properties of the theory in Kruskal-Szekeres spacetime. Some basic aspects of this theory will be discussed in the next section.

\subsection{Is there a consistent quantum scalar field theory in Kruskal-Szekeres spacetime?}
As was noted in \cite{Fulling:1987otn}, ``On the Schwarzschild-Kruskal manifold there is no ``standard'' quantization {\em a priori}.'' However, we can formally consider the canonical conjugate momentum as
\begin{equation}\label{ccmKS}
\pi(T,X,\theta,\varphi)=\frac{\partial\mathcal{L}}{\partial\left(\frac{\partial\phi(T,X,\theta,\varphi)}{\partial T}\right)}
=\sqrt{-g}\,g^{00}\frac{\partial\phi(T,X,\theta,\varphi)}{\partial T}=r^{2}(T,X)\sin\theta\frac{\partial\phi(T,X,\theta,\varphi)}{\partial T},
\end{equation}
where the metric is defined by \eqref{KS4D}, and consider the standard canonical commutation relations in Kruskal-Szekeres spacetime to see what happens in this case. We know that there exists a consistent quantum scalar field theory in Schwarzschild spacetime. Thus since the canonical commutation relations are satisfied in the Schwarzschild coordinates, the commutation relations in wedge I of Kruskal-Szekeres spacetime
\begin{align}\label{CCRKS1}
&\left[\phi\Bigl(t(T,X),r_{*}(T,X),\theta,\varphi\Bigr),\frac{\partial\phi\Bigl(t(T,X'),r_{*}(T,X'),\theta',\varphi'\Bigr)}{\partial T}\right]=i\frac{\delta(X-X')\delta(\theta-\theta')\delta(\varphi-\varphi')}{r^{2}(T,X)\sin\theta},\\\label{CCRKS2}
&\left[\phi\Bigl(t(T,X),r_{*}(T,X),\theta,\varphi\Bigr),\phi\Bigl(t(T,X'),r_{*}(T,X'),\theta',\varphi'\Bigr)\right]=0,\\\label{CCRKS3}
&\left[\frac{\partial\phi\Bigl(t(T,X),r_{*}(T,X),\theta,\varphi\Bigr)}{\partial T},\frac{\partial\phi\Bigl(t(T,X'),r_{*}(T,X'),\theta',\varphi'\Bigr)}{\partial T}\right]=0
\end{align}
are also expected to be satisfied (here $\phi(t,r_{*},\theta,\varphi)$ is the scalar field in Schwarzschild spacetime) if there exists a consistent quantum scalar field theory in Kruskal-Szekeres spacetime; in full analogy with how it happens in the case of a uniformly accelerated reference frame, see Section~\ref{CCR2Dsection}. Since the solution $\phi(t,r_{*},\theta,\varphi)$ corresponds only to wedge I of the Kruskal-Szekeres plane, the reasonings that will be presented below are not valid for the whole spacetime in the Kruskal-Szekeres coordinates. However, in principle they can be extended to the whole Kruskal-Szekeres plane if necessary (at least the analysis for wedge III is fully analogous to the one for wedge I).

There is a good reason to believe that the corresponding quantum field theory exists, at least in the case of a free field. Let us start with the simplest commutation relation \eqref{CCRKS2}. It is clear that
\begin{equation}\label{CCRKS2b}
\left[\phi\Bigl(t(T,X),r_{*}(T,X),\theta,\varphi\Bigr),\phi\Bigl(t(T,X'),r_{*}(T,X'),\theta',\varphi'\Bigr)\right]
=\left[\phi\Bigl(t,r_{*},\theta,\varphi\Bigr),\phi\Bigl(t',r_{*}',\theta',\varphi'\Bigr)\right],
\end{equation}
where
\begin{align}
&r_{*}=r_{s}\ln\left(\frac{X^{2}-T^{2}}{4r_{s}^{2}}\right),\\
&r_{*}'=r_{s}\ln\left(\frac{{X'}^{2}-T^{2}}{4r_{s}^{2}}\right),\\
&t=r_{s}\ln\left(\frac{X+T}{X-T}\right),\\
&t'=r_{s}\ln\left(\frac{X'+T}{X'-T}\right).
\end{align}
Using \eqref{Trstart} and \eqref{Xrstart}, it is not difficult to check that
\begin{equation}
\left(X-X'\right)^{2}-\left(T-T'\right)^{2}=8r_{s}^{2}e^{\frac{r_{*}+r_{*}'}{2r_{s}}}\left(\cosh\left(\frac{|r_{*}-r_{*}'|}{2r_{s}}\right)
-\cosh\left(\frac{|t-t'|}{2r_{s}}\right)\right),
\end{equation}
which means that for any $T=T'$ and $X\neq X'$
\begin{equation}
|t-t'|<|r_{*}-r_{*}'|.
\end{equation}
The latter inequality will be used below.

In explicit form commutator \eqref{CCRKS2b} looks like
\begin{align}\nonumber
&\left[\phi\Bigl(t,r_{*},\theta,\varphi\Bigr),\phi\Bigl(t',r_{*}',\theta',\varphi'\Bigr)\right]=\frac{i}{8\pi rr'}\sum\limits_{l=0}^{\infty}(2l+1)P_{l}(\cos\alpha)\\\label{CCRKS2c}
&\times
\int\limits_{0}^{\infty}\frac{dE}{E}\sum\limits_{p=1}^{2}\left(e^{-iE(t-t')}\psi_{lp}(E,r_{*})\psi_{lp}^{*}(E,r_{*}')
-e^{iE(t-t')}\psi_{lp}^{*}(E,r_{*})\psi_{lp}(E,r_{*}')\right).
\end{align}
Let us take a closer look at the integral in \eqref{CCRKS2c}. The same integral arises if we consider an analogous commutator in an auxiliary theory describing a real scalar field in two-dimensional Minkowski spacetime with an external potential:
\begin{equation}\label{actaux}
S=\int dtdx\left(\frac{1}{2}\left(\frac{\partial\varphi(t,x)}{\partial t}\right)^{2}-\frac{1}{2}\left(\frac{\partial\varphi(t,x)}{\partial x}\right)-V_{l}(x)\varphi^{2}(t,x)\right),
\end{equation}
where $V_{l}(x)$ is defined by \eqref{VSchr1}. If there were no potential $V_{l}(x)$ in action \eqref{actaux}, then the system would be Lorentz invariant. In such a case, it is easy to prove that $[\varphi(t,x),\varphi(t',x')]=0$ for $|t-t'|<|x-x'|$ either by performing direct calculations or even by using the arguments based on the Lorentz invariance of the system \cite{IZ,PS}. With $V_{l}(x)$, the situation turns out to be much more difficult --- at the moment I have no rigorous mathematical proof that the integrals in \eqref{CCRKS2c} vanish for $|t-t'|<|x-x'|$. The problem is that the solutions $\psi_{lp}(E,r_{*})$ are expressed through the Heun functions, whose properties (including possible poles and branch cuts on the plane of complex $E$) are not well known. However, since for all $l$ the potentials $V_{l}(x)$ are potential barriers (see Fig.~\ref{Vl}), which means that the existence of such a potential just gives a particle a nonzero effective mass in the area in which the potential is essentially nonzero, physical arguments based on causality suggest that $[\varphi(t,x),\varphi(t',x')]=0$ outside the light cone for any $l$. If it is so, then the integrals in \eqref{CCRKS2c} are equal to zero for all $l$, leading to
\begin{equation}\label{CCRKS2d}
\left[\phi\Bigl(t,r_{*},\theta,\varphi\Bigr),\phi\Bigl(t',r_{*}',\theta',\varphi'\Bigr)\right]=0
\end{equation}
for $|t-t'|<|r_{*}-r_{*}'|$. Of course, such an analysis cannot be considered as a proof but it can be considered as an additional argument. Also note that in the case of a four-dimensional uniformly accelerated reference frame (see Section~\ref{URF4D}) the potential in a two-dimensional auxiliary theory analogous to the one described by \eqref{actaux} is also nonzero (namely, it takes the form ${\vec k_{\perp}}^{2}e^{2ax}$) but it is possible to prove that $[\varphi(t,x),\varphi(t',x')]=0$ for $|t-t'|<|x-x'|$ using the properties of exact analytical solutions \eqref{Psi4DRind}. The latter can be also used as an argument in favor of fulfillment of \eqref{CCRKS2d}.

It is clear that the commutator
\begin{equation}\label{commeqX}
\left[\phi\Bigl(t(T,X),r_{*}(T,X),\theta,\varphi\Bigr),\frac{\partial\phi\Bigl(t(T,X'),r_{*}(T,X'),\theta',\varphi'\Bigr)}{\partial T}\right]
\end{equation}
is expressed trough the commutators
\begin{align}\label{auxccr1}
&\left[\phi\Bigl(t,r_{*},\theta,\varphi\Bigr),\frac{\partial}{\partial t'}\phi\Bigl(t',r_{*}',\theta',\varphi'\Bigr)\right]=\frac{\partial}{\partial t'}\left[\phi\Bigl(t,r_{*},\theta,\varphi\Bigr),\phi\Bigl(t',r_{*}',\theta',\varphi'\Bigr)\right]=0,\\\label{auxccr2}
&\left[\phi\Bigl(t,r_{*},\theta,\varphi\Bigr),\frac{\partial}{\partial r_{*}'}\phi\Bigl(t',r_{*}',\theta',\varphi'\Bigr)\right]=\frac{\partial}{\partial r_{*}'}\left[\phi\Bigl(t,r_{*},\theta,\varphi\Bigr),\phi\Bigl(t',r_{*}',\theta',\varphi'\Bigr)\right]=0
\end{align}
for $|t-t'|<|r_{*}-r_{*}'|$ if relation \eqref{CCRKS2d} holds. The latter leads to the fulfillment of commutation relation \eqref{CCRKS1} for $X\neq X'$. Analogous reasonings can be applied to the last commutation relation \eqref{CCRKS3}. Thus it is quite probable that the canonical commutation relations \eqref{CCRKS1}-\eqref{CCRKS3} are satisfied for $X\neq X'$.

Now we turn to the case $X=X'$. It is clear that $X=X'$ leads to $r_{*}=r_{*}'$ and $t=t'$. Thus commutator \eqref{CCRKS2c} can be rewritten as
\begin{align}\nonumber
&\left[\phi\Bigl(t,r_{*},\theta,\varphi\Bigr),\phi\Bigl(t,r_{*},\theta',\varphi'\Bigr)\right]=\frac{i}{8\pi r^{2}}\sum\limits_{l=0}^{\infty}(2l+1)P_{l}(\cos\alpha)\\
&\times
\sum\limits_{p=1}^{2}\int\limits_{0}^{\infty}\frac{dE}{E}\left(\psi_{lp}(E,r_{*})\psi_{lp}^{*}(E,r_{*})
-\psi_{lp}^{*}(E,r_{*})\psi_{lp}(E,r_{*})\right)=0,
\end{align}
leading to \eqref{CCRKS2} for $X=X'$.

The commutator in \eqref{CCRKS3} for $X=X'$ can be rewritten as
\begin{align}\nonumber
&\left[\frac{\partial\phi\Bigl(t(T,X),r_{*}(T,X),\theta,\varphi\Bigr)}{\partial T},\frac{\partial\phi\Bigl(t(T,X),r_{*}(T,X),\theta',\varphi'\Bigr)}{\partial T}\right]\\\nonumber&=\frac{i}{8\pi r^{2}}\sum\limits_{l=0}^{\infty}(2l+1)P_{l}(\cos\alpha)
\sum\limits_{p=1}^{2}\int\limits_{0}^{\infty}\frac{dE}{E}\\
&\times\left(\frac{\partial\left(e^{-iEt}\psi_{lp}(E,r_{*})\right)}{\partial T}
\frac{\partial\left(e^{iEt}\psi_{lp}^{*}(E,r_{*})\right)}{\partial T}
-\frac{\partial\left(e^{iEt}\psi_{lp}^{*}(E,r_{*})\right)}{\partial T}\frac{\partial\left(e^{-iEt}\psi_{lp}(E,r_{*})\right)}{\partial T}\right)=0.
\end{align}

It has already been mentioned that the commutator in \eqref{CCRKS1} can be expressed through the commutators in the lhs of \eqref{auxccr1} and \eqref{auxccr2}. For $t=t'$ and $r_{*}=r_{*}'$, for the commutator in \eqref{auxccr2} we get
\begin{align}\nonumber
&\left[\phi\Bigl(t,r_{*},\theta,\varphi\Bigr),\frac{\partial}{\partial r_{*}}\phi\Bigl(t,r_{*},\theta',\varphi'\Bigr)\right]
=\frac{i}{8\pi r^{2}}\sum\limits_{l=0}^{\infty}(2l+1)P_{l}(\cos\alpha)\\\label{auxccr3}
&\times
\sum\limits_{p=1}^{2}\int\limits_{0}^{\infty}\frac{dE}{E}\left(\psi_{lp}(E,r_{*})\frac{\partial\psi_{lp}^{*}(E,r_{*})}{\partial r_{*}}
-\psi_{lp}^{*}(E,r_{*})\frac{\partial\psi_{lp}(E,r_{*})}{\partial r_{*}}\right).
\end{align}
The terms in the brackets in \eqref{auxccr3} are just the Wronskians
\begin{align}
&\psi_{l1}(E,r_{*})\frac{\partial\psi_{l1}^{*}(E,r_{*})}{\partial r_{*}}-\psi_{l1}^{*}(E,r_{*})\frac{\partial\psi_{l1}(E,r_{*})}{\partial r_{*}}
=-\frac{iE}{\pi}|T_{l1}(E)|^{2},\\
&\psi_{l2}(E,r_{*})\frac{\partial\psi_{l2}^{*}(E,r_{*})}{\partial r_{*}}-\psi_{l2}^{*}(E,r_{*})\frac{\partial\psi_{l2}(E,r_{*})}{\partial r_{*}}
=\frac{iE}{\pi}|T_{l2}(E)|^{2}
\end{align}
that do not depend on $r_{*}$. Since $|T_{l1}(E)|^{2}=|T_{l2}(E)|^{2}$ (see \eqref{RT1}),
\begin{align}\nonumber
&\sum\limits_{p=1}^{2}\int\limits_{0}^{\infty}\frac{dE}{E}\left(\psi_{lp}(E,r_{*})\frac{\partial\psi_{lp}^{*}(E,r_{*})}{\partial r_{*}}
-\psi_{lp}^{*}(E,r_{*})\frac{\partial\psi_{lp}(E,r_{*})}{\partial r_{*}}\right)\\
&=\int\limits_{0}^{\infty}\frac{dE}{E}\left(-\frac{iE}{\pi}|T_{l1}(E)|^{2}+\frac{iE}{\pi}|T_{l1}(E)|^{2}\right)=0,
\end{align}
leading to
\begin{equation}
\left[\phi\Bigl(t,r_{*},\theta,\varphi\Bigr),\frac{\partial}{\partial r_{*}}\phi\Bigl(t,r_{*},\theta',\varphi'\Bigr)\right]=0.
\end{equation}
Finally, commutator \eqref{commeqX} takes the form
\begingroup
\allowdisplaybreaks
\begin{align}\nonumber
&\left[\phi\Bigl(t(T,X),r_{*}(T,X),\theta,\varphi\Bigr),\frac{\partial\phi\Bigl(t(T,X),r_{*}(T,X),\theta',\varphi'\Bigr)}{\partial T}\right]\\\nonumber
&=\frac{\partial t}{\partial T}\left[\phi\Bigl(t,r_{*},\theta,\varphi\Bigr),\frac{\partial}{\partial t}\phi\Bigl(t,r_{*},\theta',\varphi'\Bigr)\right]=\frac{\partial t}{\partial T}\frac{i}{8\pi r^{2}}\sum\limits_{l=0}^{\infty}(2l+1)P_{l}(\cos\alpha)\\
&\times
\sum\limits_{p=1}^{2}\int\limits_{0}^{\infty}dE\left(\psi_{lp}(E,r_{*})\psi_{lp}^{*}(E,r_{*})
+\psi_{lp}^{*}(E,r_{*})\psi_{lp}(E,r_{*})\right)
=i\frac{\partial t}{\partial T}\frac{\delta(r_{*}-r_{*})\delta(\theta-\theta')\delta(\varphi-\varphi')}{r^{2}\sin\theta}.
\end{align}
\endgroup
Using
\begin{equation}
\frac{\partial t}{\partial T}=\frac{2r_{s}X}{X^{2}-T^{2}}
\end{equation}
and
\begin{equation}
\delta(r_{*}-r_{*}')=\frac{X^{2}-T^{2}}{2r_{s}X}\,\delta(X-X'),
\end{equation}
one gets
\begin{equation}
\left[\phi\Bigl(t(T,X),r_{*}(T,X),\theta,\varphi\Bigr),\frac{\partial\phi\Bigl(t(T,X),r_{*}(T,X),\theta',\varphi'\Bigr)}{\partial T}\right]
=i\frac{\delta(X-X)\delta(\theta-\theta')\delta(\varphi-\varphi')}{r^{2}\sin\theta},
\end{equation}
which is \eqref{CCRKS1} for $X=X'$.

Thus there is a good reason to believe that in Kruskal-Szekeres spacetime it is possible to construct a consistent quantum field theory satisfying the canonical commutation relations at least for a free scalar field. However, we still can say nothing about the form or properties of the modes even on the horizon. Indeed, even though we can find some solutions with necessary properties in the vicinity of the horizon (like solution \eqref{Umode2}), it does not mean that these solutions are indeed represent actual modes possessing the necessary orthogonality conditions (the form of the latter is still unknown for Eq.~\eqref{KSqescf}\footnote{Note that there exist forbidden regions on the Kruskal-Szekeres plane, see Fig.~\ref{fig4}; their existence additionally complicates the problem of finding actual solutions.}). As an example, one can consider Eq.~\eqref{eqm-massiveRSch} for a four-dimensional uniformly accelerated reference frame. For $x\to-\infty$ in the leading order this equation takes the form
\begin{equation}
\epsilon^{2}\Psi(\epsilon,\vec k_{\perp},x)=-\frac{d^{2}\Psi(\epsilon,\vec k_{\perp},x)}{dx^{2}},
\end{equation}
which formally has two different solutions $e^{i\epsilon x}$ and $e^{-i\epsilon x}$. However, for a fixed $\epsilon$ the only actual solution for $x\to-\infty$ is a certain linear combination of $e^{i\epsilon x}$ and $e^{-i\epsilon x}$, the corresponding coefficients can be found only if one finds a global solution \eqref{Psi4DRind}.\footnote{In this connection I would like once again refer to the discussion in Section~III of \cite{Fulling:1972md}, in which it is written ``The quantum theory usually deals with phenomena that happen on a microscopic scale. It is hard to understand how the global structure of the universe
can affect the physics inside a small Cauchy-complete region. Nevertheless, a decomposition of a field into modes appears unavoidably to involve global integral transformations, like ...''.} Analogously, in four-dimensional Schwarzschild spacetime for $r_{*}\to-\infty$ only one of two radial solutions (namely, solution \eqref{psiTR4}) is proportional to $e^{-iEr_{*}}$, whereas another solution (namely, solution \eqref{psiTR1}) is a linear combination of $e^{iEr_{*}}$ and $e^{-iEr_{*}}$. Of course, these examples involve only spatial coordinates $x$ or $r_{*}$. However, they correspond to ``regular'' theories allowing one to perform separation of variables in the radial equations and to get stationary solutions of the form $e^{-i\omega t}f(x,\omega)$ (like in \eqref{operatordecspher}).

It is quite possible that the situation in Kruskal-Szekeres spacetime is the same: for example, one may expect that the actual solution on the horizon $T+X=0$ is not \eqref{Umode} but a linear combination of \eqref{Umode} and $e^{i\omega(T-X)}$. Indeed, it is well known that \cite{Fulling:1977zs} ``A complete set of functions on the surface $V=0$ is the family $e^{-i\lambda U}$, and those with $\lambda>0$ are the positive-frequency ones.'' However, since separation of variables (at least the standard one) is not expected in the theory defined by Eq.~\eqref{KSqescf} providing much more complicated solutions, the standard simple dependence on time $e^{-i\omega T}$ (which also allows one to define positive frequency modes in standard theories) for all $T$ is also not expected for solutions of Eq.~\eqref{KSqescf}. The latter means that one also needs to get a correct coefficient for such a solution: indeed, the factor $\frac{1}{\sqrt{\omega}}$ in \eqref{Umode} is inherent to modes in theories providing stationary solutions of the form $e^{-i\omega t}f(x,\omega)$. In such a case, $\frac{\partial(e^{-i\omega t}f(x,\omega))}{\partial t}=-i\omega e^{-i\omega t}f(x,\omega)$, so technically the coefficients $\frac{1}{\sqrt{\omega}}$ allows one to get the correct factor $\omega$ in the Hamiltonian and to obtain valid canonical commutation relations. It means that the appearance of the coefficient $\frac{1}{\sqrt{\omega}}$ like in \eqref{Umode} is rather unlikely even if the actual solution has the form $e^{-i\omega(T-X)}$ (or, more likely, the form of some linear combination of $e^{-i\omega(T-X)}$ and $e^{i\omega(T-X)}$) on the horizon $T+X=0$.

There is another small indication that the actual solution may represent itself as a combination of $e^{-i\omega(T-X)}$ and $e^{i\omega(T-X)}$ on the horizon. Indeed, apart from solution \eqref{Umode2}, there exists the solution
\begin{equation}\label{Umode2b}
\hat\phi_{l}^{(2)}(T,X,\omega)|_{T+X\to 0}=\frac{1}{r_{s}\sqrt{4\pi\omega}}\,e^{i\omega(T-X)}\left(1+\frac{T^{2}-X^{2}}{4er_{s}^{2}}+i\frac{l(l+1)+1}{4\omega er_{s}^{2}}(T+X)\right)
\end{equation}
The sum of \eqref{Umode2} and \eqref{Umode2b} has the form
\begin{align}\nonumber
&\hat\phi_{l}(T,X,\omega)+\hat\phi_{l}^{(2)}(T,X,\omega)|_{T+X\to 0}\\\label{Umode2+2b}&=\frac{2}{r_{s}\sqrt{4\pi\omega}}\left[\left(1+\frac{T^{2}-X^{2}}{4er_{s}^{2}}\right)\cos\left(\omega(T-X)\right)
-\frac{l(l+1)+1}{4\omega er_{s}^{2}}(T+X)\sin\left(\omega(T-X)\right)\right],
\end{align}
which has no singularities for $\omega\to 0$ except the one provided by the overall coefficient $\frac{1}{\sqrt{\omega}}$.

If the actual mode on the horizon is not just $e^{-i\omega(T-X)}$, all considerations (including the well-known results presented in \cite{Candelas:1980zt,Christensen:1977jc}) based on the assumption that the positive frequency function with respect to the Killing vector $\frac{\partial}{\partial(T-X)}$ on the horizon \eqref{Umode} corresponds to a solution describing the actual mode can not be considered as well justified.\footnote{Even though ``... the geometry of the horizon (more precisely, of its projection into the (t, x) plane --- the remaining coordinates playing no essential role) is the same as in Minkowski space. It therefore seems likely that the physically most natural quantization is defined by normal modes with the same analytic property as the flat-space plane waves.'' \cite{Fulling:1987otn}, it is still an assumption.} The same is valid for the mode $\sim e^{-i\omega(T+X)}$ on the horizon $T-X=0$. The fact that these reasonings work fine in the cases of uniformly accelerated reference frames in two and four dimensions and a two-dimensional Schwarzschild black hole is just a consequence of the fact that the quantum field theories in these cases are well defined at least on the basic level and that the corresponding modes on the horizon indeed have the desired form (in the two-dimensional case they are exactly $e^{-i\omega(T-X)}$ and $e^{-i\omega(T+X)}$), allowing one either to calculate the Bogolyubov coefficients explicitly or to use the method based on the consideration of linear combinations of solutions provided by the theories corresponding to wedges I and III of Fig.~\ref{fig1} (see the discussion at the end of Section~\ref{Sec2-4}).

Another problem of the theory in Kruskal-Szekeres spacetime is that the Hamiltonian defined in the standard way as \eqref{Hamilt} is not conserved over time. It means that even if there exists a consistent quantum field theory in Kruskal-Szekeres spacetime for a free scalar field satisfying the necessary canonical commutation relations, at the moment it is not clear how to define particles in such a theory or to build the corresponding Fock space. This problem still deserves a detailed analysis.

\section{Conclusion}\label{sectconclusion}
In the present paper, several aspects of quantum scalar field theories in the cases of a uniformly accelerated reference frame and a Schwarzschild black hole in two and four dimensions are discussed. In particular, a detailed description of the Fulling-Davies-Unruh effect in two and four dimensions is presented, including an explicit calculation of the Bogolyubov coefficients and a derivation of the coordinate-dependent energy densities. The latter is based on the use of specific representations of delta functions $\delta(0)$: in the general case, it is not just a volume but may correspond to certain regions of space, which seems to be quite illustrative by itself. Besides the classical Fulling-Davies-Unruh effect, a consistent theory providing the Unruh vacuum state in the case of a uniformly accelerated reference frame in two dimensions was also discussed from a purely mathematical point of view, including a verification that all canonical commutation relation are satisfied in such a theory. An interesting observation is that it is impossible to build a theory providing the Unruh vacuum state in the four-dimensional case.

Even though the foundations of quantum scalar field theories for uniformly accelerated reference frames are correct and consistent in both Minkowski and Rindler spacetimes, they still have some pathologies. Indeed, some of the well defined quantum states in Minkowski spacetime (namely, the standard one-particle states and some normalized states) have infinite energies when considered in Rindler spacetime. This effect is a consequence of the specificity of coordinate transformations \eqref{TRind} and \eqref{XRind}, the latter poses a question about the applicability of the standard quantum field theory formalism based on the use of plane waves. Of course, analogous pathologies are expected in the black hole cases too.

Although the quantum scalar field theories for uniformly accelerated reference frames and Schwarzschild black holes have a lot in common, they still have many differences. In particular, it turns out that the theory providing the Unruh vacuum state in the case of a four-dimensional Schwarzschild black hole does not satisfy the canonical commutation relations, contrary to the case of a uniformly accelerated reference frame in two dimensions (and, consequently, to the case of a two-dimensional Schwarzschild black hole that is almost identical to the case of a uniformly accelerated reference frame in two dimensions). Since the canonical commutation relations are not satisfied, the corresponding quantum field theory cannot be considered as correct and consistent, at least within the framework of the standard quantum field theory.

Finally, a possible quantum scalar field theory in Kruskal-Szekeres spacetime is discussed. It is well known that there is still no consistent quantum field theory in Kruskal-Szekeres spacetime. Apart from problems with solving the corresponding equation of motion and isolating the modes describing one-particle states (as far as I know, there are no explicit solutions for the scalar field in the Kruskal-Szekeres coordinates yet), the Hamiltonian constructed in the standard manner is not conserved over time. All these problems complicate the analysis considerably. Meanwhile, using the observation concerning the fulfillment of ``extra'' canonical commutation relations in the case of a uniformly accelerated reference frame, it is shown that the existence of a quantum scalar field theory satisfying the standard canonical commutation relations in Kruskal-Szekeres spacetime is quite possible. However, the question about the possible structure of the modes (including that on the horizon) still remains open. This problem calls for a further investigation.

\subsection*{Acknowledgments}
\addcontentsline{toc}{section}{Acknowledgments}
The author is grateful to E.E.~Boos, V.O.~Egorov, S.I.~Keizerov, E.R.~Rakhmetov, S.A.~Paston, and especially to I.P.~Volobuev for valuable discussions and useful comments.
This study was conducted within the scientific program of the National Center for Physics and Mathematics, section \#5 ``Particle Physics and Cosmology''. Stage 2026-2027.

\section*{Appendix A: Bogolyubov coefficients in the two-dimen\-sional case}\label{BCsubsect}
\addcontentsline{toc}{section}{Appendix A: Bogolyubov coefficients in the two-dimensional case}
\setcounter{equation}{0}
\renewcommand{\theequation}{A\arabic{equation}}
Let us start with $k>0$ and represent the function $e^{i\frac{k}{a}e^{-au}}$ in \eqref{PhiMRind2D} as
\begin{equation}
e^{i\frac{k}{a}e^{-au}}=\int\limits_{-\infty}^{\infty}A_{+}^{}(k,p)e^{-ipu}dp,
\end{equation}
leading to
\begin{equation}\label{A+def}
A_{+}^{}(k,p)=\frac{1}{2\pi}\int\limits_{-\infty}^{\infty}e^{i\frac{k}{a}e^{-au}}e^{ipu}du.
\end{equation}
For $k<0$ one can perform analogous steps and consider the representation
\begin{equation}
e^{i\frac{k}{a}e^{av}}=\int\limits_{-\infty}^{\infty}A_{-}^{}(k,p)e^{ipv}dp,
\end{equation}
leading to
\begin{equation}\label{A-def}
A_{-}^{}(k,p)=\frac{1}{2\pi}\int\limits_{-\infty}^{\infty}e^{i\frac{k}{a}e^{av}}e^{-ipv}dv.
\end{equation}
The latter formulas imply that the annihilation operators $b_{+}^{}(p)$ and $b_{-}^{}(p)$ can be represented through the creation and annihilation operators $a_{\pm}^{\dagger}(k)$ and $a_{\pm}^{}(k)$ as \eqref{ba1final} and \eqref{ba2final}.

\begin{figure}[ht]
\centering
\begin{minipage}{.49\textwidth}
\centering
\includegraphics[width=0.8\linewidth]{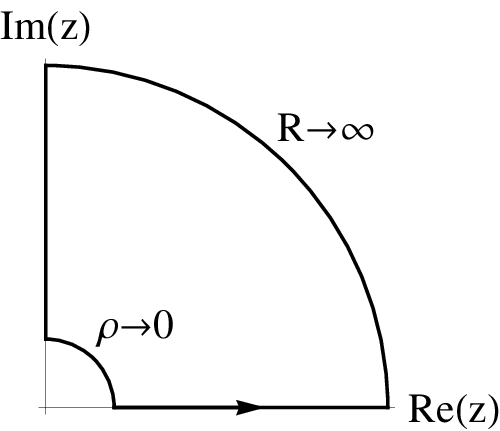}
\end{minipage}
\begin{minipage}{.49\textwidth}
\centering
\includegraphics[width=0.8\linewidth]{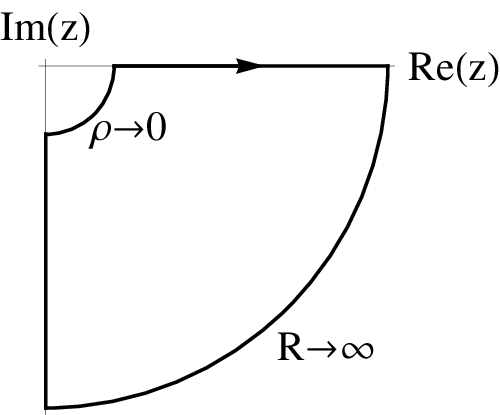}
\end{minipage}
\caption{The contours used for calculation of $A_{+}^{}(k,p)$ (left plot) and $A_{-}^{}(k,p)$ (right plot).}\label{fig2}
\end{figure}

Let us turn to calculation of the coefficients $A_{+}^{}(k,p)$ and $A_{-}^{}(k,p)$. Using the coordinate redefinition
\begin{equation}
z=\frac{k}{a}e^{-au},
\end{equation}
the integral in \eqref{A+def} can be rewritten as
\begin{equation}
\int\limits_{-\infty}^{\infty}e^{i\frac{k}{a}e^{-au}}e^{ipu}du=\frac{1}{a}\,e^{-i\frac{p}{a}\ln\left(\frac{a}{k}\right)}
\int\limits_{0}^{\infty}e^{iz}z^{-1-i\frac{p}{a}}dz.
\end{equation}
At this step it is necessary to introduce the regularization
\begin{equation}\label{regularization}
p\to p+ia\epsilon,\qquad \epsilon>0,
\end{equation}
leading to
\begin{equation}
\int\limits_{0}^{\infty}e^{iz}z^{\epsilon-1-i\frac{p}{a}}dz.
\end{equation}
In order to calculate the latter integral, it is convenient to go into the complex plane and to consider the contour presented on the left plot in Fig.~\ref{fig2},
resulting in
\begin{equation}
0=\int\limits_{\rho}^{\infty}e^{iz}z^{\epsilon-1-i\frac{p}{a}}dz+i\int\limits_{\infty}^{\rho}e^{-y}(iy)^{\epsilon-1-i\frac{p}{a}}dy
+i\int\limits_{\frac{\pi}{2}}^{0}e^{i\rho e^{i\varphi}}\left(\rho e^{i\varphi}\right)^{\epsilon-1-i\frac{p}{a}}\rho e^{i\varphi}d\varphi.
\end{equation}
One can check that for $\rho\to 0$
\begin{equation}
i\rho^{\epsilon}e^{-i\frac{p}{a}\ln(\rho)}\int\limits_{\frac{\pi}{2}}^{0}e^{\left(i\epsilon+\frac{p}{a}\right)\varphi}d\varphi\to 0,
\end{equation}
whereas
\begin{equation}
\lim\limits_{\rho\to 0}\int\limits_{\rho}^{\infty}e^{iz}z^{\epsilon-1-i\frac{p}{a}}dz=
e^{i\frac{\pi}{2}\epsilon+\frac{\pi p}{2a}}\int\limits_{0}^{\infty}e^{-y}y^{\epsilon-i\frac{p}{a}-1}dy
=e^{i\frac{\pi}{2}\epsilon+\frac{\pi p}{2a}}\,\Gamma\left(\epsilon-i\frac{p}{a}\right).
\end{equation}
Finally, for $\epsilon=0$ we get
\begin{equation}
e^{\frac{\pi p}{2a}}\Gamma\left(-i\frac{p}{a}\right),
\end{equation}
resulting in \eqref{A+kp}.

A fully analogous procedure can be used to calculate $A_{-}^{}(k,p)$. Using the coordinate redefinition
\begin{equation}
z=\frac{|k|}{a}e^{av},
\end{equation}
the integral in \eqref{A-def} can be rewritten as
\begin{equation}
\int\limits_{-\infty}^{\infty}e^{i\frac{k}{a}e^{av}}e^{-ipv}dv=\int\limits_{-\infty}^{\infty}e^{-i\frac{|k|}{a}e^{av}}e^{-ipv}dv
=\frac{1}{a}\,e^{-i\frac{p}{a}\ln\left(\frac{a}{|k|}\right)}
\int\limits_{0}^{\infty}e^{-iz}z^{-1-i\frac{p}{a}}dz.
\end{equation}
Using regularization \eqref{regularization} and the contour presented on the right plot in Fig.~\ref{fig2}, one gets
\begin{equation}
0=\int\limits_{\rho}^{\infty}e^{-iz}z^{\epsilon-1-i\frac{p}{a}}dz-i\int\limits_{\infty}^{\rho}e^{-y}(-iy)^{\epsilon-1-i\frac{p}{a}}dy
+i\int\limits_{-\frac{\pi}{2}}^{0}e^{-i\rho e^{i\varphi}}\rho^{\epsilon-i\frac{p}{a}}e^{i\varphi\left(\epsilon-i\frac{p}{a}\right)}d\varphi.
\end{equation}
One can check that for $\rho\to 0$
\begin{equation}
i\rho^{\epsilon}e^{-i\frac{p}{a}\ln(\rho)}\int\limits_{-\frac{\pi}{2}}^{0}e^{\left(i\epsilon+\frac{p}{a}\right)\varphi}d\varphi\to 0,
\end{equation}
whereas
\begin{equation}
\int\limits_{0}^{\infty}e^{-iz}z^{\epsilon-1-i\frac{p}{a}}dz
=e^{-i\frac{\pi}{2}\epsilon-\frac{\pi p}{2a}}\int\limits_{0}^{\infty}e^{-y}y^{\epsilon-1-i\frac{p}{a}}dy
=e^{-i\frac{\pi}{2}\epsilon-\frac{\pi p}{2a}}\,\Gamma\left(\epsilon-i\frac{p}{a}\right).
\end{equation}
Finally, for $\epsilon=0$ we get
\begin{equation}
e^{-\frac{\pi p}{2a}}\Gamma\left(-i\frac{p}{a}\right),
\end{equation}
resulting in \eqref{A-kp}.

\section*{Appendix~B: ``Extra'' commutation relations in the two-dimensional case}
\addcontentsline{toc}{section}{Appendix~B: ``Extra'' commutation relations in the two-dimensional case}
\setcounter{equation}{0}
\renewcommand{\theequation}{B\arabic{equation}}
Let us check that commutation relations \eqref{CCR-MR}--\eqref{CCR-MR3} indeed hold. Let us start with relation \eqref{CCR-MR2a}. After straightforward calculations, one gets (here the contributions with $k>0$ and $k<0$ are separated)
\begin{align}\nonumber
&\left[\phi_{M}(t,x),\phi_{M}(t,x')\right]=\int\limits_{0}^{\infty}\frac{dk}{4\pi k}\,e^{i\frac{k}{a}e^{-at}\left(e^{ax}-e^{ax'}\right)}
-\int\limits_{0}^{\infty}\frac{dk}{4\pi k}\,e^{-i\frac{k}{a}e^{-at}\left(e^{ax}-e^{ax'}\right)}\\\nonumber
&+\int\limits_{-\infty}^{0}\frac{dk}{4\pi k}\,e^{-i\frac{k}{a}e^{at}\left(e^{ax}-e^{ax'}\right)}
-\int\limits_{-\infty}^{0}\frac{dk}{4\pi k}\,e^{i\frac{k}{a}e^{at}\left(e^{ax}-e^{ax'}\right)}\\\label{CCR2Da1}
&=\frac{i}{2\pi}\int\limits_{0}^{\infty}\frac{dk}{k}\,\sin\left(\frac{k}{a}e^{-at}\left(e^{ax}-e^{ax'}\right)\right)
-\frac{i}{2\pi}\int\limits_{-\infty}^{0}\frac{dk}{k}\,\sin\left(\frac{k}{a}e^{at}\left(e^{ax}-e^{ax'}\right)\right).
\end{align}
By changing $k\to-k$ in the last integral of \eqref{CCR2Da1}, one gets
\begin{align}\nonumber
&\left[\phi_{M}(t,x),\phi_{M}(t,x')\right]=\frac{i}{2\pi}\int\limits_{0}^{\infty}\frac{dk}{k}\,\sin\left(\frac{k}{a}e^{-at}\left(e^{ax}-e^{ax'}\right)\right)
-\frac{i}{2\pi}\int\limits_{0}^{\infty}\frac{dk}{k}\,\sin\left(\frac{k}{a}e^{at}\left(e^{ax}-e^{ax'}\right)\right)\\\label{CCR2Da1b}
&=\frac{i}{4}\,\textrm{sign}\left(\frac{e^{-at}}{a}\left(e^{ax}-e^{ax'}\right)\right)
-\frac{i}{4}\,\textrm{sign}\left(\frac{e^{at}}{a}\left(e^{ax}-e^{ax'}\right)\right)=0,
\end{align}
where the prescription $\textrm{sign}(0)=0$ is used. We see that commutation relation \eqref{CCR-MR2a} holds.

\begin{figure}[ht]
\centering
\begin{minipage}{.49\textwidth}
\centering
\includegraphics[width=0.8\linewidth]{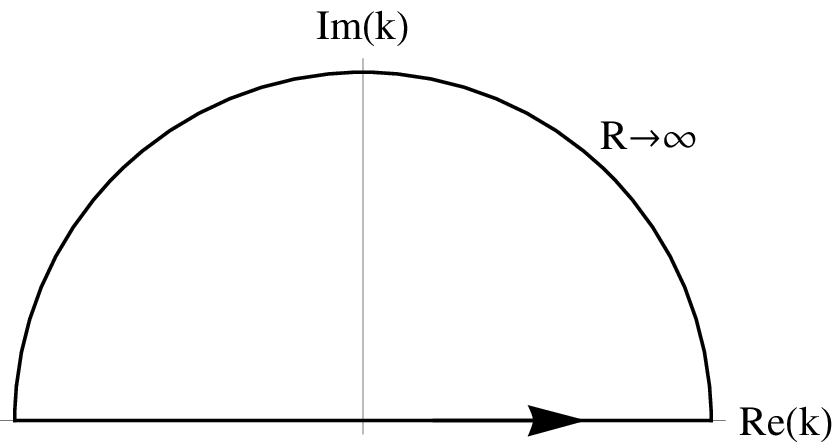}
\end{minipage}
\begin{minipage}{.49\textwidth}
\centering
\includegraphics[width=0.8\linewidth]{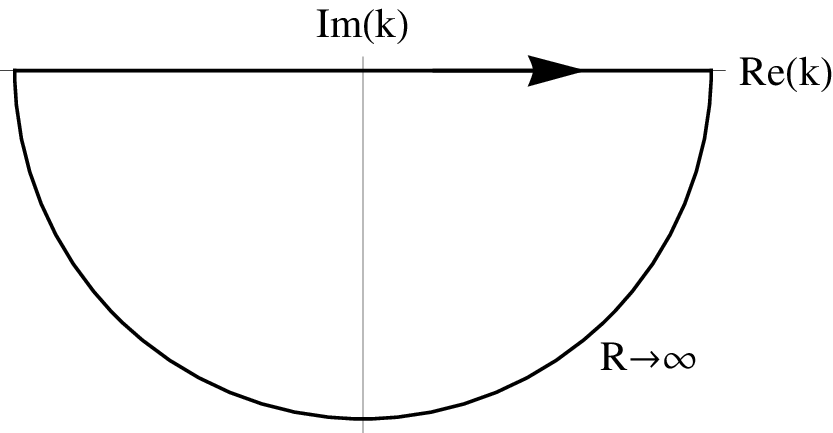}
\end{minipage}
\caption{The contours used to calculate integral \eqref{intwithC} for $C>0$ (left plot) and $C<0$ (right plot).}\label{fig3}
\end{figure}

Now let us turn to the commutator in \eqref{CCR-MR3}, which can be rewritten in explicit form as
\begin{align}\nonumber
&\left[\frac{\partial\phi_{M}(t,x)}{\partial t},\frac{\partial\phi_{M}(t,x')}{\partial t}\right]\\\nonumber
&=\frac{e^{-2at}e^{a(x+x')}}{4\pi}\left(\int\limits_{0}^{\infty}dk\,k\,e^{i\frac{k}{a}e^{-at}\left(e^{ax}-e^{ax'}\right)}
-\int\limits_{0}^{\infty}dk\,k\,e^{-i\frac{k}{a}e^{-at}\left(e^{ax}-e^{ax'}\right)}\right)\\\label{CCR2Db1}
&-\frac{e^{2at}e^{a(x+x')}}{4\pi}\left(\int\limits_{-\infty}^{0}dk\,k\,e^{i\frac{k}{a}e^{at}\left(e^{ax}-e^{ax'}\right)}
-\int\limits_{-\infty}^{0}dk\,k\,e^{-i\frac{k}{a}e^{at}\left(e^{ax}-e^{ax'}\right)}\right).
\end{align}
It is clear that for $x=x'$
\begin{equation}
\left[\frac{\partial\phi_{M}(t,x)}{\partial t},\frac{\partial\phi_{M}(t,x)}{\partial t}\right]=0.
\end{equation}
For $x\neq x'$, by changing $k\to -k$ in the second and forth integrals in \eqref{CCR2Db1}, one can get
\begin{align}\nonumber
&\left[\frac{\partial\phi_{M}(t,x)}{\partial t},\frac{\partial\phi_{M}(t,x')}{\partial t}\right]\\\label{CCR2Db2}
&=\frac{e^{-2at}e^{a(x+x')}}{4\pi}\int\limits_{-\infty}^{\infty}dk\,k\,e^{i\frac{k}{a}e^{-at}\left(e^{ax}-e^{ax'}\right)}
-\frac{e^{2at}e^{a(x+x')}}{4\pi}\int\limits_{-\infty}^{\infty}dk\,k\,e^{i\frac{k}{a}e^{at}\left(e^{ax}-e^{ax'}\right)}.
\end{align}
Calculation of the integrals in \eqref{CCR2Db2} reduces to calculation of the integral
\begin{equation}\label{intwithC}
\int\limits_{-\infty}^{\infty}dk\,k\,e^{ikC}
\end{equation}
with real $C\neq 0$. For $C>0$ the integral can be easily calculated by moving into the complex plane and taking the contour presented on the left plot in Fig.~\ref{fig3}, leading to
\begin{equation}\label{equalityC}
\int\limits_{-\infty}^{\infty}dk\,k\,e^{ikC}=0.
\end{equation}
Analogous calculation can be performed for $C<0$, but now using the contour presented on the right plot in Fig.~\ref{fig3}.
Thus we get that equality \eqref{equalityC} is valid for any real $C$ (due to the symmetry reasons, formally \eqref{equalityC} is valid for $C=0$ as well). As a consequence, one gets
\begin{equation}\label{CCR2Db3}
\left[\frac{\partial\phi_{M}(t,x)}{\partial t},\frac{\partial\phi_{M}(t,x')}{\partial t}\right]=0.
\end{equation}

Finally, let us consider the commutator in \eqref{CCR-MR}, which can be rewritten in explicit form as
\begin{align}\nonumber
&\left[\phi_{M}(t,x),\frac{\partial\phi_{M}(t,x')}{\partial t}\right]=\frac{ie^{-at}e^{ax'}}{4\pi}\left(\int\limits_{0}^{\infty}dk\,e^{i\frac{k}{a}e^{-at}\left(e^{ax}-e^{ax'}\right)}
+\int\limits_{0}^{\infty}dk\,e^{-i\frac{k}{a}e^{-at}\left(e^{ax}-e^{ax'}\right)}\right)\\\label{CCR2Dc1}
&+\frac{ie^{at}e^{ax'}}{4\pi}\left(\int\limits_{-\infty}^{0}dk\,e^{i\frac{k}{a}e^{at}\left(e^{ax}-e^{ax'}\right)}
+\int\limits_{-\infty}^{0}dk\,e^{-i\frac{k}{a}e^{at}\left(e^{ax}-e^{ax'}\right)}\right).
\end{align}
By changing $k\to -k$ in the second and forth integrals in \eqref{CCR2Dc1}, one can get
\begin{align}\nonumber
&\left[\phi_{M}(t,x),\frac{\partial\phi_{M}(t,x')}{\partial t}\right]=\frac{ie^{-at}e^{ax'}}{4\pi}\int\limits_{-\infty}^{\infty}dk\,e^{i\frac{k}{a}e^{-at}\left(e^{ax}-e^{ax'}\right)}
+\frac{ie^{at}e^{ax'}}{4\pi}\int\limits_{-\infty}^{\infty}dk\,e^{i\frac{k}{a}e^{at}\left(e^{ax}-e^{ax'}\right)}\\\nonumber
&=\frac{ie^{-at}e^{ax'}}{2}\,\delta\left(\frac{e^{-at}}{a}\left(e^{ax}-e^{ax'}\right)\right)+
\frac{ie^{at}e^{ax'}}{2}\,\delta\left(\frac{e^{at}}{a}\left(e^{ax}-e^{ax'}\right)\right)=iae^{ax'}\delta(e^{ax}-e^{ax'})\\\label{CCR2Dc2}
&=ia\delta(e^{a(x-x')}-1)=ia\frac{\delta(x-x')}{ae^{a(x-x')}}=i\delta(x-x').
\end{align}
One can see that commutation relation \eqref{CCR-MR} also holds.

\section*{Appendix~C: Bogolyubov coefficients in the four-dimen\-sional case}
\addcontentsline{toc}{section}{Appendix~C: Bogolyubov coefficients in the four-dimensional case}
\setcounter{equation}{0}
\renewcommand{\theequation}{C\arabic{equation}}
As in the two-dimensional case, we start with the equality
\begin{equation}\label{4DMRequality}
\phi_{R}(t,\vec x)\equiv \phi_{M}(t,\vec x),
\end{equation}
where $\phi_{R}(t,\vec x)$ is defined by \eqref{scfieldRmassive} and $\phi_{M}(t,\vec x)$ is defined by \eqref{phiM4Dtx}. Let us multiply \eqref{4DMRequality} by $e^{i\epsilon't}e^{-i\vec k'_{\perp}\vec x_{\perp}}$ with $\epsilon'>0$ and integrate over $t$ and $\vec x_{\perp}$. The result is
\begingroup
\allowdisplaybreaks
\begin{align}\nonumber
&\frac{\Psi(\epsilon',\vec k'_{\perp},x)}{\sqrt{2\epsilon'}}\,b^{}(\epsilon',\vec k'_{\perp})\\\nonumber&
=\int\limits_{-\infty}^{\infty}\frac{dk}{(2\pi)^{\frac{3}{2}}\sqrt{2\sqrt{k^{2}+{\vec k_{\perp}}^{'2}}}}
\int\limits_{-\infty}^{\infty}dt\Biggl(e^{-i\frac{\sqrt{k^{2}+{\vec k_{\perp}}^{'2}}}{a}\,e^{ax}\sinh(at)+
i\frac{k}{a}\,e^{ax}\cosh(at)}e^{i\epsilon't}a^{}(k,\vec k'_{\perp})\\\label{BC4D1}&
+e^{i\frac{\sqrt{k^{2}+{\vec k_{\perp}}^{'2}}}{a}\,e^{ax}\sinh(at)-i\frac{k}{a}\,e^{ax}\cosh(at)}e^{i\epsilon't}a^{\dagger}(k,-\vec k'_{\perp})\Biggr).
\end{align}
\endgroup
In what follows, just for convenience, let us change the notations $\epsilon'\to\epsilon$ and $\vec k'_{\perp}\to\vec k_{\perp}$. Let us pass from the variable $k$ to the variable $q$ defined as
\begin{equation}\label{substq14D}
\sinh(q)=\frac{k}{|\vec k_{\perp}|}.
\end{equation}
In such a case,
\begin{equation}\label{substq24D}
\cosh(q)=\frac{\sqrt{k^{2}+{\vec k_{\perp}}^{2}}}{|\vec k_{\perp}|}.
\end{equation}
Formally, formulas \eqref{substq14D} and \eqref{substq24D} are valid only for $|\vec k_{\perp}|\neq 0$. However, contributions of the integrands with $|\vec k_{\perp}|=0$ to the corresponding integrals are of the measure zero, so it is not necessary to consider the case $|\vec k_{\perp}|=0$ separately.

With \eqref{substq14D} and \eqref{substq24D}, the integral over $t$ connected with $a^{}(k,\vec k'_{\perp})$ in relation \eqref{BC4D1} takes the form
\begin{align}\nonumber
&\int\limits_{-\infty}^{\infty} e^{-i\frac{\sqrt{k^{2}+{\vec k_{\perp}}^{2}}}{a}\,e^{ax}\sinh(at)+
i\frac{k}{a}\,e^{ax}\cosh(at)}e^{i\epsilon t}\,dt
=\int\limits_{-\infty}^{\infty}e^{-i\frac{|\vec k_{\perp}|}{a}\,e^{ax}\sinh(at-q)}e^{i\epsilon t}\,dt\\\label{BCintt1}&
=\frac{e^{i\frac{\epsilon}{a}\,q}}{a}\int\limits_{-\infty}^{\infty}e^{-i\frac{|\vec k_{\perp}|}{a}\,e^{ax}\sinh(z)}e^{i\frac{\epsilon}{a}z}\,dz,
\end{align}
where $z=at-q$.

Recall that the following relations hold \cite{BE}:
\begin{align}\label{BE1}
&K_{\nu}(y)\cos\left(\frac{\nu\pi}{2}\right)=\int\limits_{0}^{\infty}\cos(y\sinh(z))\cosh(\nu z)\,dz,\\\label{BE2}
&K_{\nu}(y)\sin\left(\frac{\nu\pi}{2}\right)=\int\limits_{0}^{\infty}\sin(y\sinh(z))\sinh(\nu z)\,dz,
\end{align}
where $y>0$ and $-1<\textrm{Re}\nu<1$. Using \eqref{BE1} and \eqref{BE2}, one can easily show that
\begin{equation}\label{BCintt1K1}
K_{\nu}(y)\cos\left(\frac{\nu\pi}{2}\right)-iK_{\nu}(y)\sin\left(\frac{\nu\pi}{2}\right)
=\frac{1}{2}\int\limits_{-\infty}^{\infty}e^{-iy\sinh(z)}e^{\nu z}dz.
\end{equation}
On the other hand,
\begin{equation}\label{BCintt1K2}
K_{\nu}(y)\cos\left(\frac{\nu\pi}{2}\right)-iK_{\nu}(y)\sin\left(\frac{\nu\pi}{2}\right)
=e^{-i\frac{\nu\pi}{2}}K_{\nu}(y).
\end{equation}
Combining \eqref{BCintt1K1} and \eqref{BCintt1K2}, we get
\begin{equation}\label{BCintt1K3}
\int\limits_{-\infty}^{\infty}e^{-iy\sinh(z)}e^{\nu z}dz=2\,e^{-i\frac{\nu\pi}{2}}K_{\nu}(y).
\end{equation}
Using \eqref{BCintt1K3} and \eqref{Psi4DRind}, the integral in \eqref{BCintt1} can be rewritten as
\begin{equation}\label{BCintt12}
\frac{e^{i\frac{\epsilon}{a}\,q}}{a}\int\limits_{-\infty}^{\infty}e^{-i\frac{|\vec k_{\perp}|}{a}\,e^{ax}\sinh(z)}e^{i\frac{\epsilon}{a}z}\,dz
=\frac{2}{a}\,e^{i\frac{\epsilon}{a}\,q}e^{\frac{\pi \epsilon}{2a}}K_{i\frac{\epsilon}{a}}\left(\frac{|\vec k_{\perp}|}{a}\,e^{ax}\right)
=\frac{2\pi e^{i\frac{\epsilon}{a}\,q}e^{\frac{\pi \epsilon}{2a}}}{\sqrt{2\epsilon a\sinh\left(\pi\frac{\epsilon}{a}\right)}}\Psi(\epsilon,\vec k_{\perp},x).
\end{equation}

The integral over $t$ connected with $a^{\dagger}(k,\vec k'_{\perp})$ in relation \eqref{BC4D1} takes the form
\begin{equation}\label{BCintt2}
\int\limits_{-\infty}^{\infty} e^{i\frac{\sqrt{k^{2}+{\vec k_{\perp}}^{2}}}{a}\,e^{ax}\sinh(at)-
i\frac{k}{a}\,e^{ax}\cosh(at)}e^{i\epsilon t}\,dt
=\frac{e^{i\frac{\epsilon}{a}\,q}}{a}\int\limits_{-\infty}^{\infty}e^{i\frac{|\vec k_{\perp}|}{a}\,e^{ax}\sinh(z)}e^{i\frac{\epsilon}{a}z}\,dz.
\end{equation}
Using \eqref{BE1} and \eqref{BE2}, in the same way as how the relation \eqref{BCintt1K3} was obtained, one can easily show that
\begin{equation}\label{BCintt2K3}
\int\limits_{-\infty}^{\infty}e^{iy\sinh(z)}e^{\nu z}dz=2\,e^{i\frac{\nu\pi}{2}}K_{\nu}(y).
\end{equation}
Using \eqref{BCintt2K3} and \eqref{Psi4DRind}, the integral in \eqref{BCintt2} can be rewritten as
\begin{equation}\label{BCintt22}
\frac{e^{i\frac{\epsilon}{a}\,q}}{a}\int\limits_{-\infty}^{\infty}e^{i\frac{|\vec k_{\perp}|}{a}\,e^{ax}\sinh(z)}e^{i\frac{\epsilon}{a}z}\,dz
=\frac{2\pi e^{i\frac{\epsilon}{a}\,q}e^{-\frac{\pi\epsilon}{2a}}}{\sqrt{2\epsilon a\sinh\left(\pi\frac{\epsilon}{a}\right)}}\Psi(\epsilon,\vec k_{\perp},x).
\end{equation}
Substituting \eqref{BCintt12} and \eqref{BCintt22} into \eqref{BC4D1}, finally one can get \eqref{BC4Dfinal}.

It should be noted that a different derivation of these Bogolyubov coefficients can be found in \cite{Crispino:2007eb}.

\section*{Appendix~D: ``Extra'' commutation relations in the four-dimensional case}
\addcontentsline{toc}{section}{Appendix~D: ``Extra'' commutation relations in the four-dimensional case}
\setcounter{equation}{0}
\renewcommand{\theequation}{D\arabic{equation}}
Let us check that commutation relations \eqref{CCRExtra4D1}--\eqref{CCRExtra4D3} indeed hold. It is convenient to begin with the commutator
\begin{align}\nonumber
&\left[\phi_{M}(t,\vec x),\phi_{M}(t,\vec x')\right]\\\label{CCR-MR4Da14D}
&=\frac{1}{2(2\pi)^{3}}\iint d^{2}k_{\perp}\int\limits_{-\infty}^{\infty}\frac{dk}{E}\left(e^{\left(-\frac{iE}{a}\sinh(at)+\frac{ik}{a}\cosh(at)\right)\left(e^{ax}-e^{ax'}\right)}e^{i\vec k_{\perp}(\vec x_{\perp}-\vec x'_{\perp})}-\textrm{c.c.}\right),
\end{align}
where $E=\sqrt{k^{2}+{\vec k_{\perp}}^{2}}$. With \eqref{substq14D} and \eqref{substq24D} we get
\begin{equation}
\frac{dk}{E}=\frac{|\vec k_{\perp}|\cosh(q)\,dq}{|\vec k_{\perp}|\cosh(q)}=dq,
\end{equation}
\begin{equation}
-\frac{iE}{a}\sinh(at)+\frac{ik}{a}\cosh(at)=i\frac{|\vec k_{\perp}|}{a}\sinh(q-at),
\end{equation}
leading to
\begin{equation}\label{CCR-MR4Da24D}
\left[\phi_{M}(t,\vec x),\phi_{M}(t,\vec x')\right]
=\frac{1}{2(2\pi)^{3}}\int\limits_{-\infty}^{\infty}dq\iint d^{2}k_{\perp}\left(e^{i\frac{|\vec k_{\perp}|}{a}\sinh(q-at)\left(e^{ax}-e^{ax'}\right)}e^{i\vec k_{\perp}(\vec x_{\perp}-\vec x'_{\perp})}-\textrm{c.c.}\right).
\end{equation}
Let $q-at=\tilde q$. Then for \eqref{CCR-MR4Da24D} we get
\begin{align}\nonumber
&\left[\phi_{M}(t,\vec x),\phi_{M}(t,\vec x')\right]=\frac{1}{2(2\pi)^{3}}\int\limits_{-\infty}^{\infty}d\tilde q\iint d^{2}k_{\perp}\\\label{CCR-MR4Da34D}
&\times\left(e^{i\frac{|\vec k_{\perp}|}{a}\sinh(\tilde q)\left(e^{ax}-e^{ax'}\right)}e^{i\vec k_{\perp}(\vec x_{\perp}-\vec x'_{\perp})}-e^{-i\frac{|\vec k_{\perp}|}{a}\sinh(\tilde q)\left(e^{ax}-e^{ax'}\right)}e^{-i\vec k_{\perp}(\vec x_{\perp}-\vec x'_{\perp})}\right).
\end{align}
By changing $\tilde q\to -\tilde q$ and $\vec k_{\perp}\to -\vec k_{\perp}$ in the integrals involving the second term in the brackets in \eqref{CCR-MR4Da34D}, we arrive at
\begin{align}\nonumber
&\left[\phi_{M}(t,\vec x),\phi_{M}(t,\vec x')\right]=\frac{1}{2(2\pi)^{3}}\int\limits_{-\infty}^{\infty}d\tilde q\iint d^{2}k_{\perp}\\\label{CCR-MR4Da44D}
&\times\left(e^{i\frac{|\vec k_{\perp}|}{a}\sinh(\tilde q)\left(e^{ax}-e^{ax'}\right)}e^{i\vec k_{\perp}(\vec x_{\perp}-\vec x'_{\perp})}-e^{i\frac{|\vec k_{\perp}|}{a}\sinh(\tilde q)\left(e^{ax}-e^{ax'}\right)}e^{i\vec k_{\perp}(\vec x_{\perp}-\vec x'_{\perp})}\right)=0.
\end{align}

For the second commutation relation, performing analogous calculations (using \eqref{substq14D}, \eqref{substq24D}, and the substitution $q-at=\tilde q$), we can get
\begin{align}\nonumber
&\left[\frac{\partial\phi_{M}(t,\vec x)}{\partial t},\frac{\partial\phi_{M}(t,\vec x')}{\partial t}\right]=\frac{1}{2(2\pi)^{3}}\int\limits_{-\infty}^{\infty}d\tilde q\iint d^{2}k_{\perp}e^{a(x+x')}{\vec k_{\perp}}^{2}\cosh^{2}(\tilde q)
\\\label{CCR-MR4Db14D}&\times\left(e^{i\frac{|\vec k_{\perp}|}{a}\sinh(\tilde q)\left(e^{ax}-e^{ax'}\right)}e^{i\vec k_{\perp}(\vec x_{\perp}-\vec x'_{\perp})}
-e^{-i\frac{|\vec k_{\perp}|}{a}\sinh(\tilde q)\left(e^{ax}-e^{ax'}\right)}e^{-i\vec k_{\perp}(\vec x_{\perp}-\vec x'_{\perp})}\right).
\end{align}
As in the previous case, by changing $\tilde q\to -\tilde q$ and $\vec k_{\perp}\to -\vec k_{\perp}$ in the integrals involving the second term in the brackets in \eqref{CCR-MR4Db14D} we get
\begin{equation}\label{CCR-MR4Db24D}
\left[\frac{\partial\phi_{M}(t,\vec x)}{\partial t},\frac{\partial\phi_{M}(t,\vec x')}{\partial t}\right]=0.
\end{equation}

For the third commutation relation, performing analogous calculations (again using \eqref{substq14D}, \eqref{substq24D}, and the substitution $q-at=\tilde q$), we can get
\begin{align}\nonumber
&\left[\phi_{M}(t,\vec x),\frac{\partial\phi_{M}(t,\vec x')}{\partial t}\right]=\frac{i}{2(2\pi)^{3}}\int\limits_{-\infty}^{\infty}d\tilde q\iint d^{2}k_{\perp}e^{ax'}|\vec k_{\perp}|\cosh(\tilde q)
\\\label{CCR-MR4Dc14D}&\times\left(e^{i\frac{|\vec k_{\perp}|}{a}\sinh(\tilde q)\left(e^{ax}-e^{ax'}\right)}e^{i\vec k_{\perp}(\vec x_{\perp}-\vec x'_{\perp})}
+e^{-i\frac{|\vec k_{\perp}|}{a}\sinh(\tilde q)\left(e^{ax}-e^{ax'}\right)}e^{-i\vec k_{\perp}(\vec x_{\perp}-\vec x'_{\perp})}\right).
\end{align}
As in the previous cases, by changing $\tilde q\to -\tilde q$ and $\vec k_{\perp}\to -\vec k_{\perp}$ in the integrals involving the second term in the brackets in \eqref{CCR-MR4Dc14D} we get
\begin{equation}\label{CCR-MR4Dc24D}
\left[\phi_{M}(t,\vec x),\frac{\partial\phi_{M}(t,\vec x')}{\partial t}\right]=\frac{i}{(2\pi)^{3}}\int\limits_{-\infty}^{\infty}d\tilde q\iint d^{2}k_{\perp}e^{ax'}|\vec k_{\perp}|\cosh(\tilde q)
e^{i\frac{|\vec k_{\perp}|}{a}\sinh(\tilde q)\left(e^{ax}-e^{ax'}\right)}e^{i\vec k_{\perp}(\vec x_{\perp}-\vec x'_{\perp})}.
\end{equation}
Now let us define a new variable
\begin{equation}\label{substxi4D}
\xi=\frac{|\vec k_{\perp}|}{a}\sinh(\tilde q).
\end{equation}
With \eqref{substxi4D}, formula \eqref{CCR-MR4Dc24D} takes the form
\begin{equation}\label{CCR-MR4Dc34D}
\left[\phi_{M}(t,\vec x),\frac{\partial\phi_{M}(t,\vec x')}{\partial t}\right]=\frac{iae^{ax'}}{(2\pi)^{3}}\int\limits_{-\infty}^{\infty}d\xi\iint d^{2}k_{\perp}e^{i\xi\left(e^{ax}-e^{ax'}\right)}e^{i\vec k_{\perp}(\vec x_{\perp}-\vec x'_{\perp})}.
\end{equation}
From \eqref{CCR-MR4Dc34D} one easily gets
\begin{align}\nonumber
&\left[\phi_{M}(t,\vec x),\frac{\partial\phi_{M}(t,\vec x')}{\partial t}\right]=\frac{iae^{ax'}}{(2\pi)^{3}}\int\limits_{-\infty}^{\infty}d\xi e^{i\xi\left(e^{ax}-e^{ax'}\right)}\iint d^{2}k_{\perp}e^{i\vec k_{\perp}(\vec x_{\perp}-\vec x'_{\perp})}\\\nonumber
&=iae^{ax'}\delta(e^{ax}-e^{ax'})\delta^{(2)}(\vec x_{\perp}-\vec x'_{\perp})
=ia\delta(e^{a(x-x')}-1)\delta^{(2)}(\vec x_{\perp}-\vec x'_{\perp})\\\label{CCR-MR4Dc44D}&=ia\frac{\delta(x-x')}{ae^{a(x-x')}}\delta^{(2)}(\vec x_{\perp}-\vec x'_{\perp})
=\delta^{(3)}(\vec x-\vec x').
\end{align}

Thus we see that all the necessary commutation relations are satisfied.


\begin{thebibliography}{99}
\bibitem{Boulware:1974dm}
D.G.~Boulware, {\em ``Quantum field theory in Schwarzschild and Rindler spaces''},
Phys. Rev. D \textbf{11} (1975) 1404.

\bibitem{HH}
J.B.~Hartle, S.W.~Hawking, {\em ``Path-integral derivation of black-hole radiance''}, Phys. Rev. D \textbf{13} (1976) 2188.

\bibitem{Fulling:1972md}
S.A.~Fulling, {\em ``Nonuniqueness of canonical field quantization in Riemannian space-time''}, Phys. Rev. D \textbf{7} (1973) 2850-2862.

\bibitem{Davies:1974th}
P.C.W.~Davies, {\em ``Scalar particle production in Schwarzschild and Rindler metrics''}, J. Phys. A \textbf{8} (1975) 609-616.

\bibitem{Unruh:1976db}
W.G.~Unruh, {\em ``Notes on black hole evaporation''}, Phys. Rev. D \textbf{14} (1976) 870.

\bibitem{BD}
N.D.~Birrell, P.C.W.~Davies, {\em ``Quantum fields in curved space''}, Cambridge Univ. Press, 1984.

\bibitem{Traschen:1999zr}
J.H.~Traschen, {\em ``An Introduction to black hole evaporation''}, [arXiv:gr-qc/0010055 [gr-qc]].

\bibitem{Jacobson:2003vx}
T.~Jacobson, {\em ``Introduction to quantum fields in curved space-time and the Hawking effect''}, [arXiv:gr-qc/0308048 [gr-qc]].

\bibitem{Crispino:2007eb}
L.C.B.~Crispino, A.~Higuchi, G.E.A.~Matsas, {\em ``The Unruh effect and its applications''}, Rev. Mod. Phys. \textbf{80} (2008) 787-838 [arXiv:0710.5373 [gr-qc]].

\bibitem{Buoninfante:2024oxl}
L.~Buoninfante, R.~Carballo-Rubio, V.~Cardoso, F.~Di Filippo, A.~Eichhorn, N.~Afshordi, A.~Ashtekar, E.~Barausse, E.~Berti, R.~Brito, \textit{et al.}
{\em ``Black Holes Inside and Out 2024: visions for the future of black hole physics''}, [arXiv:2410.14414 [gr-qc]].

\bibitem{LL-FT}
L.D.~Landau, E.M.~Lifshitz, {\em ``The classical theory of fields''}, Butterworth-Heinemann (1987).

\bibitem{Lee:1985rp}
T.D.~Lee, {\em ``Are black holes black bodies?''}, Nucl. Phys. B \textbf{264} (1986) 437.

\bibitem{Korn-Korn}
G.A.~Korn, T.M.~Korn, {\em ``Mathematical Handbook for Scientists and Engineers''}, McGraw-Hill, Inc., New York (1968).

\bibitem{DeVos}
A.~De~Vos, {\em ``Thermodynamics of radiation energy conversion in one and in three physical dimensions''}, Phys. Chem. Solids, \textbf{49} (1988) 725.

\bibitem{LandsbergDeVos}
P.T.~Landsberg, A.~De~Vos, {\em ``The Stefan-Boltzmann constant in n-dimensional space''}, J. Phys. A: Math. Gen. \textbf{22} (1989) 1073.

\bibitem{Unruh:1983ms}
W.G.~Unruh, R.M.~Wald, {\em ``What happens when an accelerating observer detects a Rindler particle''}, Phys. Rev. D \textbf{29} (1984) 1047-1056.

\bibitem{Kruskal}
M.D.~Kruskal, {\em ``Maximal extension of Schwarzschild metric''}, Phys. Rev. \textbf{119} (1960) 1743.

\bibitem{Szekeres}
G.~Szekeres, {\em ``On the singularities of a Riemannian manifold''}, Publ. Math. Debrecen \textbf{7} (1960) 285.

\bibitem{Anderson:2022icz}
P.R.~Anderson, S.G.~Siahmazgi, Z.P.~Scofield, {\em ``Infrared effects and the Unruh state''}, Class. Quant. Grav. \textbf{40} (2023) 135004 [arXiv:2210.16397 [gr-qc]].

\bibitem{Balbinot:2023vcm}
R.~Balbinot, A.~Fabbri, {\em ``The Unruh Vacuum and the In-Vacuum in Reissner-Nordstr{\"o}m Spacetime''}, Universe \textbf{10} (2024) 18 [arXiv:2311.09943 [gr-qc]].

\bibitem{Sciama:1981hr}
D.W.~Sciama, P.~Candelas, D.~Deutsch, {\em ``Quantum field theory, horizons and thermodynamics''}, Adv. Phys. \textbf{30} (1981) 327.

\bibitem{Pena:2013zfd}
I.~Pe{\~n}a, D.~Sudarsky, {\em ``On the possibility of measuring the Unruh effect''}, Found. Phys. \textbf{44} (2014) 689-708 [arXiv:1306.6621 [quant-ph]].

\bibitem{Candelas:1977zza}
P.~Candelas, D.~Deutsch, {\em ``On the vacuum stress induced by uniform acceleration or supporting the ether''}, Proc. Roy. Soc. Lond. A \textbf{354} (1977) 79-99.

\bibitem{PBM}
A.P.~Prudnikov, Yu.A.~Brychkov, O.I.~Marichev, {\em ``Integrals and Series. Volume 1: Elementary Functions''}, Gordon and Breach, New York (1986).

\bibitem{Higuchi:1993fn}
A.~Higuchi, G.E.A.~Matsas, {\em ``Fulling-Davies-Unruh effect in classical field theory''}, Phys. Rev. D \textbf{48} (1993) 689-697.

\bibitem{Dwight}
H.B.~Dwight, {\em ``Tables of Integrals and other Mathematical Data''}, The Macmillan Company, New York (1961).

\bibitem{Christensen:1977jc}
S.~M.~Christensen, S.A.~Fulling, {\em ``Trace anomalies and the Hawking effect''}, Phys. Rev. D \textbf{15} (1977) 2088.

\bibitem{Barranco:2011eyw}
J.~Barranco, A.~Bernal, J.C.~Degollado, A.~Diez-Tejedor, M.~Megevand, M.~Alcubierre, D.~Nunez, O.~Sarbach,
{\em ``Are black holes a serious threat to scalar field dark matter models?''},
Phys. Rev. D \textbf{84} (2011) 083008 [arXiv:1108.0931 [gr-qc]].

\bibitem{Zecca3}
A.~Zecca, {\em ``Properties of radial equation of scalar field in Schwarzschild space-time''},
Il Nuovo Cim. B \textbf{124} (2009) 1251.

\bibitem{KRSV}
S.I.~Keizerov, E.R.~Rakhmetov, M.N.~Smolyakov, I.P.~Volobuev, {\em ``Exact solutions for the salar field in the gravitational field of spherically symmetric black holes''}, Phys. Part. Nuclei \textbf{56} (2025) 162.

\bibitem{LL-QM}
L.D.~Landau, E.M.~Lifshitz, {\em ``Quantum mechanics. Non-relativistic theory'', Second edition}, Pergamon press (1965).

\bibitem{Messiah}
A.~Messiah, {\em ``Quantum Mechanics. Volume 1''}, North-Holland Publishing Company, Amsterdam (1961).

\bibitem{Egorov:2022hgg}
V.~Egorov, M.~Smolyakov, I.~Volobuev, {\em ``Doubling of physical states in the quantum scalar field theory for a remote observer in the Schwarzschild spacetime''}, Phys. Rev. D \textbf{107} (2023) 025001 [arXiv:2209.02067 [gr-qc]].

\bibitem{Smolyakov:2023pml}
M.N.~Smolyakov, {\em ``Asymptotic behavior of solutions and spectrum of states in the quantum scalar field theory in the Schwarzschild spacetime''},
Phys. Rev. D \textbf{108} (2023) 105006 [arXiv:2309.06249 [gr-qc]].

\bibitem{Candelas:1980zt}
P.~Candelas, {\em ``Vacuum polarization in Schwarzschild space-time''}, Phys. Rev. D \textbf{21} (1980) 2185.

\bibitem{Fulling:1987otn}
S.A.~Fulling, S.N.M.~Ruijsenaars, {\em ``Temperature, periodicity and horizons''}, Phys. Rept. \textbf{152} (1987) 135-176.

\bibitem{IZ}
C.~Itzykson, J.-B.~Zuber, {\em ``Quantum Field Theory''}, McGraw-Hill, Inc., New York (1980).

\bibitem{PS}
M.~Peskin, D.~Schroeder, {\em ``An Introduction to Quantum Field Theory''}, Addison-Wesley, Reading (1995).

\bibitem{Fulling:1977zs}
S.A.~Fulling, {\em ``Alternative vacuum states in static space-times with horizons''},  J. Phys. A \textbf{10} (1977) 917.

\bibitem{BE}
H.~Bateman, A.~Erdelyi, {\em ``Higher Transcendental Functions. Volume 2''}, McGraw-Hill, Inc., New York (1953).

\end{thebibliography}
\end{document}